\documentclass[showpacs,preprint,preprintnumbers,amsmath,amssymb,floatfix]{revtex4}
\usepackage{graphicx}
\usepackage{dcolumn}
\usepackage{bm}

\begin{document}

\title{Magnetic field tuned quantum criticality of heavy fermion system YbPtBi}

\author{E. D. Mun$^{1}$, S. L. Bud'ko$^{1}$, C. Martin$^{1}$, H. Kim$^{1}$, M. A. Tanatar$^{1}$, J.-H. Park$^{2}$, T. Murphy$^{2}$, G. M. Schmiedeshoff$^{3}$, N.
Dilley$^{4}$, R. Prozorov$^{1}$, P. C. Canfield$^{1}$}
\affiliation{$^{1}$Ames Laboratory US DOE and Department of Physics and Astronomy, Iowa State University, Ames, Iowa 50011, USA}%
\affiliation{$^{2}$National High Magnetic Field Laboratory, Florida State University, Tallahassee, Florida 32310, USA}%
\affiliation{$^{3}$Department of Physics, Occidental College, Los Angeles, California 90041, USA}%
\affiliation{$^{4}$Quantum Design, 6325 Lusk Boulevard, San Diego, California 92121, USA}%


\begin{abstract}

In this paper, we present systematic measurements of the temperature and magnetic field dependences of the thermodynamic and transport properties of the Yb-based heavy fermion YbPtBi
for temperatures down to 0.02 K with magnetic fields up to 140\,kOe to address the possible existence of a field-tuned quantum critical point. Measurements of magnetic field and
temperature dependent resistivity, specific heat, thermal expansion, Hall effect, and thermoelectric power indicate that the AFM order can be suppressed by applied magnetic field of
$H_{c}\,\sim$ 4\,kOe. In the $H-T$ phase diagram of YbPtBi, three regimes of its low temperature states emerges: (I) AFM state, characterized by spin density wave (SDW) like feature,
which can be suppressed to $T$ = 0 by the relatively small magnetic field of $H_{c}$ $\sim$ 4\,kOe, (II) field induced anomalous state in which the electrical resistivity follows
$\Delta\rho(T)\,\propto\,T^{1.5}$ between $H_{c}$ and $\sim$\,8\,kOe, and (III) Fermi liquid (FL) state in which $\Delta\rho(T)\,\propto\,T^{2}$ for $H\,\geq$ 8\,kOe. Regions I and II
are separated at $T$ = 0 by what appears to be a quantum critical point. Whereas region III appears to be a FL associated with the hybridized 4$f$ states of Yb, region II may be a
manifestation of a spin liquid state.

\end{abstract}



\maketitle

\section{Introduction}

The face-centered cubic (fcc) YbPtBi is a member of $R$PtBi ($R$ = rare-earth) systems and one of the few stoichiometric Yb-based heavy fermion compounds \cite{Fisk1991,
Canfield1991}. An enormous low temperature Sommerfeld coefficient, $\gamma$ $\simeq$ 8\,J/mol$\cdot$K$^{2}$ \cite{Fisk1991}, which corresponds to one of highest effective mass values
among heavy fermion (HF) systems, is a characteristic of YbPtBi. This system manifests what is thought to be antiferromagnetic (AFM) ordering below $T_{N}$ = 0.4\,K, below the
estimated Kondo temperature of $T_{K}\,\sim$ 1\,K \cite{Fisk1991}. The results of electrical resistivity and specific heat measurements suggested that a spin density wave transition
occurs below $T_{N}$ \cite{Movshovich1994} with a small ordered moment of only $\sim$\,0.1\,$\mu_{B}$ or less that so far has prevented determination of the ordering wave vector
\cite{Amato1992, Robinson1994}. It has been proposed that the massive electronic state manages to appear due to either (1) the frustrated (for nearest neighbors) fcc crystal structure
suppressing long range order to below the Kondo temperature or (2) the low carrier density, metallic nature leading to an unusually low Kondo temperature \cite{Fisk1991, Hundley1997},
or both.

For an AFM quantum critical point (QCP) in HF systems the conventional theory, so-called spin density wave (SDW) scenario, considers itinerant $f$-electrons on both the ordered and
the paramagnetic side of the QCP \cite{Hertz1976, Millis1993, Moriya1995}. The critical SDW fluctuations are responsible for non-Fermi liquid (nFL) behavior in which the electrical
resistivity follows $\Delta\rho(T)\,\propto\,T^{n}$ with $n\,<$\,2 ($n$ = 1.5 for $d$ = 3 and $n$ = 1 for $d$ = 2). In this scenario, the quasi-particle effective mass is finite
$C(T)/T\,\propto\,-\sqrt{T}$ at QCP for $d$ = 3 critical fluctuations. For $d$ = 2 critical fluctuations, the theory predicts a logarithmic divergence of the effective mass
$C(T)/T\,\propto$\,-$\log$($T$). An essential aspect of the SDW scenario is that the characteristic energy scale, $T_{K}$, remains finite across the QCP, thus the quasi-particles
survive in the vicinity of the QCP \cite{Gegenwart2008}. An alternate scenario, so-called Kondo breakdown scenario, has proposed that a localization of the $f$-electrons at the QCP
gives rise to a breakdown of the local Kondo energy scale and a dramatic change of the Fermi surface topology \cite{Coleman2001, Si2001, Si2003, Senthil2004, Paul2008}. The SDW
scenario has been applied to several HF compounds such as CeCu$_{2}$Si$_{2}$ \cite{Gegenwart1998} and CeNi$_{2}$Ge$_{2}$ \cite{Julian1996} and the Kondo breakdown model seems to be
applicable to Au-doped CeCu$_{6-x}$Au$_{x}$ \cite{Schroder1998, Schroder2000} (specially called a local quantum criticality \cite{Si2001, Si2003}) and YbRh$_{2}$Si$_{2}$
\cite{Paschen2004, Gegenwart2008}. However, unfortunately, neither SDW nor the Kondo break down model are sufficient to explain the observed experimental results from these systems.

Magnetic field-induced AFM QCP systems have been limited to relatively few examples, only among stoichiometric compounds, in particular YbRh$_{2}$Si$_{2}$ \cite{Trovarelli2000,
Gegenwart2002, Paschen2004, Gegenwart2007, Friedemann2009} and YbAgGe \cite{Budko2004, Budko2005, Budko2005a, Niklowitz2006, Tokiwa2006, Schmiedeshoff2011}. In addition to strong
quantum fluctuations in the vicinity of the QCP, the existence of a new crossover field scale, apparently associated with the QCP, detected by several thermodynamic and transport
measurements, has emerged from the extensive study of YbRh$_{2}$Si$_{2}$ \cite{Paschen2004, Gegenwart2007, Friedemann2009} and YbAgGe \cite{Budko2004, Budko2005}. This crossover field
scale was associated with changes in Hall effect measurements \cite{Paschen2004}, interpreted as a change of the Fermi surface at the QCP, and more clearly seen in the other HF
antiferromagnet, YbAgGe, in an applied magnetic field of $\sim$\,45\,kOe \cite{Budko2005a, Niklowitz2006, Tokiwa2006}, particularly in Hall resistivity data \cite{Budko2005a} and
extended to higher temperatures via thermoelectric power \cite{Mun2010b} measurements. Among magnetic field-tuned QCP systems, YbAgGe shows a wide nFL region characterized by the
linear temperature dependence of the resistivity, $\Delta\rho\propto T$ \cite{Niklowitz2006}. Recently a similar range of nFL behavior has also been observed in Ge-doped
YbRh$_{2}$Si$_{2}$ \cite{Custers2010}. Mainly based on the magnetic field-tuned QCP systems, a new mechanism for quantum criticality has been proposed, one that considers two tuning
parameters \cite{Si2006, Si2010, Coleman2010, Custers2010}: (i) the ratio between the Kondo temperature and the Ruderman-Kittel-Kasuya-Yosida (RKKY) interaction and (ii) the quantum
zero-point fluctuations which can be tuned by increasing the amount of frustration. A Doniach-like \cite{Doniach1977}, two dimensional, phase diagram can be established with these two
tuning parameters. In order to better understand, and test, the details of this and other models of magnetic field-tuned quantum criticality, new, hopefully simpler, model
stoichiometric Yb-based systems are needed.

In this manuscript, we present systematic measurements of the thermodynamic and transport properties of YbPtBi down to 0.02 K with magnetic fields up to 140\,kOe to establish,
delineate, and understand the nature of magnetic field-induced QCP in this canonical system. In the constructed $H-T$ phase diagram for YbPtBi three low temperature regimes emerge:
(i) AFM state, characterized by signatures of a SDW, which can be suppressed to $T$ = 0 with a relatively small, external magnetic field of $H_{c}\,\sim$ 4\,kOe, (ii) a field induced,
anomalous state in which the electrical resistivity follows $\rho(T)\,\propto\,T^{1.5}$ between $H_{c}$ and $\sim$\,8\,kOe, and (iii) a Fermi liquid (FL) state in which
$\rho(T)\,\propto\,T^{2}$ for $H\,\geq$ 8\,kOe. Associated with these regions are two crossover scales, emerging near $H_{c}$ $\sim$ 4\,kOe and $H^{*}\,\sim$\,7.8\,kOe at $T$ = 0. For
$H\,>\,H^{*}$, the FL coefficient $A$ of the temperature dependence of resistivity and $\gamma$ the linear component of the temperature dependence of specific heat are drastically
enhanced as $\sim$ 1/($H - H_{c}$) and $\sim$ 1/($H - H_{c}$)$^{2}$, respectively, when approaching $H_{c}$ from the high magnetic field side. In contrast to the resistivity results,
the electronic specific heat coefficient, $C(T)/T$, does not show any pronounced nFL behavior as either $C(T)/T\,\propto\,-\sqrt{T}$ or -$\log(T)$ down to 0.05 K near $H_{c}$ and
$H^{*}$.

\section{Experimental}

Single crystals of YbPtBi and LuPtBi were grown out of a Bi-rich ternary melt \cite{Canfield1991, Canfield1992, Canfield2010}. The crystals were characterized by powder x-ray
diffraction measurements, collected at room temperature on a Rigaku MiniFlex. The determined lattice parameters and space group are in agreement with earlier studies
\cite{Canfield1991, Robinson1994}; MgAgAs structure type, space group $F$$\overline{4}$3m, Z = 4. The electrical resistivity, $\rho(T, H)$, and Hall resistivity, $\rho_{H}(T, H)$,
measurements as functions of temperature (0.02 - 300\,K) and magnetic field (0 - 140\,kOe) were performed by ordinary ac ($f$ = 16\,Hz) four-probe methods. Below 1\,K, $\rho(T, H)$
and $\rho_{H}(T, H)$ were measured in an Oxford Instrument $^{3}$He-$^{4}$He dilution refrigerator with a Lakeshore LS370 and a Linear Research LR700 ac resistance bridges. In order
to reduce heating effects, the excitation current, $I$, was selected as low as possible, 10-30 $\mu$A, and the magnetic field was swept very slowly, 100-500 Oe/min. Above 0.4\,K,
$\rho(T, H)$ and $\rho_{H}(T, H)$ were measured in a Quantum Design (QD) Physical Property Measurements System (PPMS) with ${^3}$He option. The magnetoresistance (MR) measurements
were performed in a transverse configuration (\textbf{I} $\perp$ \textbf{H}): \textbf{I} $\parallel$ [010] and \textbf{H} $\parallel$ [100]. The Hall resistivity was measured with the
following configuration: the Hall voltage is perpendicular to the current and magnetic field (\textbf{V}$_{H}$ $\perp$ \textbf{I} $\perp$ \textbf{H}), where \textbf{I} $\parallel$
[010] and \textbf{H} $\parallel$ [100]. In order to remove MR contributions in $\rho_{H}$ due to the misalignments of the Hall voltage wires, the polarity of magnetic field was
switched. For LuPtBi $\rho(T)$ and $\rho_{H}(T)$ measurements were performed with \textbf{H} $\parallel$ [111], \textbf{I} $\perp$ [111], and \textbf{H} $\perp$ \textbf{I} $\perp$
\textbf{V}$_{H}$ configuration.

The specific heat was measured in a PPMS with a $^3$He option by the relaxation method in the temperature range between 0.4 and 100\,K and magnetic fields up to 140\,kOe, applied along
the [100] direction. The specific heat measurements were extended down to 0.05\,K using a PPMS with dilution refrigerator option, at the Quantum Design headquarters in San Diego,
California. The DC magnetization as a function of temperature from 1.8 to 300 K, and magnetic fields, up to 70 kOe, was measured in a QD Magnetic Property Measurement System. Thermal
expansion and magnetostriction were measured using a capacitive dilatometer \cite{Schmiedeshoff2006} constructed of copper for the ${^3}$He-setup and titanium for the dilution
refrigerator-setup. The dilatometer was mounted in a ${^3}$He cryostat and was operated over a temperature range of 0.3 - 300\,K and magnetic fields up to 90\,kOe. The
magnetostriction measurements were extended to temperatures down to 0.02\,K and magnetic fields up to 180\,kOe in a $^{3}$He-$^{4}$He dilution refrigerator at the Millikelvin Facility,
High Magnetic Field National Laboratory, Tallahassee, Florida. The variation of the sample length was measured in the longitudinal configuration, $\Delta L$ $\parallel$ \textbf{H}
$\parallel$ [100]. Thermoelectric power (TEP) measurements were carried out using a dc, alternating heating, (two-heater-two-thermometer) technique \cite{Mun2010} over the temperature
range from 0.35 to 300 K and magnetic fields up to 140 kOe. The heat current was generated along $\Delta T\,\parallel$ [010] and the temperature difference, $\Delta T$, was kept
between 0.03 to 0.05\,K below 2\,K. The magnetic field was applied along \textbf{H} $\parallel$ [100], maintaining a transverse configuration with heat current; \textbf{H} $\perp$
$\Delta T$. For LuPtBi TEP was measured in a transverse configuration with $\Delta T$ $\perp$ [111] and \textbf{H} $\parallel$ [111] configuration.

\section{RESULTS}

\subsection{Magnetization}

\begin{figure*}%
\centering
\includegraphics[width=0.5\linewidth]{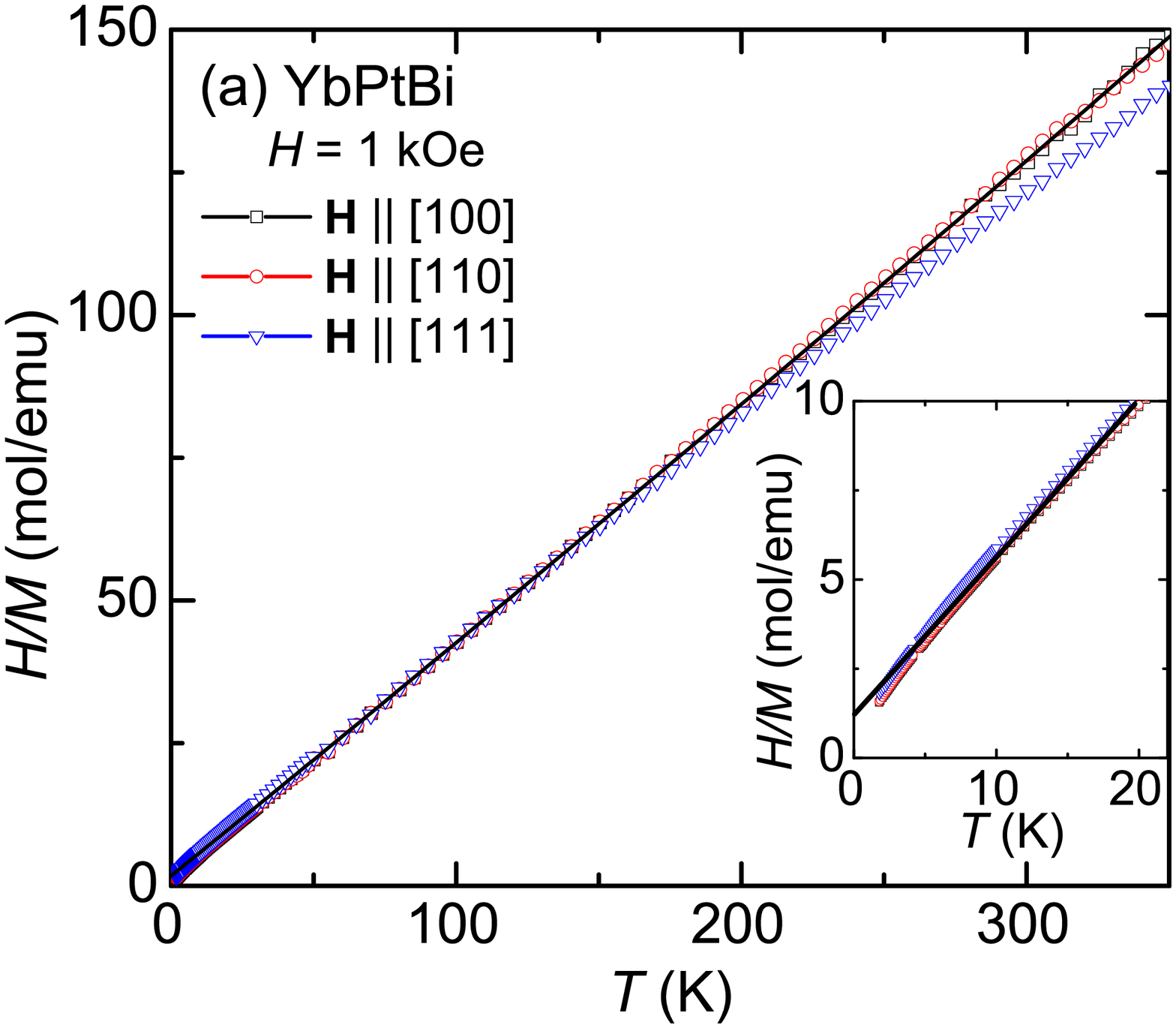}\includegraphics[width=0.5\linewidth]{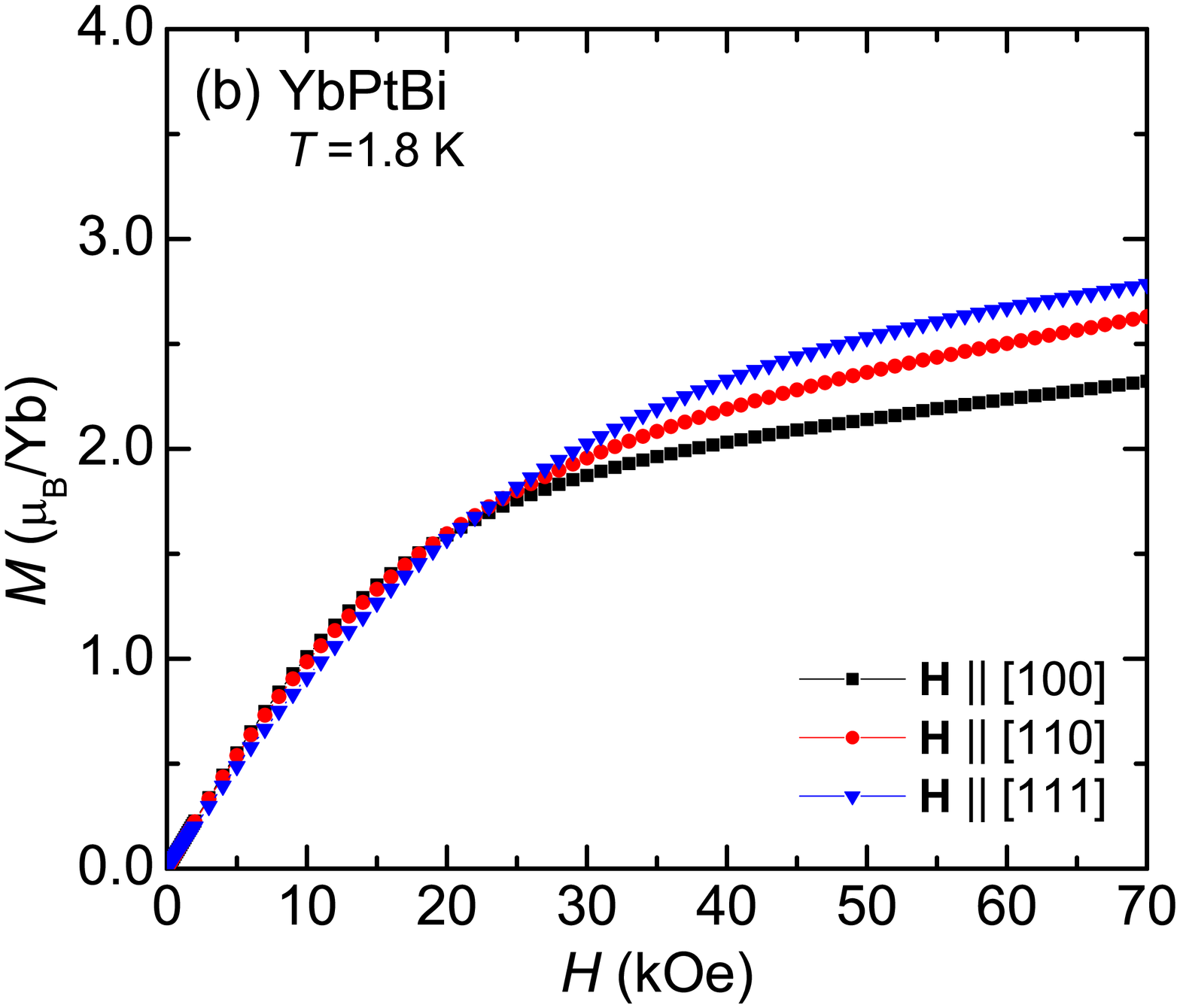}
\caption[Magnetic susceptibilities and magnetization isotherms of YbPtBi]{(a) Inverse magnetic susceptibility, $H/M(T)$, of YbPtBi, where the magnetic field was applied along [100],
[110], and [111] directions. The solid line represents the Curie-Weiss fit to the data for \textbf{H}\,$\parallel$\,[100]. Inset displays $H/M(T)$ at low temperatures. (b)
Magnetization isotherms of YbPtBi at $T$ = 1.8\,K for \textbf{H}\,$\parallel$\,[100], [110], and [111] direction.}
\label{YbPtBiMTMH}%
\end{figure*}

The anisotropic inverse magnetic susceptibilities, $H/M(T)$, of YbPtBi are shown in Fig. \ref{YbPtBiMTMH} (a), where the magnetic field was applied along the [100], [110], and [111]
directions. The observed magnetic susceptibility is essentially isotropic down to 2\,K. Between 10\,K and 250\,K, $H/M(T)$ obeys the Curie-Weiss law, $\chi(T)$ = C/($T - \theta_{p}$),
with $\theta_{p}$ $\simeq$ -2\,K and $\mu_{eff}$ $\simeq$ 4.3\,$\mu_{B}$/Yb$^{3+}$ which is close to the free ion value of 4.5$\mu_{B}$ and consistent with earlier studies
\cite{Fisk1991}. Magnetization isotherms, $M(H)$, of YbPtBi were measured at 1.8\,K for the magnetic field applied along the [100], [110], and [111] directions as shown in Fig.
\ref{YbPtBiMTMH} (b). The magnetic moment develops a modest anisotropy for $H$\,$>$ 25\,kOe at 1.8\,K and reaches values between 2.3-2.8 $\mu_{B}$/Yb$^{3+}$ at 70\,kOe, depending on
the magnetic field orientations, all of which are below the theoretical saturated value of 4\,$\mu_{B}$ for the free Yb$^{3+}$ ion that is expected due to the Kondo and crystalline
electric field (CEF) effects.

\subsection{Resistivity}

Earlier studies of the low temperature resistivity of YbPtBi found that below $T_{N}$ $\sim$ 0.4\,K an unexpected sample-to-sample variation of the resistive anomaly, and even an
apparent anisotropy, could develop. It was speculated that strain associated with the sample mounting and hypothesized magnetoelastic effects could be responsible for these
observations \cite{Lacerda1993}. Before embarking on extensive detailed transport measurements, we decided to examine this in detail.

\begin{figure}%
\centering
\includegraphics[width=0.5\linewidth]{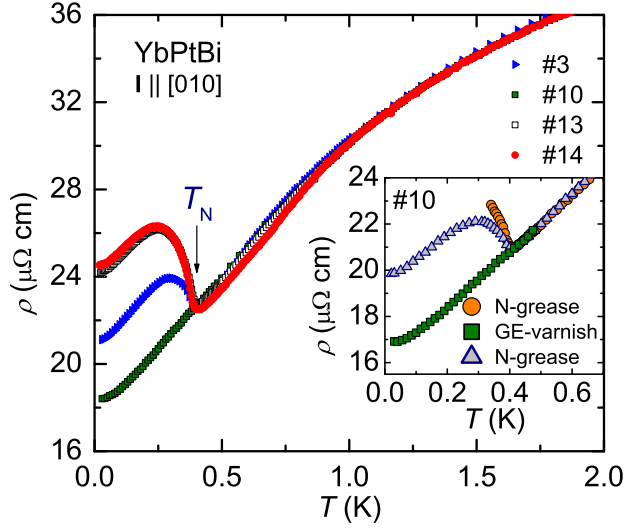}
\caption[$\rho(T)$ of YbPtBi for several different measurements conditions]{Temperature-dependent electrical resistivity, $\rho(T)$, of YbPtBi measured for different sample mount
conditions for cooling. $\rho(T)$ curves are normalized to the sample \#13 curve at $T$ = 1\,K. Sample \#13 and \#14 were hanging in vacuum, thus cooled only through high purity,
platinum voltage and current lead wires. Samples \#3 and \#10 were attached to the thermal bath by GE 7301 varnish. The inset shows $\rho(T)$ of sample \#10, measured by the following
temporal procedure; (i) initially sample was mounted with Apiezon N-grease in $^{3}$He cryostat and $\rho(T)$ was measured down to 0.34\,K (circles) in order to see the onset of a
sharp phase transition. After cleaning the N-grease (ii) sample was attached to the dilution refrigerator with GE-varnish and $\rho(T)$ was measured (squares, inset and main figure).
After cleaning the GE-varnish (iii) sample was mounted with N-grease again in dilution refrigerator and $\rho(T)$ was measured (triangles).}
\label{YbPtBiRT1}%
\end{figure}

Figure \ref{YbPtBiRT1} shows the electrical resistivity, $\rho(T)$, for several differently mounted samples of YbPtBi as a function of temperature for cooling. The electrical
resistivity curves of samples \#3, \#10, and \#14 are normalized at 1\,K to the resistivity of sample \#13, for clarity. The detailed shape of the $\rho(T)$ curve below the AFM
ordering temperature, $T_{N}$, turns out to be very sensitive to the details of how the sample is attached to heat sink for cooling. Samples \#13 and \#14, both of which show a sharp
increase of $\rho(T)$ below the phase transition, were measured with the sample hanging in vacuum (without being directly affixed to the thermal bath). Thus, these samples were cooled
down to 0.02\,K primarily through the platinum voltage and current wires. On the other hand, the electrical resistivity measurements, taken on samples that were mechanically attached
to the heat sink, showed less reproducible behavior. Samples \#3 and \#10 were attached to the heat sink with GE 7301 varnish (GE-varnish). The $\rho(T)$ curve for sample \#3 shows a
relatively weak jump below $T_{N}$, compared to the results from samples \#13 or \#14, and no obvious anomaly, corresponding to the AFM phase transition, was observed for sample \#10,
which manifests a weak slope change, best seen in a d$\rho(T)$/d$T$ plot.

The degree of sensitivity to mounting conditions can be illustrated in further detail by the measurement sequence illustrated in the inset of Fig. \ref{YbPtBiRT1}. Initially $\rho(T)$
data on several samples of YbPtBi were measured down to 0.34\,K in $^{3}$He cryostat in order to confirm a sharp onset of the phase transition below 0.4\,K; Apiezon N-grease was used
to secure the sample to the heat sink. Most of the samples showed a sharp rise of $\rho(T)$ below 0.4\,K in which the slope of $\rho(T)$ below 0.4\,K was comparable to that of sample
\#13 in Fig. \ref{YbPtBiRT1}. The $\rho(T)$ data for sample \#10 is representative and is shown as circles in the inset to Fig. \ref{YbPtBiRT1}. Next, from these samples, after
cleaning the N-grease off using toluene, eight samples were mounted on a dilution refrigerator cold stage with the GE-varnish and $\rho(T)$ was measured down to 0.02\,K. The $\rho(T)$
data obtained for sample \#10 in this measurement are presented as squares in Fig. \ref{YbPtBiRT1} (and its inset) in which the phase transition is no longer discernible, due to the
complete suppression of the $\rho(T)$ feature below 0.4\,K. All eight samples showed $\rho(T)$ behavior similar that of sample \#10. Lastly, after cleaning of the GE-varnish, using
ethanol, samples were remounted with Apiezon N-grease to the cold stage of the dilution refrigerator. The $\rho(T)$ data obtained in this measurement for sample \#10 are plotted in
the inset of Fig. \ref{YbPtBiRT1} as triangles. Interestingly, $\rho(T)$ shows sharp rise below 0.4\,K, which is similar to the result of the sample \#3. The magnitude of enhancement
of $\rho(T)$ below $T_{N}$ is still smaller than that for the results for free hanging samples, \#13 and \#14, however much bigger than that for GE-varnish results, where among the
eight remounted samples, five of them indicate a sharply rising $\rho(T)$ below 0.4\,K. The observed $\rho(T)$ response for different sample mounting methods may be related to the
anisotropic local pressure (strain), generated by different thermal contraction between sample and heat sink via thermal bond (GE-varnish) combined with changes associated with the
AFM transition.

\begin{figure}%
\centering
\includegraphics[width=0.5\linewidth]{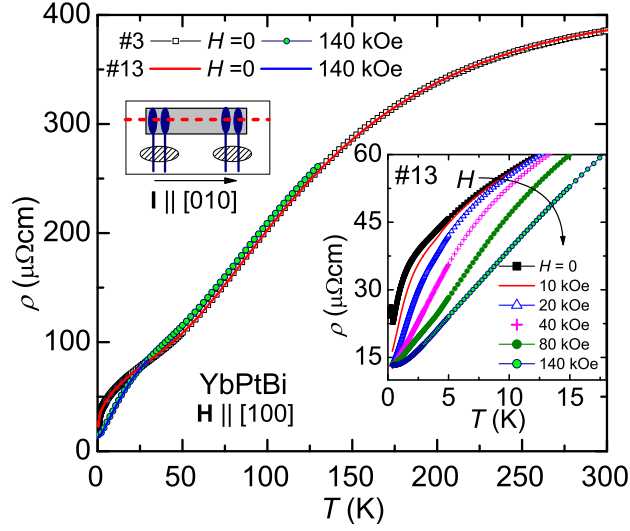}
\caption[High temperature $\rho(T)$ of YbPtBi at selected magnetic fields]{Temperature-dependent electrical resistivity, $\rho(T)$, of YbPtBi (for samples \#3 and \#13) at $H$ = 0 and
140\,kOe. For $T\,>$ 0.35\,K data, samples were mounted in PPMS $^{3}$He option with Apiezon N-grease. Below 1 K, in a dilution refrigerator sample \#3 was mounted to the heat sink
with GE-varnish and sample \#13 was measured with the sample hanging in vacuum. The sample mounting configuration for the sample \#13 in a dilution refrigerator is illustrated in the
upper left side. The sample \#13 was held onto the heat sink with four Pt electrical contact wires and very thin dental floss (dashed line) to address the torque on sample; in order
to cool down the sample through contact wires, Pt wires were glued to the heat sink using dilute Ge-Varnish (scratched area). Inset: $\rho(T)$ of sample \#13 for several selected
magnetic fields.}
\label{YbPtBiRT1a}%
\end{figure}

In the paramagnetic region, $T\,>\,T_{N}$, the electrical resistivity of YbPtBi is not sensitive to either the sample mounting methods for cooling or the sample growth conditions,
which was tested with more than 20 samples. All resistivity curves, normalized at 300\,K, collapse to a single curve, where the resistivity values at 300\,K range between 350 $\sim$
420 $\mu\Omega$cm (reflecting our geometric error in evaluating sample dimensions). In Fig. \ref{YbPtBiRT1a}, as an example, the $\rho(T)$ data of the samples \#3 and \#13 are plotted
for $H$ = 0 (down to 0.02 K) and 140\,kOe (down to 0.35 K), where the $\rho(T)$ curve of the sample \#3 is scaled at 300\,K to the sample \#13. For $T$ $>$ 0.35 K, both samples were
mounted in PPMS $^{3}$He option with Apizon N-grease. For measurements below 1 K, in a dilution refrigerator, sample \#3 was mounted to the heat sink with GE-varnish and sample \#13
was hanging in vacuum. Two curves between 0.35 K and 1 K overlap very well within instrumental error range. In zero field the two $\rho(T)$ curves are identical above 0.4\,K. For $H$
= 140\,kOe data, two curves also show virtually identical temperature dependencies with an approximately 10\% (1.6\,$\mu\Omega$cm) difference at 0.4\,K. In zero field, $\rho(T)$
decreases with decreasing temperature, displayed an inflection around 85\,K (a maximum in d$\rho(T)$/d$T$, not shown in the figure), and shows a shoulder-like feature below 5\,K as it
drops sharply until $T$ = $T_{N}$. These two characteristic features in $\rho(T)$, around 5 and 85\,K, are probably due to the Kondo and CEF effects. Without correction for the phonon
contribution to the resistivity, the local maximum associated with the coherence effect in a Kondo lattice and the logarithmic temperature dependence of $\rho(T)$ can not be resolved.
In the inset of Fig. \ref{YbPtBiRT1a}, $\rho(T)$ data from sample \#13 are plotted down to 0.02 K for $H$ = 0 and down to 0.4 K in various applied magnetic fields. As magnetic field
increases $\rho(T)$ shows a continuous suppression of the low temperature anomaly, developed near 5\,K, which is no longer visible at least for $H$ = 140 kOe. The observed
magnetoresistance (MR) for $H$ = 140 kOe changes from negative below to positive above approximately 25\,K. In the following, we will mainly present the resistivity results of sample
\#13 and the results will be compared to those of samples \#3 and \#14.

It should be noted we were aware of the possibility that torque on free hanging samples in vacuum (\#13 and \#14) might affect the measurements under magnetic fields. As shown in Fig.
\ref{YbPtBiMTMH} (b), $M(H)$ has an anisotropy although small for $H$ $>$ 25 kOe. This small anisotropy can affect the resistivity measurements when the sample is hanging with only
current and voltage wires (without glue). Thus, samples were secured by very thin dental floss across the silver paste contacts as shown in the upper left side of Fig.
\ref{YbPtBiRT1a}. Dilution refrigerator based measurements of the resistivity under magnetic field for samples \#13 and \#14 was made only up to 50 kOe due to the concern of potential
torque on sample and the resistivity was measured in $^{3}$He setup fixed with Apizon grease. Based on the several measurements, a detailed analysis leads us to the conclusion that
the torque on sample is not an issue at least up to 50 kOe when holding samples with dental floss and four electrical contact wires. The two sets of temperature dependent resistivity
data between the data below 1 K without glue and down to 0.35 K with Apizon grease, are well matched with each other above 0.35 K. In addition, the magnetic field dependence of
resistivity down to 0.4 K measured with Apizon grease overlap well with the curve with GE-varnish and no noticeable difference was observed between the up- and down-sweeps of magnetic
fields. As we will show below, the power law analysis of the resistivity, $\rho(T)$ = $\rho_{0}$ + $AT^{n}$, indicates virtually same behavior of $A$ and $n$ between the sample \#3
(GE-varnish) and \#13 (free hanging).

\begin{figure*}%
\centering
\includegraphics[width=0.5\linewidth]{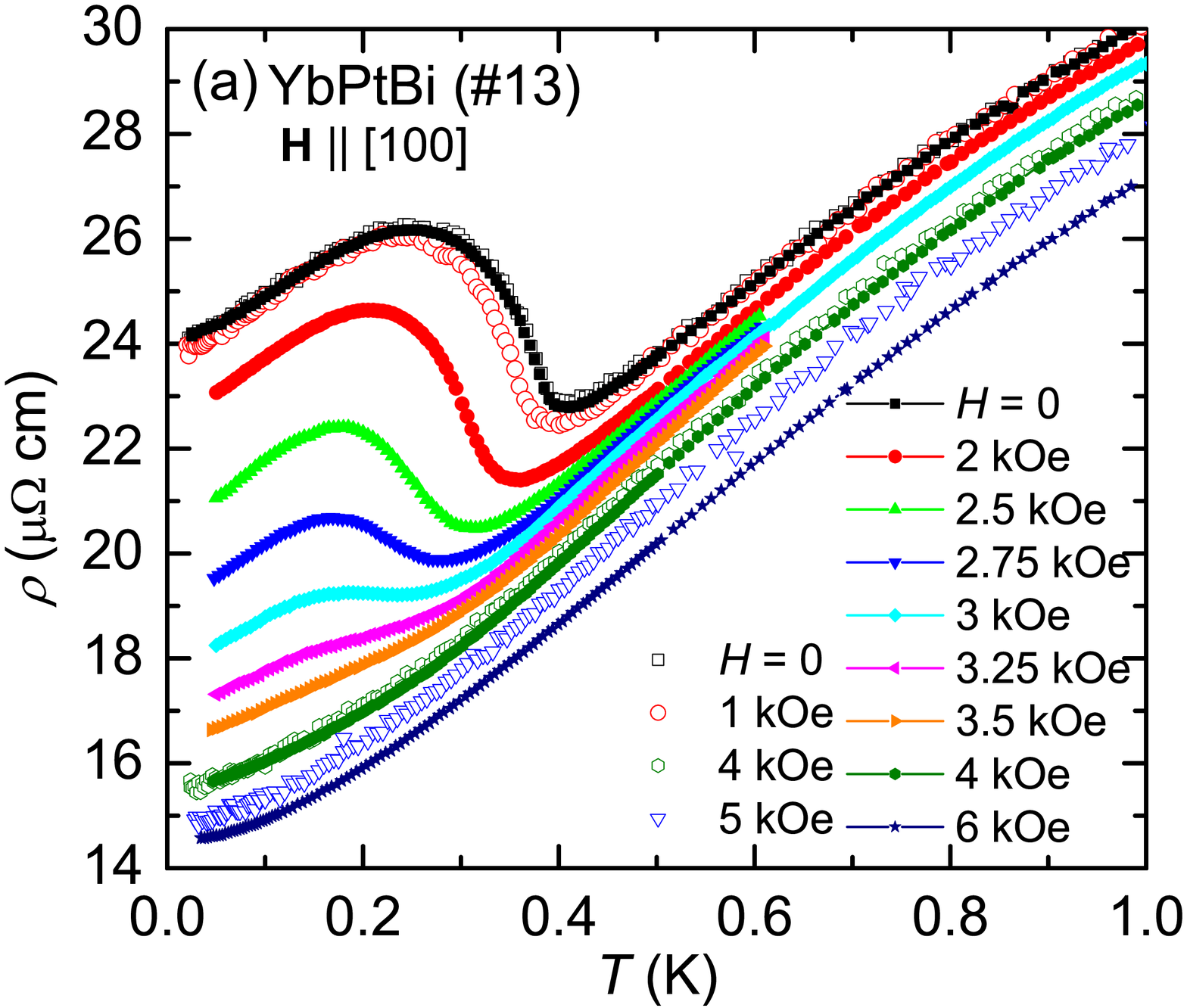}\includegraphics[width=0.5\linewidth]{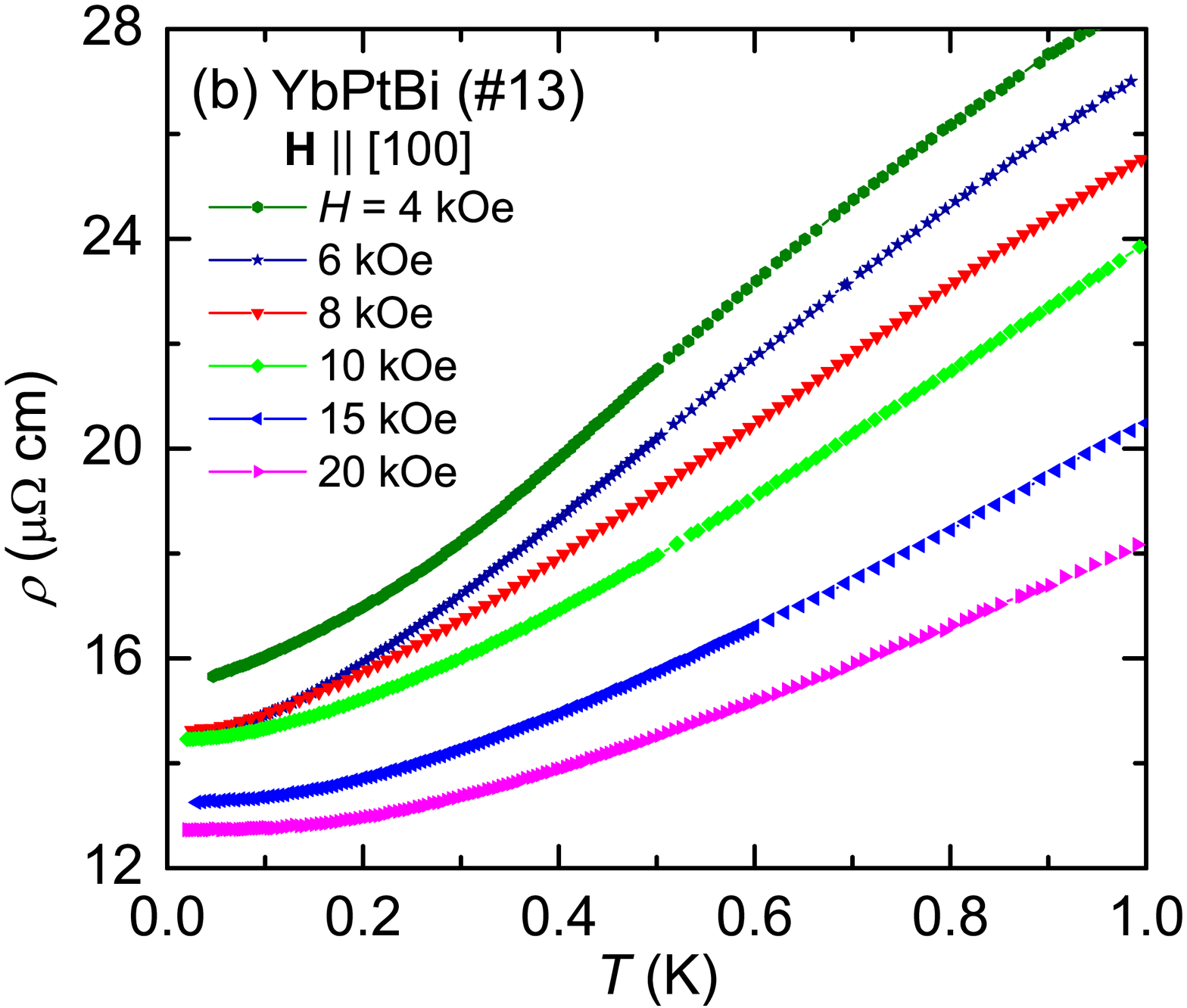}
\caption[Low temperature $\rho(T)$ of YbPtBi at selected magnetic fields]{Low temperature electrical resistivity ($\rho(T)$, sample \#13) of YbPtBi in various magnetic fields applied
along the [100] direction (a) for $H$ $\leq$ 6\,kOe and (b) for 4\,kOe $\leq$ $H$ $\leq$ 20\,kOe. For comparison, $\rho(T)$ data at $H$ = 4 and 6\,kOe are plotted in both figures. (a)
Open- and closed-symbols correspond to the data taken with 3\,$\mu$A and 30\,$\mu$A excitation current, respectively.}
\label{YbPtBiRT2}%
\end{figure*}

Figures \ref{YbPtBiRT2} (a) and (b) show the low temperature $\rho(T)$ of YbPtBi for sample \#13. In zero field there is a monotonic quasi-linear decrease with temperature from 1\,K
down to just above 0.4\,K, followed by a sharp increase of $\rho(T)$ is observed below 0.4\,K (which is consistent with earlier results \cite{Movshovich1994}). This behavior is not
consistent with that observed for simple, local moment AFM ordering for which $\rho(T)$ decreases below $T_{N}$ due to a loss of spin disorder scattering. A sharp rise of the
resistivity below 0.4\,K is reminiscent of the resistivity signature of charge density wave (CDW) \cite{Myers1999} and spin density wave (SDW) materials \cite{Fawcett1988}, and of
that in AFM materials which form a magnetic superzone gap below $T_{N}$ \cite{Elliott1972}. The ac magnetic susceptiblity suggests that YbPtBi exhibits an AFM order below 0.4\,K
\cite{Fisk1991} but the $\mu$SR \cite{Amato1992} and neutron scattering experiments \cite{Robinson1994} indicate that if there is an ordered moment it is 0.1\,$\mu_{B}$ or less. Thus,
the $\rho(T)$ data are not inconsistent with an increase of $\rho(T)$ along the direction of the SDW modulation, indicating a partial gapping of the Fermi surface, similar to what is
observed for a number of SDW systems. For $H\,>$ 4\,kOe the resistive anomaly is completely suppressed and a monotonic increase of $\rho(T)$ is observed as temperature increases as
shown in Fig. \ref{YbPtBiRT2} (b). Interestingly, an anomalous behavior of the resistivity in the zero temperature limit, $\rho(0)$, is observed around 8\,kOe at which $\rho(0)$ seems
to have a local maximum with varying magnetic field (see below).

\begin{figure*}%
\centering
\includegraphics[width=0.5\linewidth]{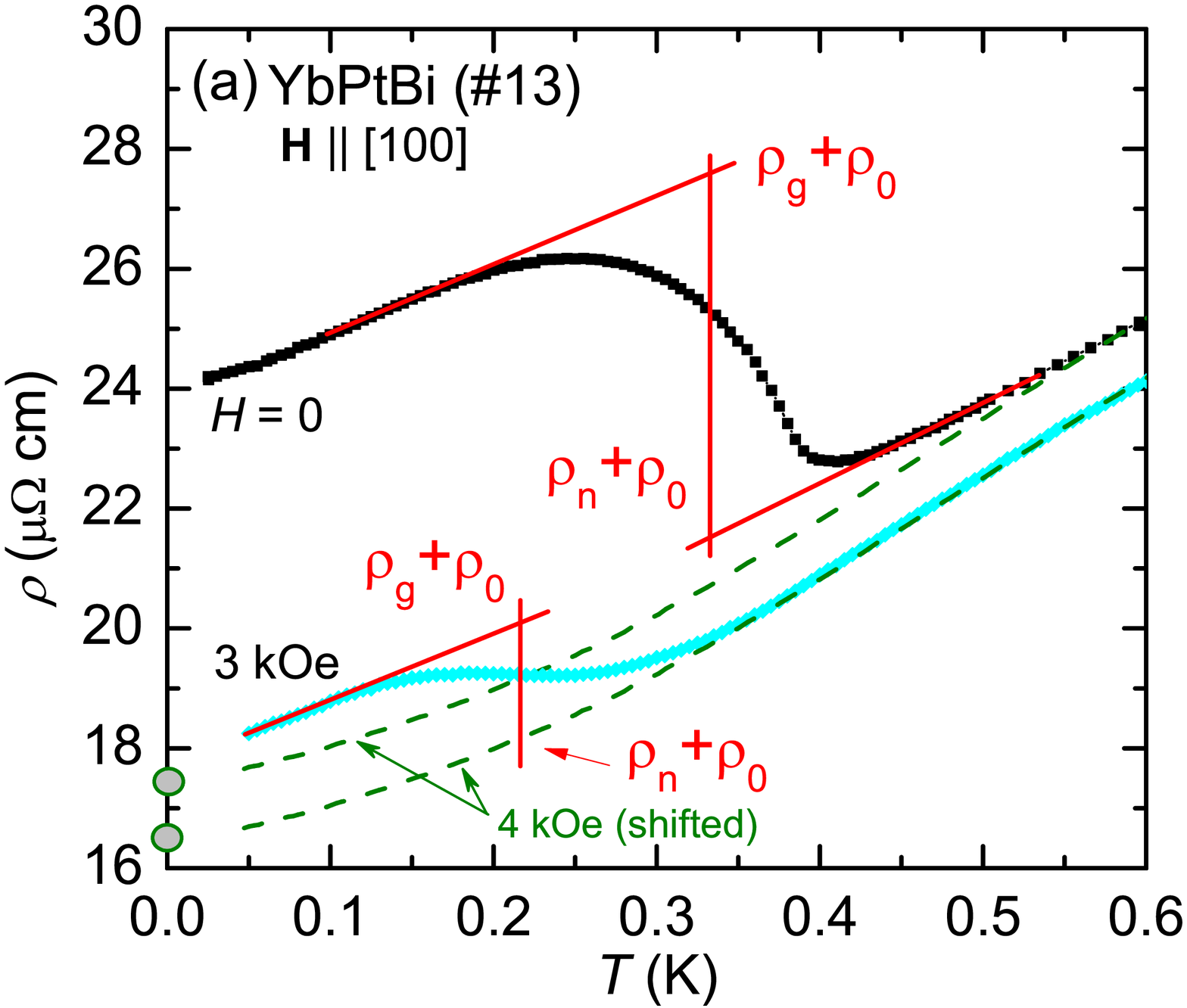}\includegraphics[width=0.5\linewidth]{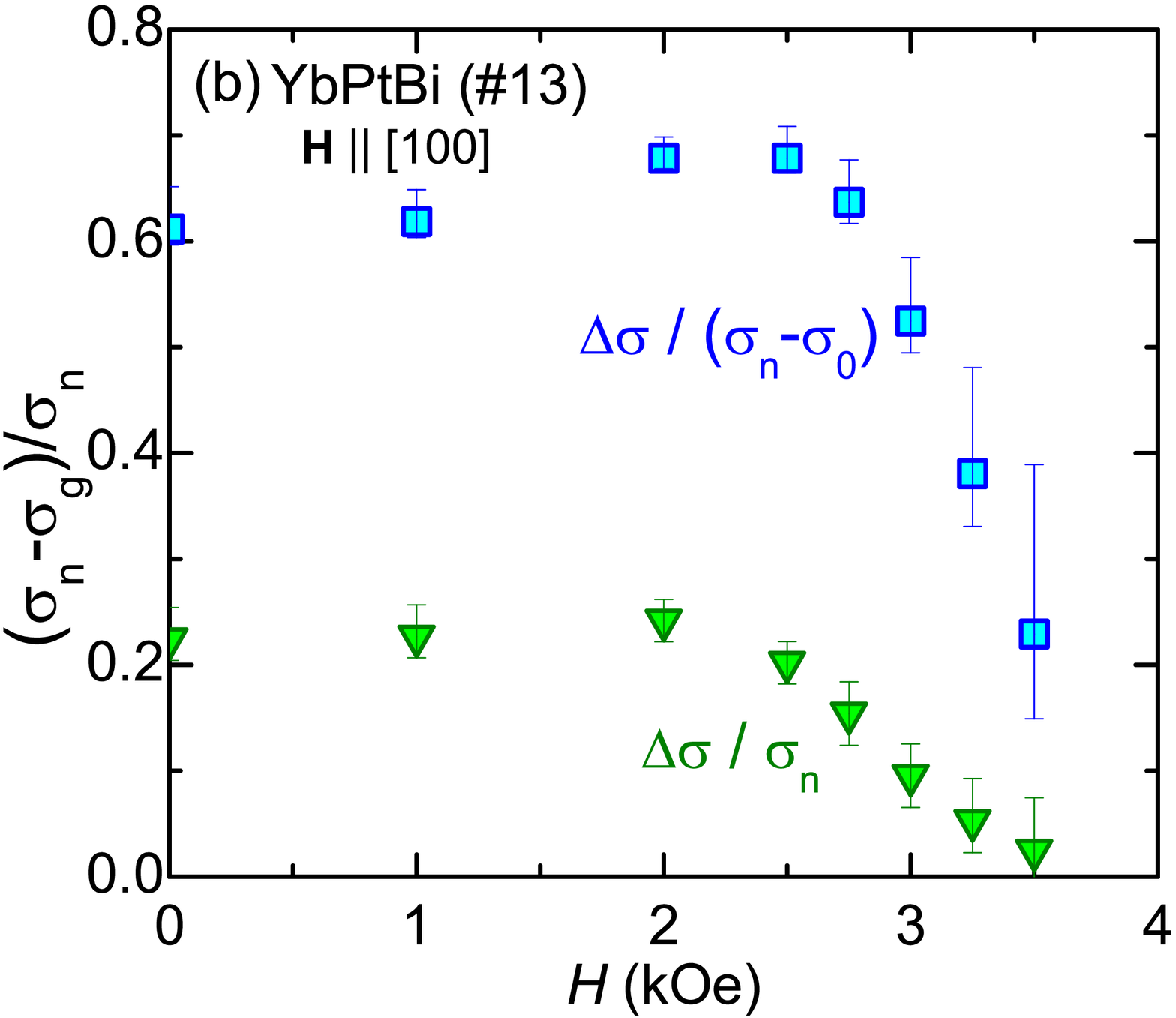}
\caption{(a) Resistivity for sample \#13 at $H$ = 0 and 3 kOe, where, in order to subtract residual resistivity ($\rho_{0}$), shown as filled circles on the \textbf{y} - axis, 4 kOe
curves (dashed-lines) are shifted to match with curves at 0.6 K for $H$ = 0 and 3 kOe. $\rho_n$ and $\rho_g$, the resistivities associated with the normal and gapped states are
inferred as shown. (b) The degree of Fermi surface gapping can be estimated from the relative change in conductivity, $(\sigma_n - \sigma_g)/\sigma_n$ = $\Delta \sigma/\sigma_n$, where
$\sigma_n = 1/\rho_n$, $\sigma_g = 1/\rho_g$, and $\sigma_0 = 1/\rho_0$. Square and triangle symbols are based on $\Delta \sigma/(\sigma_n - \sigma_0)$ (subtracting the residual
resistivity) and $\Delta \sigma/\sigma_n$ (including the residual resistivity), respectively. Error bars represent uncertainty, primarily associated with determination of $\rho_0$ and
magnetoresistive effects.}
\label{YbPtBiRT2a}%
\end{figure*}

As magnetic field increases from $H = 0$, the resistive anomaly associated with $T_{N}$ is not only reduced in height but also shifts to lower temperature as shown in Fig.
\ref{YbPtBiRT2} (a). In order to estimate of how much Fermi surface is being gapped and its magnetic field dependence, the relative change in conductivity, $(\sigma_{n} -
\sigma_{g})/\sigma_{n}$, is determined, where the subscripts $g$ and $n$ refer to the gapped and normal state, respectively \cite{Movshovich1994, McWhan1967}. The criteria for
determination of resistivity values above ($\rho_{n}$ + $\rho_{0}$) and below ($\rho_{g}$ + $\rho_{0}$) the SDW transition are shown in Fig. \ref{YbPtBiRT2a} (a). Since the residual
resistivity ($\rho_{0}$) of HF compounds is often dependent on magnetic field and pressure, especially close to the magnetic instability \cite{Flouquet1988, Jaccard1998, Stewart2001},
the $\rho_{0}$ value is not solely due to impurity or defect scattering. Thus, the deconvolution of contributions to $\rho_{0}$ for HF compounds is complex and not trivial. For this
reason, we estimate, how much FS is gapped, based on two extremes; one subtracting off the $\rho_{0} = \sigma_{0}$ term and the other leaving it in. First,
$(\sigma_{n}-\sigma_{g})/\sigma_{n}$ is estimated by subtracting $\rho_{0}$; [$(\sigma_{n} - \sigma_{0}) - (\sigma_{g} - \sigma_{0})$] / $(\sigma_{n} - \sigma_{0})$. The $\rho_{0}$
for $H$ $<$ 4 kOe curves is determined by shifting the 4 kOe curve (as a reference, dashed-lines) to match the resistivity value of each curve at 0.6 K and 1 K as shown in Fig.
\ref{YbPtBiRT2a} (a). Then conductivities in the normal and gapped states are determined as $\sigma_{n} = 1/\rho_{n}$ and $\sigma_{g} = 1/\rho_{g}$, respectively. The relative change
in conductivity data, $(\sigma_{n}-\sigma_{g})/\sigma_{n}$, are plotted as square symbols in Fig. \ref{YbPtBiRT2a} (b), where the error bars are determined by considering the two
criteria for determining $\rho_0$ (shifting 4 kOe curves to 0.6 and 1 K). Second, $(\sigma_{n}-\sigma_{g})/\sigma_{n}$ values are determined by without any subtraction of $\sigma_{0}$
term and plotted as triangle symbols in the same figure. As can be clearly seen in the Fig. \ref{YbPtBiRT2a} (b), the FS gapping due to the formation of SDW is about 60 \% or 20 \% as
estimated by subtracting $\rho_0$ or including it, respectively. For both cases, the ratio $(\sigma_{n}-\sigma_{g})/\sigma_{n}$ weakly depends on applied magnetic field up to 2.5 kOe
and decreases with further increasing magnetic field. Although $T_N$ is suppressed by applied magnetic field in a continuous manner, the degree of Fermi surface gapping is fairly
independent of field up to 2.5 kOe. This implies that the mechanism suppressing $T_N$ does not significantly depend on, or effect, the degree of Fermi surface gapping, at least
initially.

\begin{figure*}%
\centering
\includegraphics[width=0.5\linewidth]{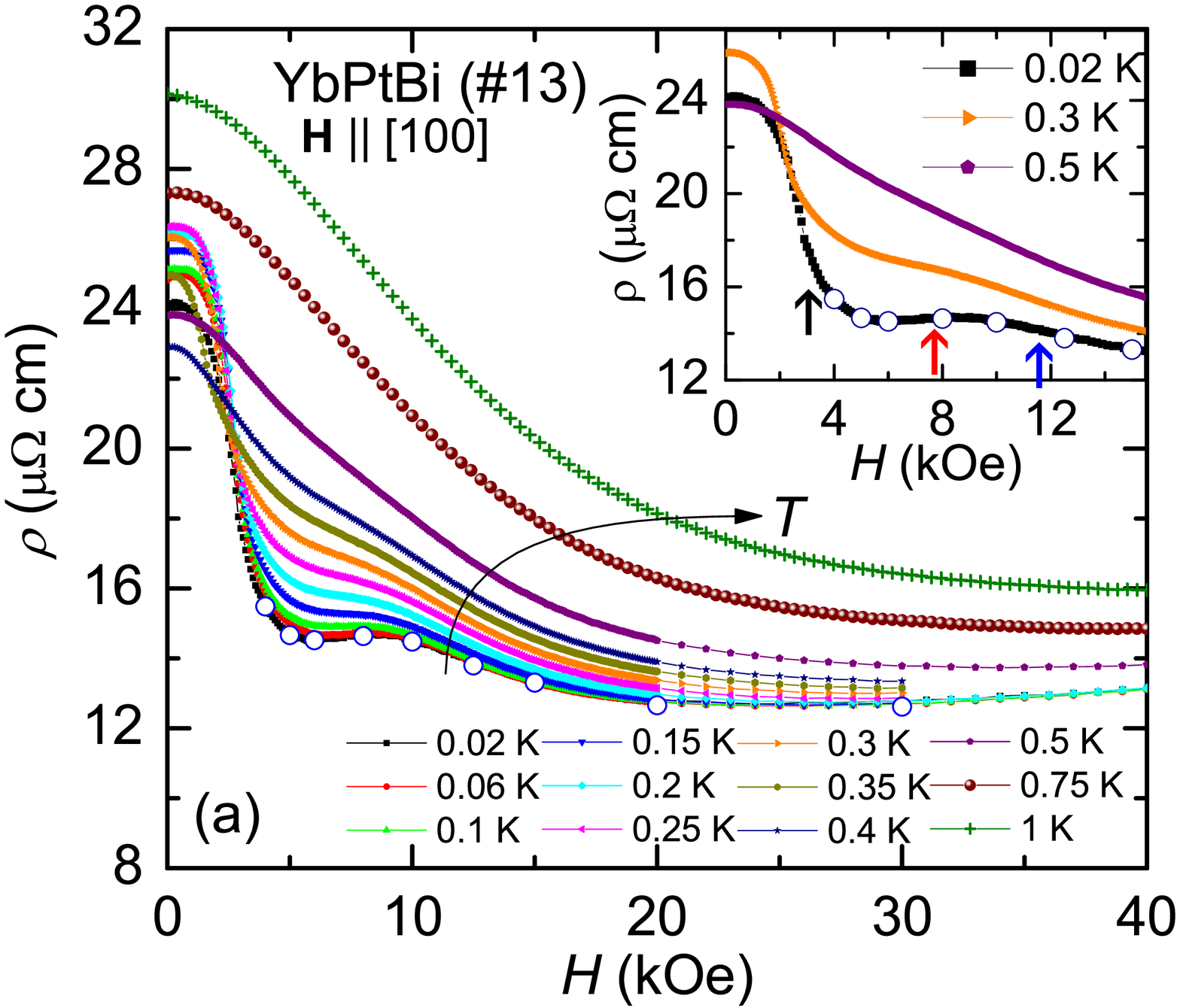}\includegraphics[width=0.5\linewidth]{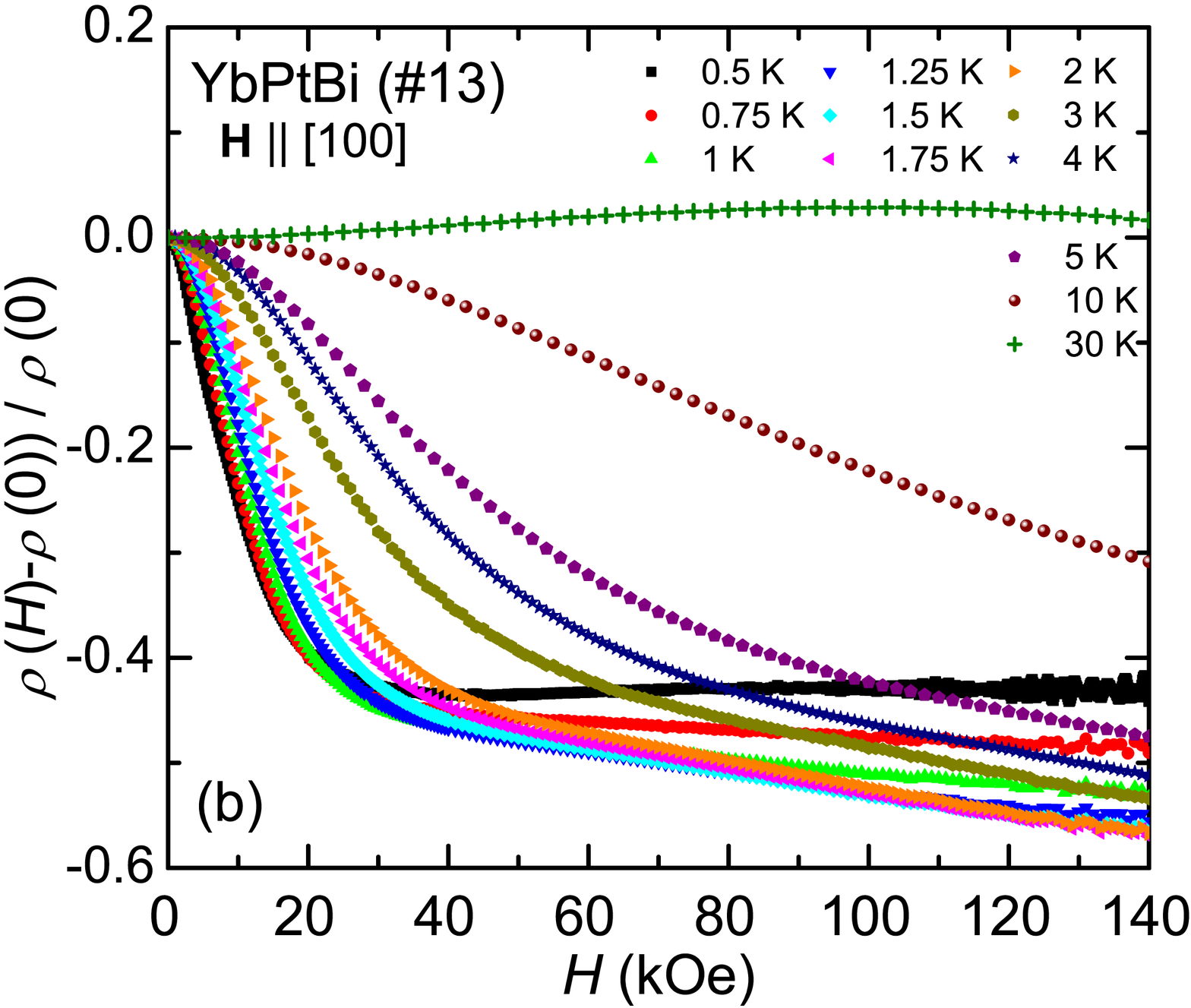}
\caption{(a) Transverse magnetoresistivity of YbPtBi (sample \#13) as plotted $\rho$ vs. $H$ at various temperatures; \textbf{H}\,$\parallel$\,[100] and \textbf{I}\,$\parallel$\,[010]
(\textbf{H}\,$\perp$\,\textbf{I}). Inset shows a expanded plot in the low field regime for $T$ = 0.02, 0.3, and 0.5\,K. Open circles in both main figure and inset represent the
residual resistivity taken from the power law fit to $\rho(T)$ data (Fig. \ref{YbPtBiRT2}); $T^{1.5}$-fit for $H\,<$ 8\,kOe and $T^{2}$-fit for $H\,\geq$ 8\,kOe. Vertical arrows in
the inset indicate slope changes in d$\rho(H)$/d$H$ curve. (b) Transverse magnetoresistance of YbPtBi (sample \#13) as plotted $[\rho(H)-\rho(0)]/\rho(0)$ vs. $H$ at various
temperatures.}
\label{YbPtBiRH1}%
\end{figure*}

Figure \ref{YbPtBiRH1} (a) shows the transverse magnetoresistivity, $\rho(H)$, of sample \#13 at various temperatures, data taken with a configuration; \textbf{H}\,$\parallel$\,[100]
and \textbf{I}\,$\parallel$\,[010] (\textbf{H}\,$\perp$\,\textbf{I}). At $T$ = 0.02\,K $\rho(H)$ steeply decreases with increasing magnetic field, has a local minimum near 5.6\,kOe,
exhibits a hump around 8\,kOe, and then decreases with further increasing magnetic field. As temperature is increased, the maximum around 8\,kOe at $T$ = 0.02\,K broadens further and
turns into a weak slope change as temperature increases up to 0.5\,K, above which the anomaly is no longer noticeable. The steep decrease of $\rho(H)$ as magnetic field increases from
zero to 5\,kOe can be related to the boundary of the AFM state. It is not clear at present whether the additional signature around 8\,kOe represents a phase transition, or some kind
of crossover. For $T\,>\,T_{N}$ a negative MR appears, only without an $\sim$ 8\,kOe hump, up to 40\,kOe. Figure \ref{YbPtBiRH1} (b) shows the higher temperature MR, plotted as
[$\rho(H)-\rho(0)$]/$\rho(0)$ vs. $H$. The MR decreases without any noticeable anomaly as magnetic field increases and the sign of the MR change from negative to positive for $T\,>$
20\,K. In the high magnetic field regime ($H\,>$ 100\,kOe), quantum oscillations are visible at low temperatures, consistent with well ordered, high quality samples.

\begin{figure}%
\centering
\includegraphics[width=0.5\linewidth]{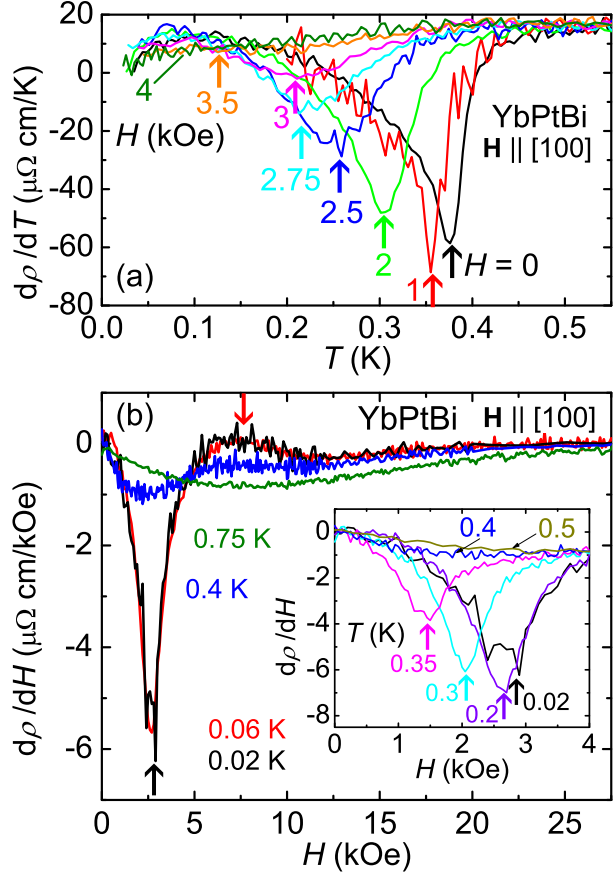}
\caption[Criteria for determination of $T_{N}$ from $\rho(T, H)$]{(a) d$\rho(T)$/d$T$ at various magnetic fields up to 4\,kOe. Vertical arrows indicate the determined AFM phase
transition temperature. (b) d$\rho(H)$/d$H$ at various temperatures. Up-arrow indicates the AFM phase boundary and down-arrow correspond to a local maximum. Inset shows
d$\rho(H)$/d$H$ up to 0.5\,K, where vertical arrows indicate the determined phase transition field.}
\label{YbPtBiRTPhase}%
\end{figure}

The AFM phase boundary was determined from the peak position in d$\rho(T)$/d$T$ because the steep rise, seen in the zero field $\rho(T)$ below $T_{N}$, broadens as magnetic field
increases. Figure \ref{YbPtBiRTPhase} (a) shows d$\rho(T)$/d$T$ of sample \#13 for selected magnetic fields. As magnetic field increases, the peak height at $T_{N}$ decreases and the
peak in d$\rho(T)/$d$T$ becomes wider, indicating that the phase transition broadens. The peak in d$\rho(T)$/d$T$ is fairly sharp for $H\,\leq$ 3\,kOe curves, whereas it is no longer
visible, down to 0.02\,K, for $H$\,$\geq$ 4\,kOe. Thus, with increasing magnetic field, the AFM phase transition shifts to lower temperatures and vanishes at around 4\,kOe. The arrows
in Fig. \ref{YbPtBiRTPhase} (a) illustrate the criterion used to determine $T_{N}$.

Figure \ref{YbPtBiRTPhase} (b) shows the magnetic field dependence of the derivatives, d$\rho(H)$/d$H$, obtained from the $\rho(H)$ curves presented in Fig. \ref{YbPtBiRH1}. The sharp
peak positions of d$\rho(H)$/d$H$ were selected as the critical field of the phase transition. The sharp peak at 2.9\,kOe, shown in 0.02\,K curve, shifts to lower field as temperature
increases (inset) and turns into a broad minimum for $T\,\geq$ 0.4\,K. The higher field broad maximum near 7.6\,kOe for 0.02\,K curve broadens as temperature increases. For $T\,>$
0.75\,K, the lower field broad minimum and a slope change near 6\,kOe shown for $T$ = 0.4\,K curve are no longer visible and instead d$\rho(H)$/d$H$ shows a single minimum near $\sim$
10\,kOe. As will be discussed below, the positions of the sharp peak and the local maximum agree with the observed anomalies
in the magnetostriction, Hall resistivity, and thermoelectric power measurements.

\begin{figure}
\centering
\includegraphics[width=0.5\linewidth]{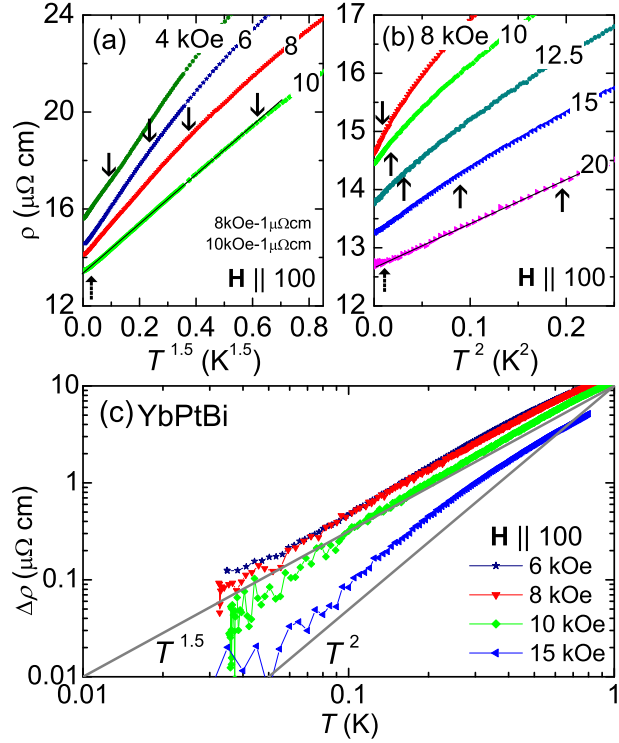}
\caption{(a) $\rho(T)$ vs. $T^{1.5}$ at various magnetic fields, where $\rho(T)$ curves for $H$ = 8 and 10\,kOe are shifted by -1$\mu\Omega$cm each for clarity. Down-arrows indicate
the temperature below which $\Delta\rho(T)$ = $AT^{1.5}$, determined from a power law fit ($\Delta\rho(T)$ = $AT^{n}$) to the data. For $H$ = 10\,kOe the line is the fit of the power
law to the data and up-arrow indicates a deviation from $T^{1.5}$-dependence of $\Delta\rho(T)$. (b) $\rho(T)$ vs. $T^{2}$ at various magnetic fields. The arrows mark the temperature
where the fits ($\Delta\rho(T)$ = $AT^{2}$) deviates from the data. These temperatures, $T_{FL}$, are plotted in the $H-T$ phase diagram (see Fig. \ref{YbPtBiRTPhasediagram}). For $H$
= 20\,kOe the line is the fit of the power law to the data and up-arrow in the low temperature side indicates a deviation from $T^{2}$-dependence of $\Delta\rho(T)$. (c)
Double-logarithmic plots of $\Delta\rho(T)$ vs. $T$ for $H$ = 6, 8, 10, and 15\,kOe. The solid lines represent the temperature dependence expected for the exponent $n$ = 1.5 and $n$ =
2.}
\label{YbPtBiRTPhase1}%
\end{figure}

To get further insight from the low temperature transport data from YbPtBi, $\rho(T)$ data are analyzed in terms of a power law; $\Delta\rho(T)$ = $\rho(T)$ - $\rho_{0}$ = $AT^{n}$,
where $\rho_{0}$ is the residual resistivity and $A$ is the coefficient. The coefficient $A$ can be interpreted as the quasi-particle scattering cross-section. The exponent, $n$,
indicates whether the system is in a Fermi Liquid (FL) regime ($n$ = 2) with dominant electron-electron scattering or whether strong quantum fluctuation effects dominate, generally
$n$\,$<$ 2, in the vicinity of a QCP \cite{Stewart2001}. Figures \ref{YbPtBiRTPhase1} (a) and (b) show plots of $\rho(T)$ vs. $T^{1.5}$ and $T^{2}$, respectively, at various magnetic
fields. In Fig. \ref{YbPtBiRTPhase1} (a) $\rho(T)$ for $H$ = 8 and 10\,kOe data are shifted by -1\,$\mu\Omega$cm each for clarity. Since the anomaly in $\rho(T)$ below the SDW phase
transition for $H\,<$ 4\,kOe prevents the power law fit to the data, the fit was performed for $H$\,$\geq$ 4\,kOe at which no sharp feature in d$\rho(T)$/d$T$ was observed down to
0.02\,K (see Fig. \ref{YbPtBiRTPhase} (a)).

For 4\,kOe $\leq$\,$H$\,$\leq$ 8\,kOe, $\rho(T)$ can be well described by a $T^{1.5}$-dependence down to the lowest accessible temperature of 0.02\,K, where the exponent $n$ ranges
between 1.45 $\sim$ 1.6 depending on the fit range. The maximum temperature below which $\Delta\rho(T)$ = $AT^{1.5}$ shifts to higher temperature as magnetic field increases,
indicated by down-arrows in Fig \ref{YbPtBiRTPhase1} (a). For $H$ = 8 and 10\,kOe, plotted in both Figs. \ref{YbPtBiRTPhase1} (a) and (b), $\rho(T)$ can be described by a
$T^{2}$-dependence at low temperatures above which $T^{1.5}$-dependence is predominant. For $H\,>$ 10\,kOe a characteristic of FL state is clearly evidenced by the relation
$\Delta\rho(T)$ = $AT^{2}$ at low temperatures as indicated by the arrow in Fig. \ref{YbPtBiRTPhase1} (b). Note for $H\,\geq$ 20\,kOe that as temperature decreases $\rho(T)$ follows
$T^{2}$-dependence and then flattens, revealing the deviation of FL behavior with $n\,>\,2$. In Fig. \ref{YbPtBiRTPhase1} (b) the up-arrow in the low temperature side on $\rho(T)$
curve for $H$ = 20\,kOe curve indicates a deviation of $T^{2}$-dependence.

Since the difference of the exponent between $n$ = 1.5 and 2 is very small, the results based on the power law analysis are also visualized in Fig. \ref{YbPtBiRTPhase1} (c) as
$\log$-$\log$ plot of $\Delta\rho(T)$ vs. $T$ at selected magnetic fields. $\Delta\rho(T)$ for $H$ = 6\,kOe is a straight line at least up to 0.4\,K, which is parallel to the
$T^{1.5}$-line, whereas $\Delta\rho(T)$ for $H$ = 10\,kOe deviates from a straight line parallel to the $T^{1.5}$-line near 0.1\,K below which the slope is parallel to the
$T^{2}$-line. Note that at low temperatures the slope in $\log$\,-\,$\log$ plot depends on the $\rho_{0}$ value. When $\rho(T)$ is corrected by the $\rho_{0}$ value obtained from the
fit of $T^{1.5}$-dependence above $\sim$ 0.1\,K, the slope for $H$ = 10\,kOe is parallel to the $T^{1.5}$-line above 0.1\,K. For $H$ = 15\,kOe curve, $\Delta\rho$ is a straight line
parallel to the $T^{2}$-line below $\sim$ 0.25\,K, which clearly indicate a quadratic temperature dependence down to lowest temperature measured.

\begin{figure}
\centering
\includegraphics[width=0.5\linewidth]{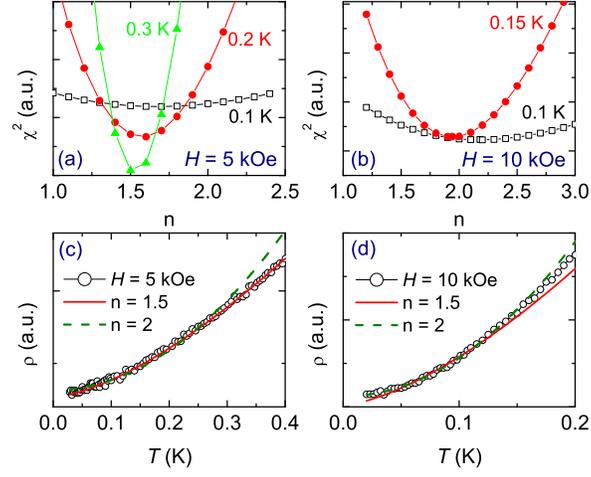}
\caption{Plots of the sum of the squares of the offsets (residual), $\chi^{2}$, from least square fitting of the power law, $\rho(T)$ = $\rho_{0}$ + $AT^{n}$, to the data with fixed
$n$ values for (a) $H$ = 5 kOe and (b) 10 kOe. The calculated resistivity with $n$ = 1.5 (solid line) and 2 (dashed line) for (c) $H$ = 5 kOe and (d) 10 kOe. The calculated
resistivity curves are based on the fitting up to 0.2 K and 0.15 K for $H$ = 5 kOe and 10 kOe, respectively. }
\label{YbPtBiRTpower}%
\end{figure}

In addition, the exponent of power law analysis depends on the fitting temperature range. In order to further quantify the robustness of the exponent value, we tried a least square
fitting of the power law to the data with fixed $n$ values between 1 and 3. The results of $\chi^{2}$ of the least square fit as a function of the power $n$ are plotted in Figs.
\ref{YbPtBiRTpower} (a)-(d), where several temperature ranges for fitting are selected; the $\chi^{2}$ and the power law fit with $n$ = 1.5 and 2 are plotted only for $H$ = 5 kOe (a
and c) and $H$ = 10 kOe (b and d) as representative data sets. For $H$ = 5 kOe, Fig. \ref{YbPtBiRTpower} (a), the $\chi^{2}$ obtained by fitting from base temperature, $T_{B}$ $\sim$
0.04 K, up to 0.2 K or 0.3 K clearly indicates a deep minimum near $n$ = 1.5. The $\chi^{2}$ data for fitting from $T_{B}$ to 0.1 K (a very limited range) show a shallow minimum
around $n$ $\sim$ 1.7, but also indicates that this temperature range is at the edge of being too small for such analysis. For $H$ = 10 kOe, the fit up to 0.1 K and 0.15 K shows a
minimum around $n$ = 2.2 and $n$ = 2, respectively, as shown in Fig. \ref{YbPtBiRTpower} (b). Based on the minimum of $\chi^{2}$, low temperature fits of the resistivity of the form
$\rho(T) = \rho_{0} + AT^{n}$ with $n$ = 1.5 and 2 are shown by the solid and dashed line, respectively, in Figs. \ref{YbPtBiRTpower} (c) and (d). The resistivity for $H$ = 5 kOe can
be better described with $n$ = 1.5 than 2, whereas the resistivity for $H$ = 10 kOe shows a good agreement with $n$ = 2 curve. Similar $\chi^{2}$ analysis clearly suggests that the
best exponent is $n$ = 1.5 for 4 kOe $\leq$ $H$ $<$ 8 kOe and $n$ = 2 for $H$ $\geq$ 8 kOe. It should be noted, though, that when the maximum fitting temperature range is set to 0.1 K
or lower, the very shallow minimum in $\chi^{2}$ curve does move toward $n$ $\sim$ 2 for $H$ $<$ 8 kOe. Thus, there is a possibility to have exponent $n$ = 2 by selecting the fitting
temperature range below 0.1 K for $H$ $<$ 8 kOe.

\begin{figure}%
\centering
\includegraphics[width=0.5\linewidth]{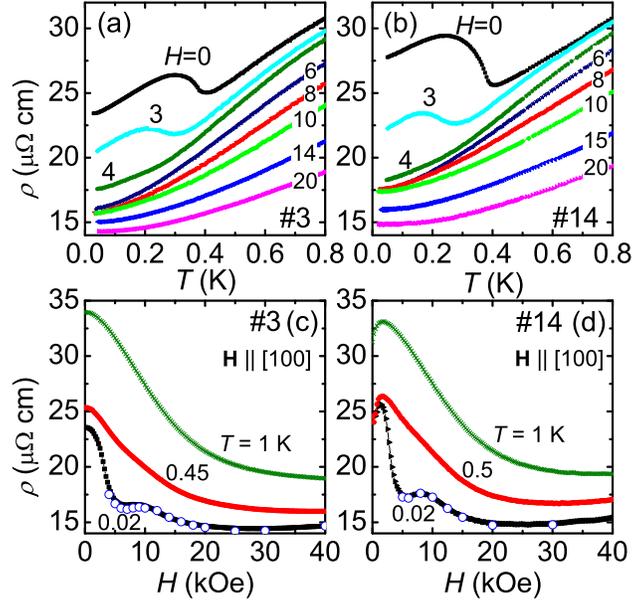}
\caption[$\rho(T, H)$ of YbPtBi for various samples]{$\rho(T)$ of (a) sample \#3 and (b) sample \#14 at selected magnetic fields. $\rho(H)$ of (c) sample \#3 and (d) sample \#14 at
selected temperatures. (b) Zero field $\rho(T)$ of sample \#14 was shifted by +3$\mu\Omega$cm for clarity. Open circles in (c) and (d) represent the residual resistivity obtained from
the power law fit to the $\rho(T)$ data. As discussed in text and caption of Fig. \ref{YbPtBiRT1}, sample \#14 was suspended in vacuum by its four Pt leads and sample \#3 was mounted
to the thermal bath with N-Grease.}
\label{YbPtBiRT3}%
\end{figure}

\begin{figure}%
\centering
\includegraphics[width=0.5\linewidth]{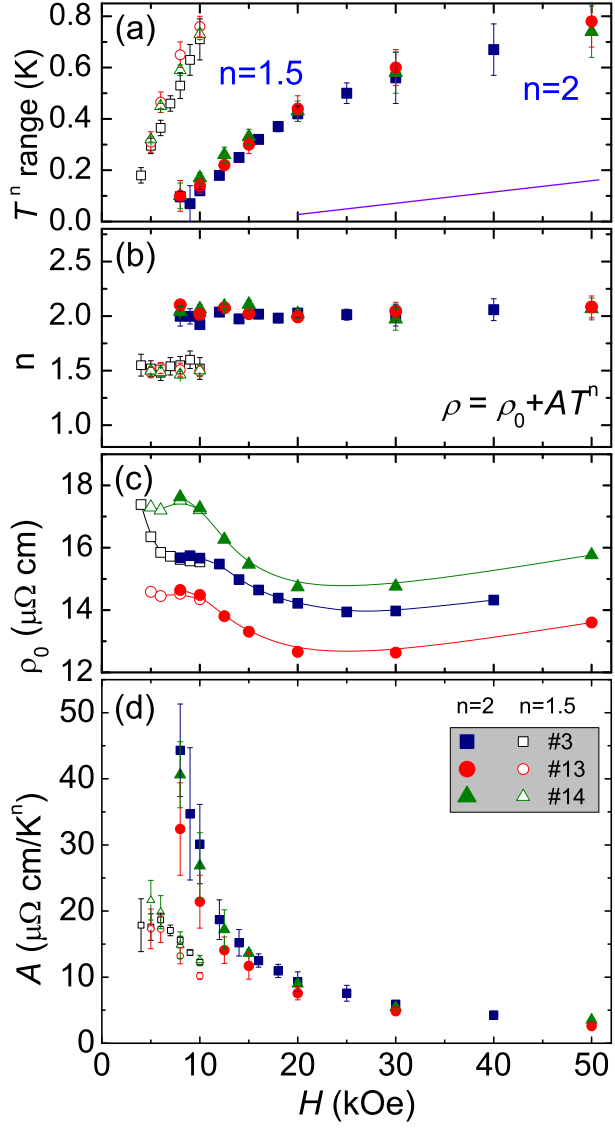}
\caption[Parameters obtained from power law analysis for various samples]{Parameters obtained from power law fits, $\rho(T)$ = $\rho_{0}$ + $AT^{n}$, to the data for three different
samples. Open- and solid-symbols correspond to fits with $n$ = 1.5 and $n$ = 2, respectively. For $n$ = 1.5, the obtained parameters are plotted only up to 10 kOe. (a) Temperatures of
the fit range below which a $T^{n}$-dependence of $\rho(T)$ fit data well. For 8\,kOe $\leq$\,$H$\,$\leq$ 10\,kOe the fit of $T^{1.5}$-dependence was performed above the temperature,
satisfying $T^{2}$-dependence. The horizontal line for $H\,\geq$ 20\,kOe indicates the temperature below which $\rho(T)$ flattens. (b) Determined exponents $n$ from least square fits
to the data. (c) Obtained $\rho_{0}$. Solid lines are guide to the eye. (d) Field dependencies of the coefficients, $A = (\rho(T) - \rho_{0})/T^{n}$ with $n$ = 1.5 and 2, for three
different samples.}
\label{YbPtBiRTPhase2}%
\end{figure}

\begin{figure}%
\centering
\includegraphics[width=0.5\linewidth]{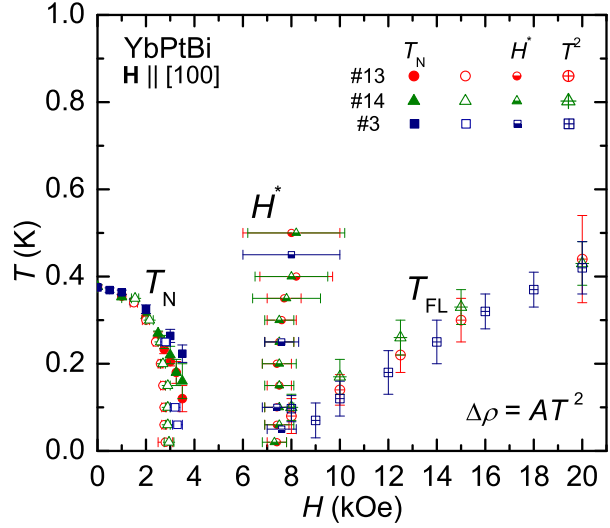}
\caption[$H-T$ phase diagram of YbPtBi constructed from $\rho(T, H)$]{$H-T$ phase diagram of YbPtBi constructed from the $\rho(T, H)$ results for three different samples; all circles,
triangles, and squares correspond to the results of samples \#13, \#14, and \#3, respectively. $T_{N}$ was derived from the sharp minimum in d$\rho(T)$/d$T$ (solid-symbols) and
d$\rho(H)$/d$H$ (open-symbols). $H^{*}$ was derived from the broad local maximum in d$\rho(H)$/d$H$ (Fig. \ref{YbPtBiRTPhase}). $T_{FL}$ represents the upper limit of the
$T^{2}$-dependence of $\rho(T)$.}
\label{YbPtBiRTPhasediagram}%
\end{figure}

The electrical resistivity data for samples \#3 and \#14 are plotted in Figs \ref{YbPtBiRT3} (a)-(d), respectively, at selected temperatures and magnetic fields as representative
data. For \textbf{H}\,$\parallel$\,[100], the overall temperature and magnetic field dependences of the resistivity for both samples \#3 and \#14 are the same as those for sample \#13
(Figs. \ref{YbPtBiRT2} and \ref{YbPtBiRH1}). These data were analyzed by the same methods, applied to sample \#13, to determine phase transitions and power law dependences of
$\rho(T)$. These results together with those of sample \#13 are summarized in Fig. \ref{YbPtBiRTPhase2} and Fig. \ref{YbPtBiRTPhasediagram}.

In Fig. \ref{YbPtBiRTPhase2} parameters of $\rho_{0}$, $A$, $n$, and the maximum temperature range satisfying $T^{1.5}$ and $T^{2}$, obtained from the power law fits, are summarized
for $H\,\geq$ 4\,kOe. All open- and solid-symbols correspond to the fits with $n$ = 1.5 and $n$ = 2 (Fig. \ref{YbPtBiRTPhase2} (b)), respectively, for three different samples. The
obtained $\rho_{0}$ shows a broad local maximum around 8 kOe as shown in Fig. \ref{YbPtBiRTPhase2} (c). For comparison with magnetoresistivity at 0.02\,K, the obtained $\rho_{0}$
values for sample \#13 are plotted in Fig. \ref{YbPtBiRH1} as open-circles which track well the magnetoresistivity at $T$ = 0.02\,K. $\rho_{0}$ values for samples \#3 and \#10 also
track the low temperature $\rho(H)$ well (Figs. \ref{YbPtBiRT3} (c) and (d), respectively). As shown in Fig. \ref{YbPtBiRTPhase2} (a) for magnetic fields above 4\,kOe the temperature
range, following $T^{1.5}$-dependences of $\rho(T)$, increases monotonically and for magnetic fields higher than 8\,kOe the FL region, $\Delta\rho(T)$ = $AT^{2}$, gradually increases.
The field dependences of the coefficients, $A\,=\,(\rho(T)-\rho_{0})/T^{n}$ with $n$ = 1.5 and 2, are plotted Fig. \ref{YbPtBiRTPhase2} (d). A strong enhancement of the
$T^{2}$-coefficient is observed as magnetic field approaches 8\,kOe from higher magnetic fields.

The various characteristics (field-temperature points) observed from sample \#13, together with those from samples \#3 and \#14, are collected in the $H-T$ phase diagram displayed in
Fig. \ref{YbPtBiRTPhasediagram}. The magnetic field dependence of $T_{N}$ was determined from the sharp peak position in d$\rho(T)$/d$T$ and d$\rho(H)$/d$H$ (Fig.
\ref{YbPtBiRTPhase}). The crossover scale, $H^{*}$, was obtained from the maximum of d$\rho(H)$/d$H$ (Fig. \ref{YbPtBiRTPhase} (b)). The FL region, $T_{FL}$, marks the upper limit of
$T^{2}$-dependence of $\rho(T)$ (Fig. \ref{YbPtBiRTPhase1}). The AFM phase boundary of $T_{N}$ and the crossover of $H^{*}$ and $T_{FL}$, obtained from the results of three different
samples, agree well each other. Therefore, it seems to be reasonable to assume that the strength of the anomaly in $\rho(T)$ below $T_{N}$ is only sensitive to the strain generated
through bonding agent for sample cooling (see Fig. \ref{YbPtBiRT1}), but the relevant physics of the samples remains the same. The AFM boundary determined from d$\rho(T)$/d$T$ does
not fully agree with the one obtained from d$\rho(H)$/d$H$ at low temperatures; the AFM phase boundary below 0.2\,K spreads significantly. It is most likely that this inconsistency is
based on the criteria used to determine phase transition coordinates, but it is possible that there are two closely spaced phase boundaries.

From the $H-T$ phase diagram for the applied magnetic field parallel to the [100] direction, it is clear that the AFM ordering can be suppressed to zero for $H_{c}$ $\lesssim$ 4\,kOe.
For $H\,>\,H_{c}$ a field induced anomalous state, characterized by $\Delta\rho(T)$ = $AT^{1.5}$, is established up to $\sim$ 8\,kOe, and a FL state, characterized by $\Delta\rho(T)$
= $AT^{2}$, is induced for $H\,\geq$ 8\,kOe.  The $T_{FL}$ region enlarges monotonically with increasing magnetic field. It is apparent that at lowest temperature measured ($T$ =
0.02\,K) a crossover from $T^{1.5}$- to $T^{2}$-dependence of $\rho(T)$ occurs near 8\,kOe. At higher magnetic fields, for $H\,\geq$ 8\,kOe, a crossover from $T^{1.5}$- to
$T^{2}$-dependence of $\rho(T)$ is observed with decreasing temperature. Note that for $H\,<$ 8\,kOe, because of the poor signal to noise ratio, below 0.08\,K, where the exponent $n$
is ill-defined if only this small range as used, $\rho(T)$ can be described with the exponent $n$ = 2, depending on the fit region.

As magnetic field decreases from the higher magnetic field (paramagnetic) side, the temperature range, $T_{FL}$, becomes smaller, while the coefficient $A$ of $T^{2}$-dependence
increases rapidly and shows a tendency of diverging as $H\rightarrow H_{c}$. A divergent nature of this coefficient, when approaching to the critical field from paramagnetic side, is
considered strong evidence for a field-induced quantum phase transition \cite{Gegenwart2002}, which will be discussed below together with the field dependence of the electronic
specific heat coefficient ($\gamma$). In addition, the exponent $n$=1.5 near a QCP was predicted by the traditional SDW scenario of quantum criticality with $d$ = 3 and $z$ = 2
\cite{Hertz1976, Millis1993}. From the phase diagram it is apparent that $H^{*}$ separates $T_{FL}$ region from the AFM phase boundary $T_{N}$.

\subsection{Specific heat}

\begin{figure*}%
\centering
\includegraphics[width=0.5\linewidth]{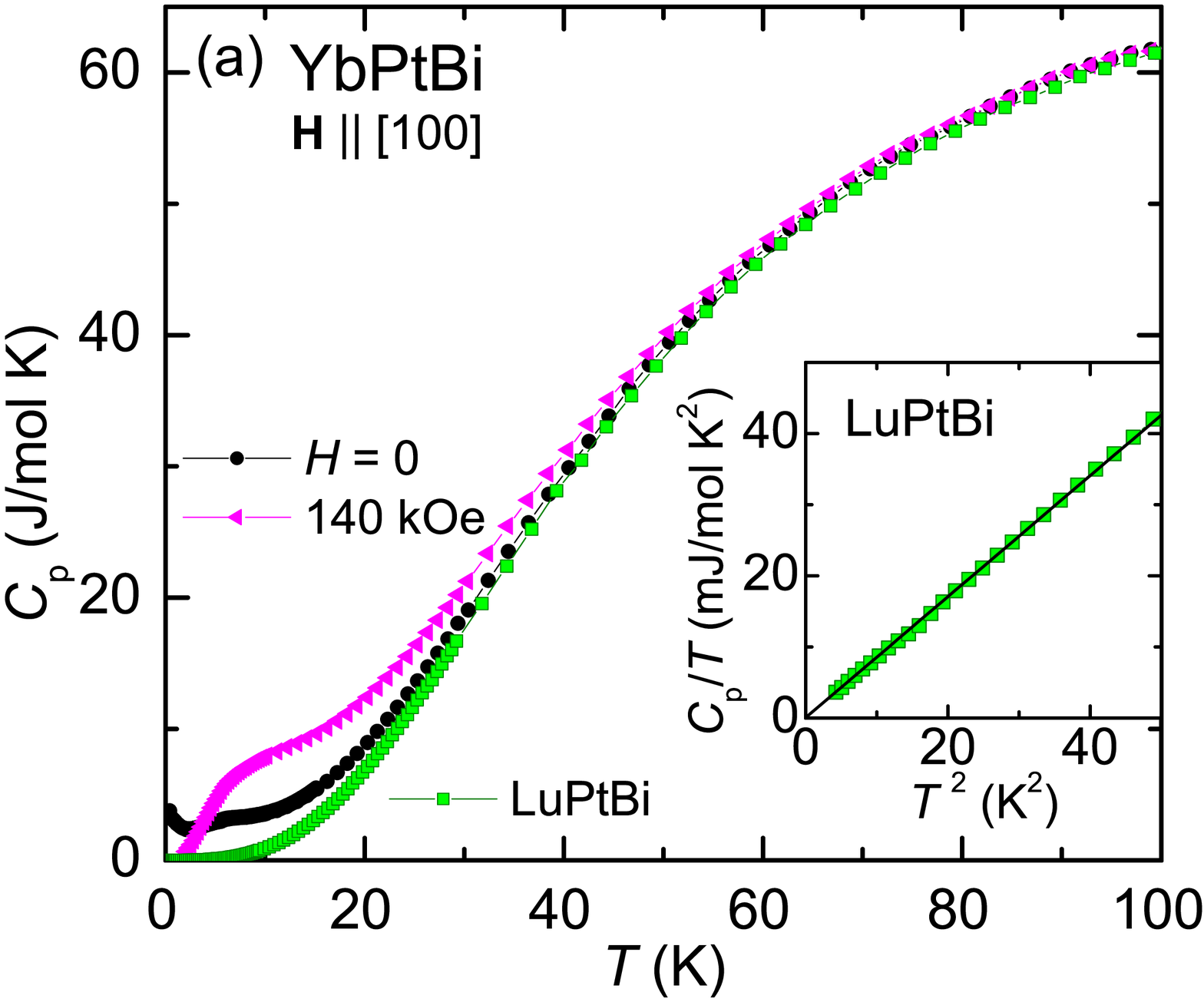}\includegraphics[width=0.5\linewidth]{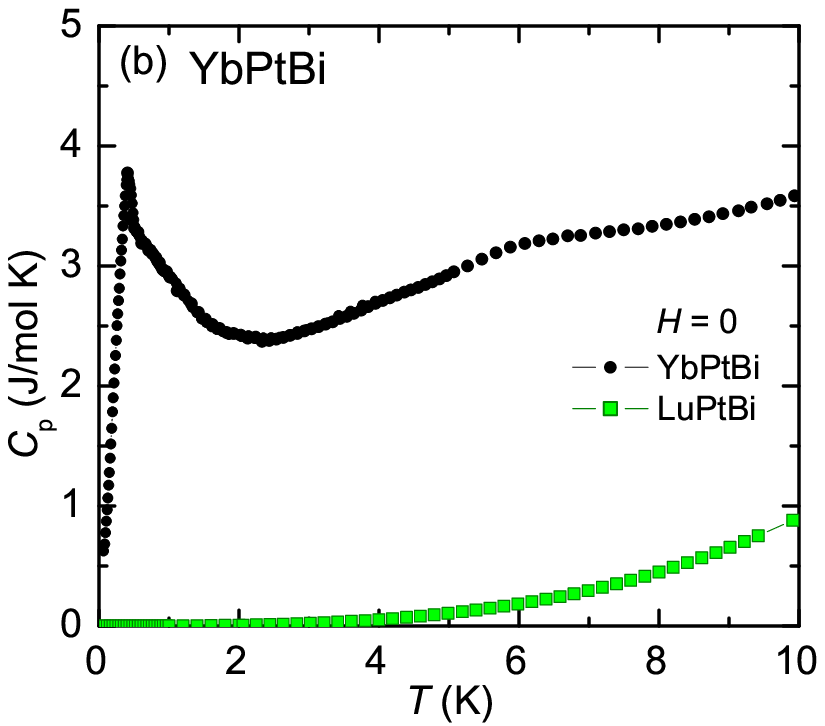}
\caption[Specific heat results for YbPtBi and LuPtBi]{(a) Specific heat of YbPtBi as $C_{p}(T)$ vs. $T$ for $H$ = 0 (circles) and 140\,kOe (triangles), applied magnetic field along
the [100] direction, and zero-field $C_{p}(T)$ of LuPtBi (squares). Inset displays $C_{p}/T$ vs. $T^{2}$ for LuPtBi. The solid line is a fit of the equation $C_{p}(T)$ = $\gamma T$ +
$\beta T^{3}$ to the data. (b) Zero field specific heat for YbPtBi and LuPtBi below 10\,K.}
\label{YbPtBiCp}%
\end{figure*}

Figure \ref{YbPtBiCp} (a) displays the temperature dependences of the specific heat, $C_{p}(T)$, of YbPtBi for $H$ = 0 and 140\,kOe, applied along the [100] direction, together with
zero field $C_{p}(T)$ of its nonmagnetic isostructural counterpart, LuPtBi. The overall shape of $C_{p}(T)$ for LuPtBi is typical for a nonmagnetic systems. In particular, below 8\,K
it is easily described by the relation, $C_{p}(T)$ = $\gamma T$ + $\beta T^{3}$, in which the first term is a conventional conduction electron contribution to the specific heat with
the Sommerfeld coefficient, $\gamma$, and the second term is a low temperature phonon contribution in a form of the Debye-$T^{3}$ law with the Debye temperature, $\Theta_{D}$. For
LuPtBi, shown in the inset of Fig. \ref{YbPtBiCp} (a), least-square fitting of this formula to the experimental data yields the $\gamma$ $\simeq$ 0 (6$\times$10$^{-5}$
J/mol$\cdot$K$^{2}$) and from $\beta$, the $\Theta_{D}$ $\simeq$ 190\,K. Since $\gamma$ is negligible, which is consistent with a low carrier density system, $C_{p}(T)$ of LuPtBi is
dominated by the phonon specific heat.

The zero field, $C_{p}(T)$ of YbPtBi indicates a distinct anomaly at about 0.41\,K as shown in Fig. \ref{YbPtBiCp} (b) which is consistent with earlier results \cite{Fisk1991}. Since
$C_{p}(T)$ of YbPtBi shows a broad hump around 6\,K and a peak at $T_{N}$, we were not able to extract $\gamma$ and $\Theta_{D}$ from a fit of $C_{p}(T)/T$ = $\gamma$ + $\beta T^{2}$
to the data. The result of $C_{p}(T)$ for $H$ = 140\,kOe shows the development of a large, broad peak structure, centered near 10\,K, probably related to a magnetic Schottky anomaly.
At high temperatures ($T$ $>$ 60 K) the $C_{p}(T)$ data are essentially the same for all curves shown in Fig. \ref{YbPtBiCp} (a).

\begin{figure*}%
\centering
\includegraphics[width=0.5\linewidth]{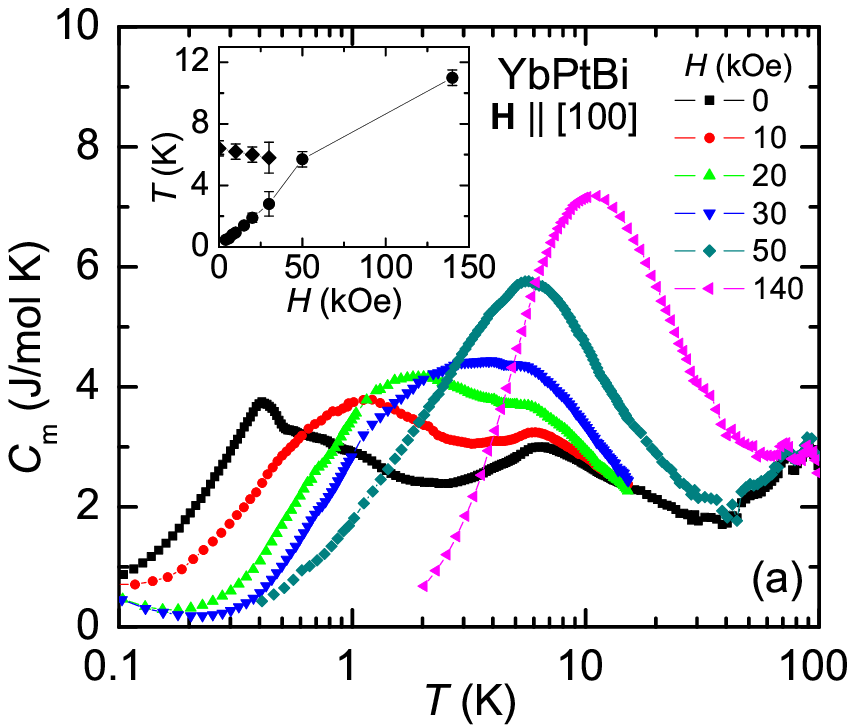}\includegraphics[width=0.5\linewidth]{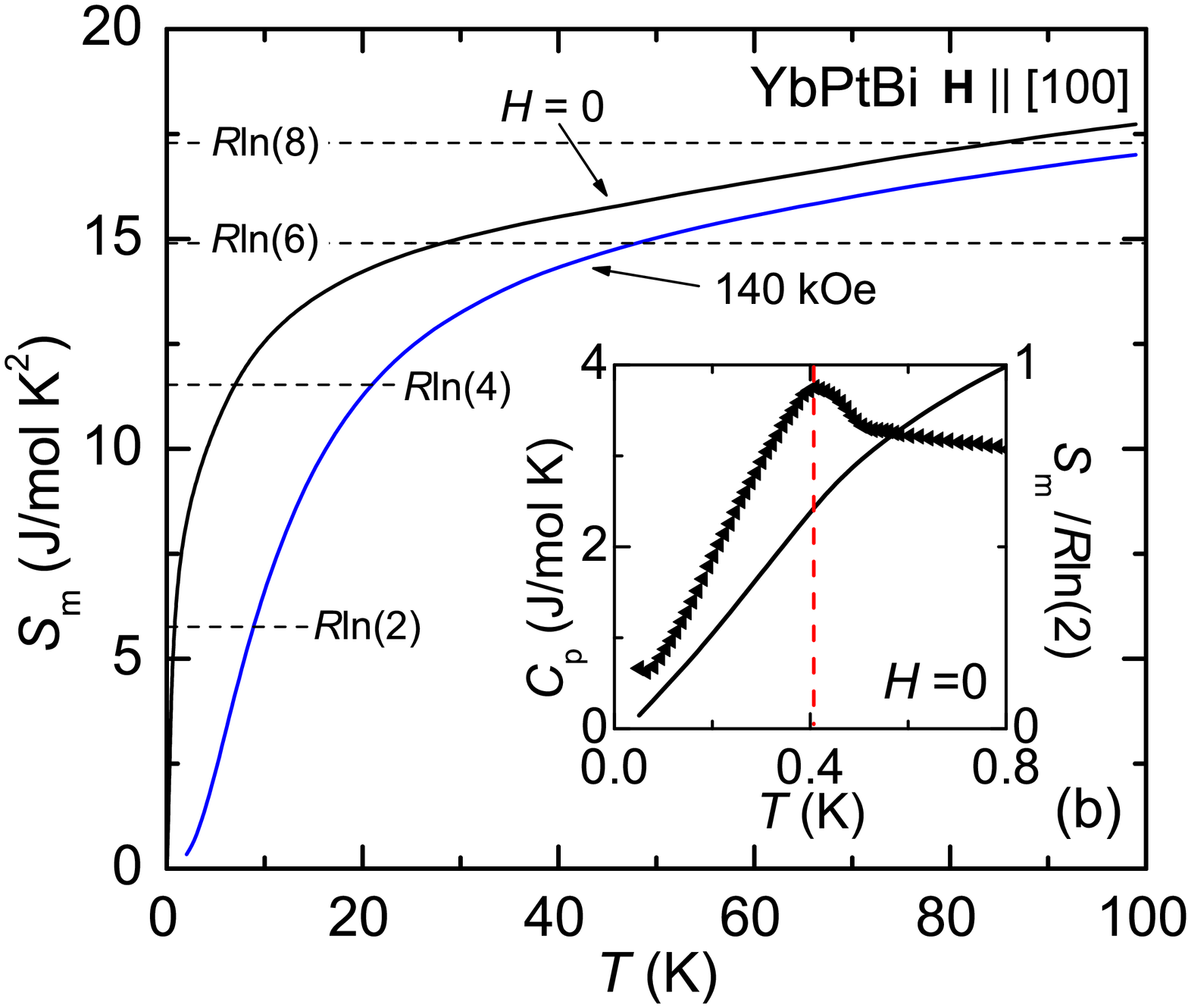}
\caption[$C_{m}(T)$ and $S_{m}(T)$ for YbPtBi]{(a) Logarithmic temperature variation of the magnetic contribution $C_{m}(T)$ to the specific heat of YbPtBi at selected magnetic
fields; $C_{m}(T)$ = $C_{p}(T)$(YbPtBi) - $C_{p}(T)$(LuPtBi). Inset displays positions of maxima developed in $C_{m}(T)$. (b) Magnetic entropy, $S_{m}(T)$, for $H$\,= 0 and 140\,kOe,
inferred by integrating $C_{m}/T$ starting from the lowest temperature measured. Inset shows the low temperature $C_{p}(T)$ of YbPtBi (left axis, symbols) and the magnetic entropy
($S_{m}$) divided by $R$ln(2) (right axis, line). Dashed vertical line marks the peak position of the $\lambda$-shaped anomaly.}
\label{YbPtBiCp1}%
\end{figure*}

The total specific heat obtained for YbPtBi can be assumed to consist of the nuclear Schottky ($C_{N}$), electronic ($C_{el}$), phonon ($C_{lattice}$), and magnetic ($C_{m}$)
contributions. At higher temperatures, where $C_{N}(T)$ contribution can be ignored, $C_{p}(T)$ consists of $C_{el}$, $C_{lattice}$, and $C_{m}$ contributions. Thus, $C_{m}(T)$ of
YbPtBi was estimated by subtracting $C_{p}(T)$ of LuPtBi and plotted as $C_{m}(T)$ vs. $\log$($T$) in Fig. \ref{YbPtBiCp1} (a) for selected magnetic fields.

In zero field, in addition to a distinct anomaly at $T_{N}$, the two anomalies, which can be expected due to the Schottky contributions (associated with the splitting of the Hund's
rule, ground state multiplet of the Yb$^{3+}$ - by the crystalline electric field), are visible near 6\,K and higher than 50\,K. For $H\,>$ 4\,kOe which is high enough to suppress
$T_{N}$, as shown by $\rho(T, H)$ results, a broad peak developes in the low temperature data (around 1 K in the $H$ = 10\,kOe data). The position of the maximum of this low
temperature anomaly continuously shifts to higher temperature as magnetic field increases to 140\,kOe. The anomaly, shown near 6\,K in zero field, merges into lower temperature
anomaly around 40\,kOe, causing significant broadening of the combined feature. The evolution of these two anomalies as a function of magnetic field is plotted in the inset of Fig.
\ref{YbPtBiCp1} (a), where the position of maximum was determined from the Gaussian fit to the data.

For $H$ = 0 and 140\,kOe, the magnetic entropy, $S_{m}(T)$, was inferred by integrating $C_{m}(T)/T$ starting from the lowest temperature measured and plotted in Fig. \ref{YbPtBiCp1}
(b). For $H$ = 140\,kOe, since the $C_{p}(T)$ data were taken above 2\,K and no up-turn in $C_{p}(T)$ data at low temperatures was observed, the nuclear contribution was ignored in
the evaluation of the magnetic entropy. For $H$ = 0, $S_{m}(T)$ reaches about 55\% of $R$ln(2) at $T_{N}$ and recovers the full doublet, $R$ln(2), entropy by $\sim$ 0.8\,K (inset),
which suggests that the ordered moment at $T_{N}$ is compensated (reduced) by Kondo screening. The calculated $S_{m}(T)$ reaches a value of $R$ln(4) by 7\,K and $R$ln(6) by 28\,K, and
the recovered $S_{m}(T)$ at $T$ = 100\,K is close to the full $R$ln(8), which suggests that the highest CEF energy levles are separated by approximately 100\,K from the ground state.
The inferred $S_{m}(T)$ data for $H$ = 140\,kOe is released slower than that for $H$\,=\,0.

\begin{figure*}%
\centering
\includegraphics[width=0.5\linewidth]{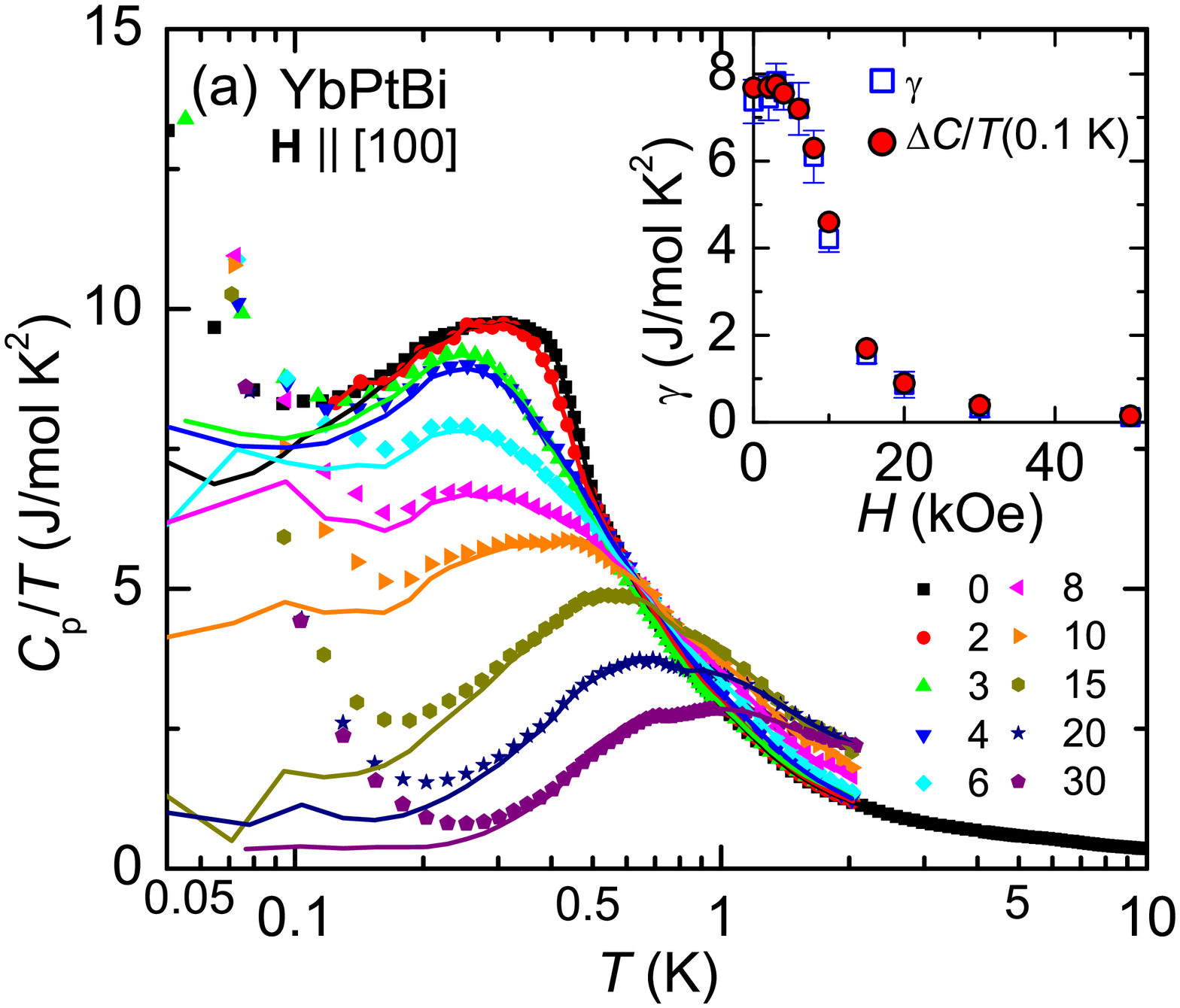}\includegraphics[width=0.5\linewidth]{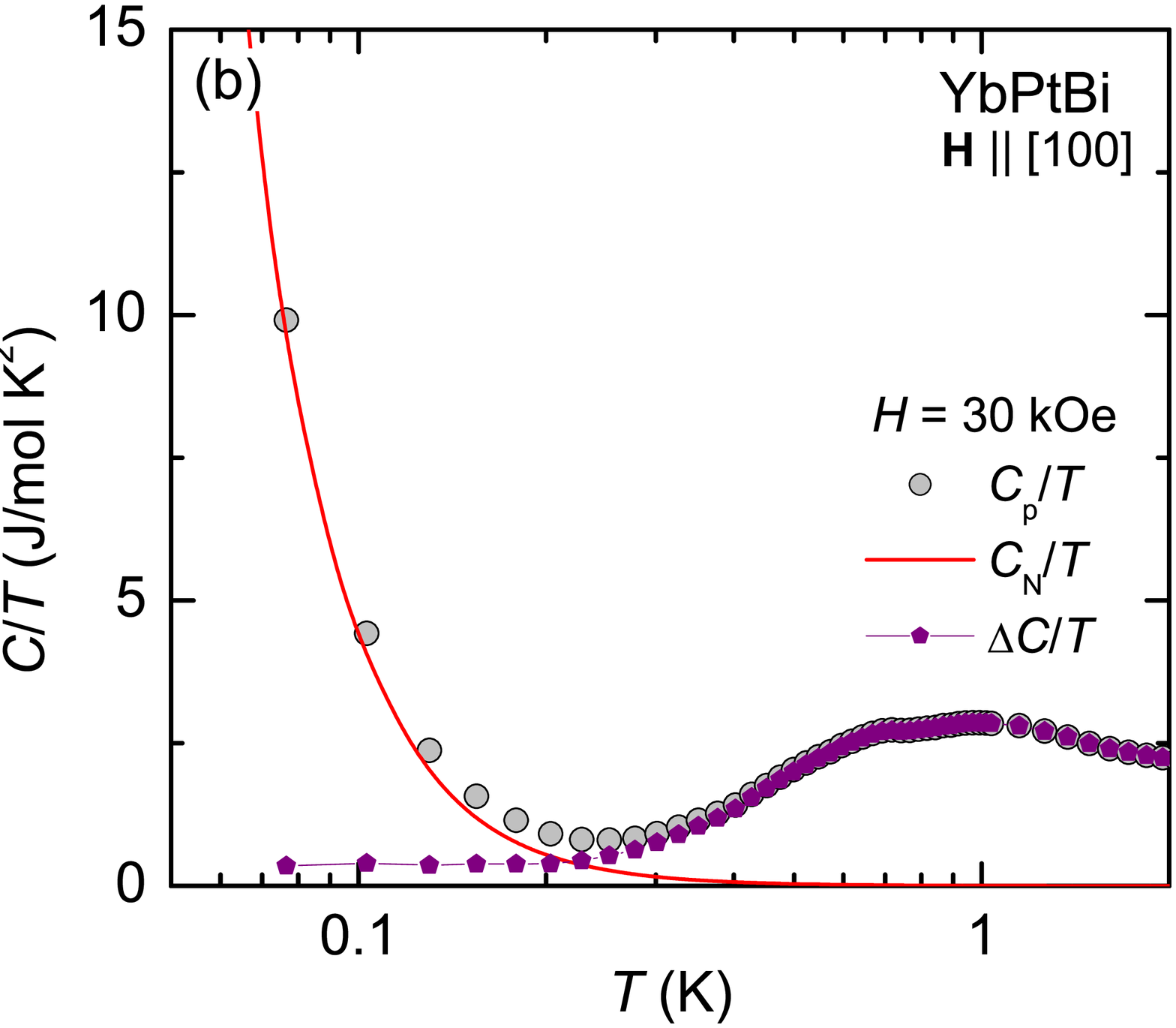}
\caption[Low temperature specific heat of YbPtBi]{(a) Low temperature specific heat as $C_{p}(T)/T$ (solid symbols) and $\Delta C(T)/T$ (solid lines) vs. $\log(T)$ for YbPtBi at
various magnetic fields, applied along the [100] direction; $\Delta C(T)$ = $C_{p}(T)$ - $C_{N}(T)$, where the nuclear Schottky contribution was subtracted by using $C_{N}(T) \propto
1/T^{2}$. Inset shows the electronic specific heat coefficient, $\gamma$ = $\Delta C(T)/T|_{T\rightarrow 0}$ (open squares), and $\Delta C(T)/T$ at $T$ = 0.1\,K (solid circles) as a
function of magnetic field. (b) $C_{p}/T$, $C_{N}/T$, and $\Delta C/T$ for $H$ = 30\,kOe, plotted in a $\log(T)$ scale.}
\label{YbPtBiCp2}%
\end{figure*}

The results of low temperature specific heat measurements shed light on the HF state of YbPtBi, where the evolution of the quasi-particle mass can be inferred as the system is tuned
by external magnetic field. The specific heat data divided by temperature are plotted in Fig. \ref{YbPtBiCp2} (a) (solid symbols) as $C_{p}(T)/T$ vs. $\log$($T$) for $T\,\leq$ 2\,K
and $H\,\leq$ 30\,kOe, where the $C(T)/T$ data for $H$ = 0 are plotted below 10\,K. When magnetic field is applied, the well defined anomaly at $T_{N}$ is no longer visible for
$H\,>$\,3\,kOe and instead the data show a broad maximum. This broad maximum decreases in magnitude and shifts to higher temperature with increasing magnetic field, indicating that
the magnetic entropy is removed at higher temperature for larger applied magnetic fields (see for $H$ = 140\,kOe curve in Fig. \ref{YbPtBiCp1} (b)). At the lowest temperatures, a
slight up-turn in $C_{p}(T)$, associated with a nuclear Schottky anomaly, becomes increasingly visible as field increases. This nuclear Schottky anomaly is much more pronounced in the
$C_{p}(T)/T$ plots.

Below 2\,K, where the $C_{lattice}$ contribution can be safely ignored, the electronic specific heat coefficient was estimated by subtracting the nuclear contribution, using $C_{N}(T)$
$\propto$ $1/T^{2}$; $\Delta C(T)$ = $C_{p}(T)$ - $C_{N}(T)$. As an example, the $C_{p}(T)$, the estimated $C_{N}(T)$, and the $\Delta C(T)$ for $H$ = 30\,kOe are plotted as circles,
line, and pentagons, respectively, in Fig. \ref{YbPtBiCp2} (b). Above $\sim$ 0.4\,K, the $C_{N}(T)$ contribution to the total $C(T)/T$ is very small, however, below $\sim$ 0.2\,K,
$C(T)/T$ is dominated by $C_{N}(T)$ contribution. The obtained $\Delta C(T)$ data for several magnetic fields are plotted as $\Delta C(T)/T$ vs. $\log(T)$ in Fig. \ref{YbPtBiCp2} (a)
(solid lines). In zero field, by extrapolating $\Delta C(T)/T$ to zero temperature ($\gamma$ = $\Delta C(T)/T|_{T\rightarrow 0}$), $\gamma$ is estimated to be 7.4 J/mol$\cdot$K$^{2}$,
which is consistent with earlier result ($\sim$ 8 J/mol$\cdot$K$^{2}$ \cite{Fisk1991}) and is one of the highest effective mass values observed among HF compounds. Note that recently
a similar $\gamma$ value has been observed in face centered cubic YbCo$_{2}$Zn$_{20}$ compound, where no magnetic order was detected down to 20 mK \cite{Torikachvili2007}. The
magnetic field dependence of $\gamma$ is plotted in the inset of Fig. \ref{YbPtBiCp2} (a) as open squares. For comparison, the $\Delta C(T)/T$ data at $T$\,=\,0.1\,K are also plotted
as solid circles, which are essentially the same as $\gamma$. At magnetic fields below 8\,kOe, $\gamma$ is approximately constant within about 1\,J/mol$\cdot$K$^{2}$. A strong
decrease of $\gamma$ is observed for $H$\,$\geq$ 8\,kOe, implying that the quasi-particle mass diverges when approaching the critical field from higher magnetic fields. For magnetic
fields larger than 8\,kOe, $\gamma$ shows a very similar field dependence as $A$ (see Fig. \ref{YbPtBiRTPhase2} (c) and the discussion below). For any of the specific heat data,
measured in magnetic fields up to 30\,kOe, $\Delta C(T)/T$ shows no clear indication of a nFL-like behavior either as a logarithmic (-log($T$)) or non-analytic (-$\sqrt{T}$)
temperature dependence. It should be noted that with such a small $T_K$ value, a simple temperature dependence of $\Delta C/T$ may be convoluted with the field dependence of the Kondo
scale.  A -$\log$($T$) dependence of $\Delta C(T)/T$ is observed over only a limited temperature range; for example, $\Delta C(T)/T$ shows such a -$\log$($T$) dependence between 0.3
$\sim$ 0.8\,K near 4\,kOe and between 0.45 $\sim$ 1.6\,K near 8\,kOe.

\subsection{Thermal expansion and magnetostriction}

\begin{figure*}%
\centering
\includegraphics[width=0.5\linewidth]{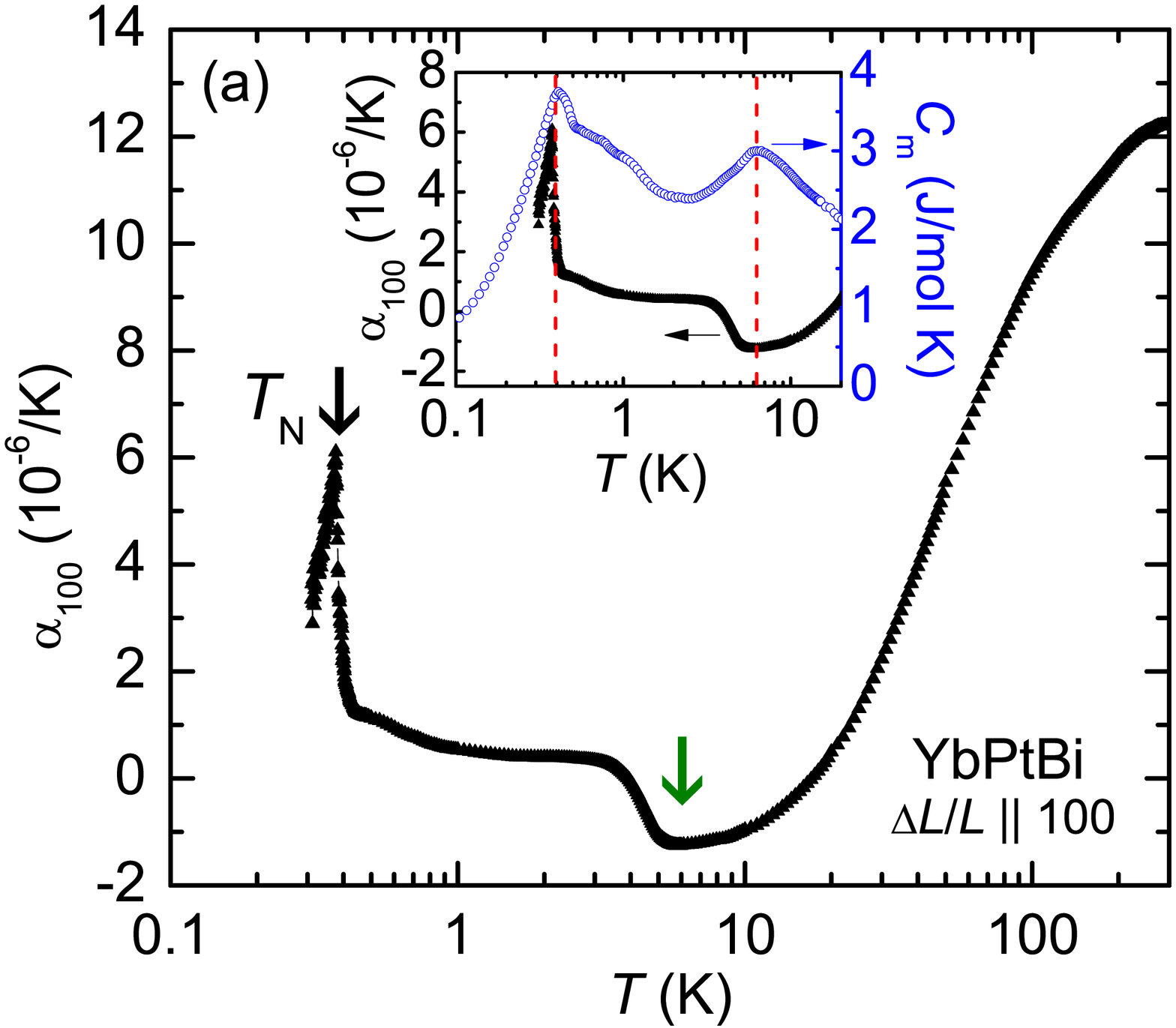}\includegraphics[width=0.5\linewidth]{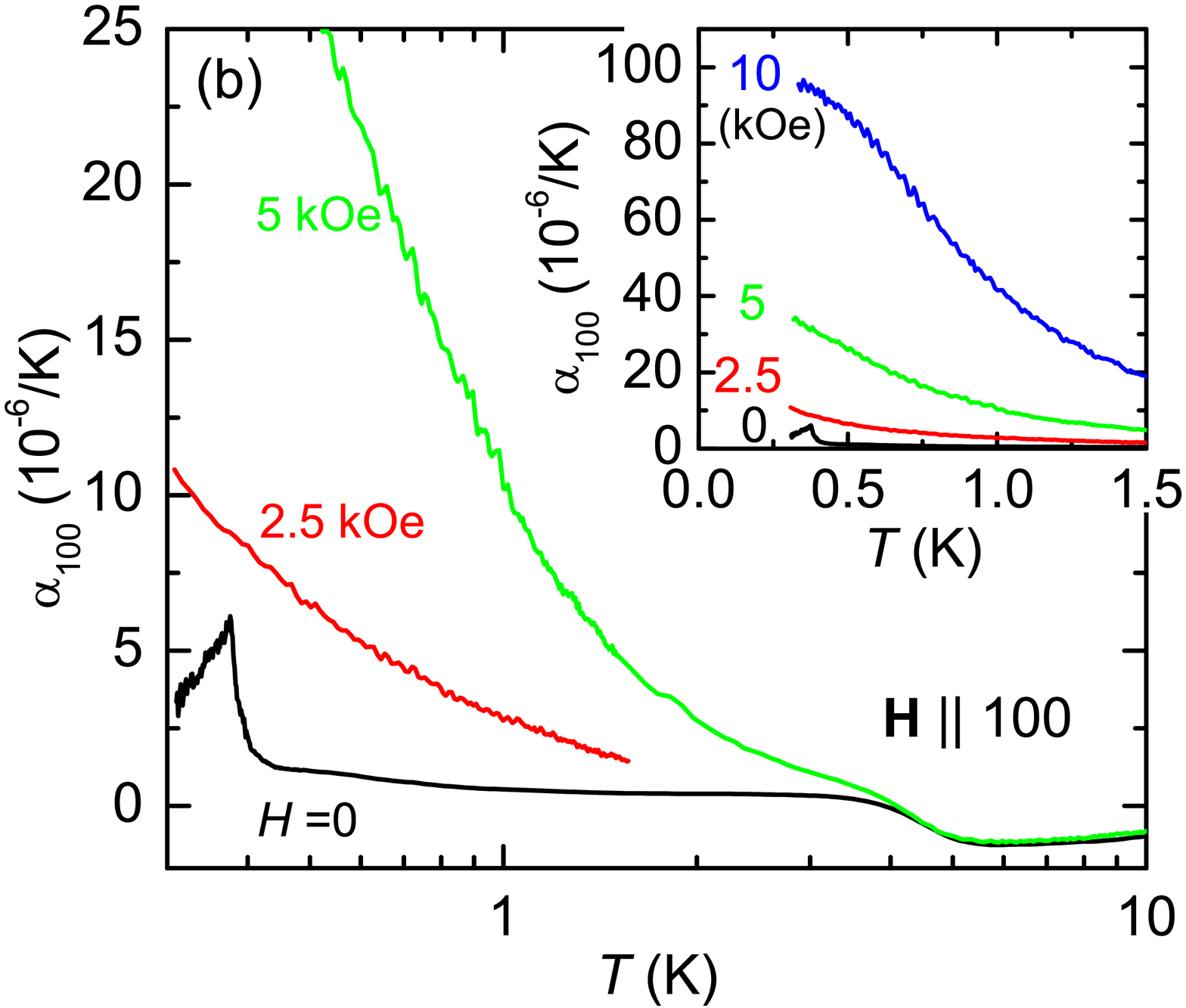}
\caption[Linear thermal expansion coefficient of YbPtBi]{(a) The linear thermal expansion coefficient, $\alpha_{100}$ = d($\Delta L/L_{100}$)/d$T$, of YbPtBi, where $L$ is the sample
length along the [100] direction. The AFM ordering temperature is indicated by the arrow at 0.38\,K. The arrow at 6\,K represents the maximum observed in the $H$ = 0 specific heat
data. Inset shows $\alpha_{100}$ and $C_{m}$ at $H$ = 0. Dashed vertical lines indicate maximum positions observed in $C_{m}$. (b) $\alpha_{100}$ at selected magnetic fields up to 10
K for $H$ = 0, 2.5, and 5 kOe, measured with a longitudinal configuration $\Delta L/L$ $\parallel$ \textbf{H} $\parallel$ [100]. Inset shows $\alpha_{100}$ for $H$ = 0, 2.5, 5, and
10\,kOe (bottom to top) below 1.5 K.}
\label{YbPtBiTE}%
\end{figure*}

Figure \ref{YbPtBiTE} (a) shows the linear thermal expansion coefficient, $\alpha_{100}$ = d($\Delta L/L$)/d$T$, where $\Delta L$ is the length variation along the [100] direction
($\Delta L/L$ $\parallel$ [100]). At high temperatures, $\alpha_{100}$ gradually decreases with lowering temperature and then, below 100 K, $\alpha_{100}$ decreases more rapidly down
to $\sim$ 6\,K. With decreasing temperature further, $\alpha_{100}$ shows a sudden enhancement below 5\,K, followed by a sharp peak at $T$ = 0.38\,K. The observed characteristics in
the temperature dependence of the zero field $\alpha_{100}$ are very similar to that shown in the magnetic specific heat (see inset). The AFM transition manifests itself as a sharp
peak in $\alpha_{100}$ at $T_{N}$ = 0.38\,K, where $C_{m}(T)$ exhibits the AFM transition as a maximum at $T_{N}$ = 0.41\,K. If the thermal expansion, $\Delta L/L$, was composed of
only the lattice contribution, it will only decrease monotonically with decreasing temperature. Thus, the two features, at which $\alpha_{100}$ shows a decrease with warming, at about
5\,K and a saturation for $T\,>$\,100\,K, can be related to a substantial magnetic CEF contribution associated with Yb$^{3+}$ ions, which is in agreement with the broad peak positions
centered at about 6\,K and higher than 50\,K in $C_{m}(T)$. The saturation of $\alpha_{100}$ for $T\,>$ 100\,K is most likely due to CEF effects of higher energy levels combined with
simple lattice effects. Similar $\alpha (T)$ behavior at high temperatures has been shown in YbAl$_{3}$ and YbNi$_{2}$B$_{2}$C \cite{Budko2008, Schmiedeshoff2009}. The anomaly near
5\,K can be related to the first excited state due to CEF effects, where the lattice contribution can be ignored at low temperatures. In order to examine the magnetic field effect on
$\alpha_{100}$ at low temperatures, the temperature-dependent, constant field, thermal expansion was measured in the magnetic field parallel to [100], i.e. $\Delta L$ $\parallel$
\textbf{H} $\parallel$ [100]. The results are plotted in the Fig. \ref{YbPtBiTE} (b). The peak at $T_{N}$ is suppressed below 0.3\,K for $H\,>$ 2.5\,kOe. Below $\sim$5 K anomaly, low
temperature $\alpha_{100}$ increases rapidly with application of magnetic field.

\begin{figure}%
\centering
\includegraphics[width=0.5\linewidth]{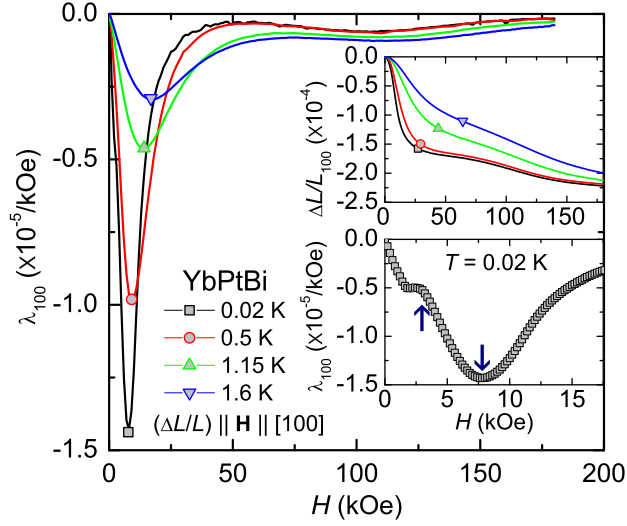}
\caption[Linear magnetostriction and the coefficient of YbPtBi]{Magnetostriction and the coefficient of YbPtBi. The linear magnetostriction coefficient, $\lambda_{100}$ = d($\Delta
L/L_{100}$)/d$H$ vs. $H$, at selected temperatures, where $L$ is the sample length along the [100] direction parallel to the magnetic field applied along the [100] (longitudinal
configuration $\Delta L/L$ $\parallel$ \textbf{H} $\parallel$ [100]). The upper inset shows the magnetic field dependence of the magnetostriction $\Delta L/L_{100}$. The lower inset
shows $\lambda_{100}$ at 0.02\,K. The up- and down-arrow indicate the phase transition $T_{N}$ and the local minimum $H^{*}$, respectively.}
\label{YbPtBiMS}%
\end{figure}

Figure \ref{YbPtBiMS} shows the linear magnetostriction coefficient, $\lambda_{100}$ = d($\Delta L/L_{100}$)/d$H$, and the linear magnetostriction, $\Delta L/L_{100}$ (upper inset), of
YbPtBi for selected temperatures, where the longitudinal linear magnetostriction has been measured parallel to the [100] direction, i.e., $\Delta L$ $\parallel$ \textbf{H} $\parallel$
[100]. The magnetic field was swept with a rate of between 5\,$\sim$\,10 Oe/sec for temperatures up to 10\,K. No hysteresis larger than $\sim$ 100\,Oe could be detected. In the low
magnetic field regime $\Delta L/L$ at $T$ = 0.02\,K shows weak slope changes and then decreases rapidly as magnetic field increases, which manifests in $\lambda_{100}$ as sharp slope
changes below 3\,kOe and a minimum around 7.8\,kOe (see arrows in the lower inset). As temperature is raised, the sharp slope changes are no longer visible for $T >$ 0.4\,K and the
minimum shifts to higher magnetic field. At high magnetic fields, there are broad features: a shoulder near 50\,kOe and a shallow minimum near 100\,kOe in $\lambda_{100}$.

\begin{figure*}%
\centering
\includegraphics[width=1\linewidth]{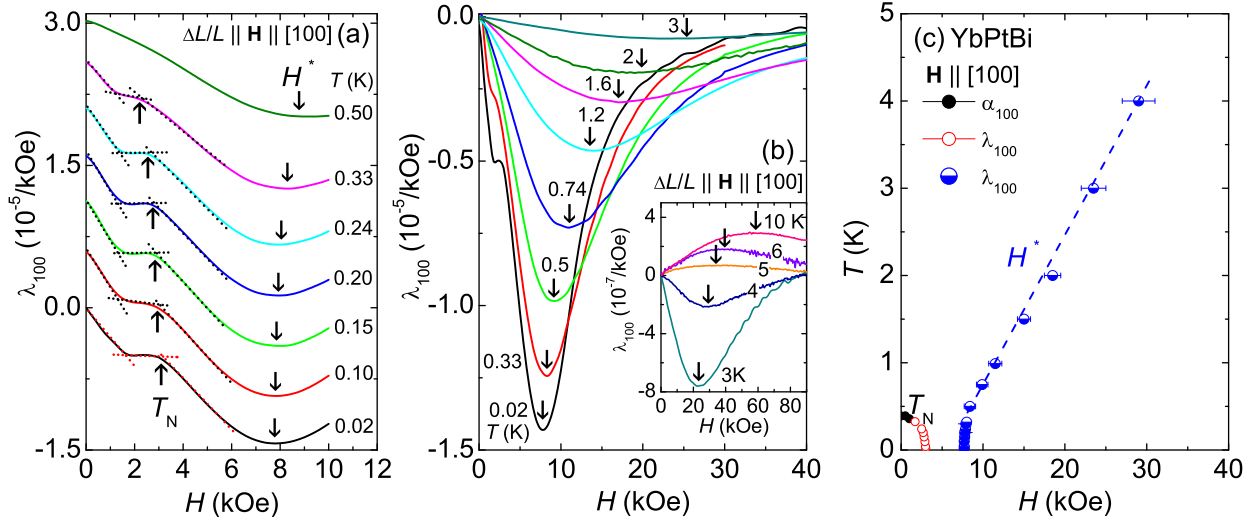}
\caption[$H-T$ phase diagram of YbPtBi constructed from $\alpha_{100}$ and $\lambda_{100}$]{(a) The linear magnetostriction coefficient, $\lambda_{100}$ = d($\Delta L/L_{100}$)/d$H$
vs. $H$, at selected temperatures. For clarity the data sets are vertically shifted each by 5$\times$10$^{-5}$/kOe. The up- and down-arrows indicate the phase transition $T_{N}$ and
the local minimum $H^{*}$, respectively. (b) $\lambda_{100}$ up to 3\,K. Inset shows $\lambda_{100}$ up to 10\,K. Arrows indicate the minimum ($H^{*}$) and maximum in $\lambda_{100}$.
(c) $H-T$ phase diagram of YbPtBi, constructed form $\alpha_{100}$ and $\lambda_{100}$. Dashed-line is guide to eye.}
\label{YbPtBiMSPhase}%
\end{figure*}

Figure \ref{YbPtBiMSPhase} (a) shows a plot of the magnetic field variation of $\lambda_{100}$ at selected temperatures. For $T$ = 0.02\,K data, the two slope changes in
$\lambda_{100}$ are visible at about 1.5 and 3\,kOe. These anomalies shift to lower magnetic field as temperature increases. The phase transition field was selected for the higher
field slope change because the higher field one is well matched with the sharp peak position in d$\rho(H)$/d$H$ (see discussion below). The determined phase transition fields are
indicated by up-arrow in Fig. \ref{YbPtBiMSPhase} (a). The local minimum, observed from $T$ = 0.02\,K curve at $H^{*}$ $\sim$ 7.8\,kOe, is not very sensitive to temperature up to
0.5\,K ($H^{*}$ = 8.4\,kOe), above which $H^{*}$ shifts, almost linearly, to higher magnetic field with further increase of temperature up to 10\,K, which can be clearly seen when
this position is plotted in the $H-T$ plane in Fig. \ref{YbPtBiMSPhase} (c). A negative $\lambda_{100}$ is observed up to 4\,K and it changes to positive for $T\,>$ 5\,K, shown in the
inset of Fig. \ref{YbPtBiMSPhase} (b). Figure \ref{YbPtBiMSPhase} (c) displays a $H-T$ phase diagram constructed from both $\alpha_{100}$ and $\lambda_{100}$: The AFM phase boundary,
$T_{N}$, corresponds to the sharp peak position in $\alpha_{100}$ and the higher field slope change in $\lambda_{100}$, and a crossover scale, $H^{*}$, corresponds to the position of
the minimum for $T$\,$\leq$ 4\,K in $\lambda_{100}$.

\subsection{Hall effect}

\begin{figure}%
\centering
\includegraphics[width=0.5\linewidth]{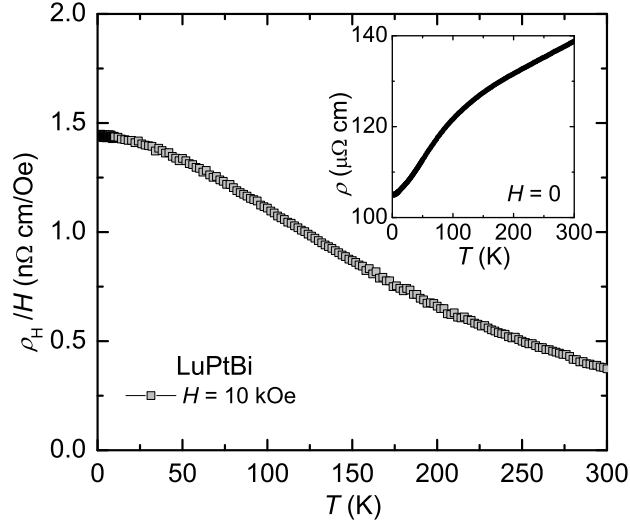}
\caption[Hall coefficient and electrical resistivity of LuPtBi]{Temperature dependence of the Hall coefficient, $R_{H}$\,=\,$\rho_{H}/H$, of LuPtBi for $H$\,=\,10\,kOe, applied along
the [111] direction. Inset shows the zero field resistivity.}
\label{YbPtBiHallLu}%
\end{figure}

Figure \ref{YbPtBiHallLu} shows the temperature-dependent Hall coefficient, $R_{H}$ = $\rho_{H}/H$, of LuPtBi at $H$ = 10\,kOe, applied along the [111] direction. The positive $R_{H}$
of LuPtBi, suggesting that the dominant carriers are holes, monotonically increases as temperature decreases. Assuming a single band model, the carrier concentration at 300\,K is
estimated to be $n$ = 1.7$\times$10$^{26}$ m$^{-3}$ ($R_{H}$ = 0.37 n$\Omega$cm/Oe) corresponding to $\sim$\,0.02 hole per formula unit. As shown in the inset of Fig.
\ref{YbPtBiHallLu} $\rho(T)$ of LuPtBi decreases as temperature is lowered. Thus, LuPtBi can be characterized as a low carrier density metallic (or semimetallic) system. The carrier
concentration of LuPtBi is approximately 100 times smaller than that for copper \cite{Ziman1960}, comparable to that for earlier result of isostructural YbPtBi \cite{Hundley1997}, and
2 orders of magnitude larger than that of NdPtBi \cite{Morelli1996} and LaPtBi \cite{Jung2001}. This trend is consistent with the earlier resistivity results \cite{Canfield1991} in
which the resistivity systematically changes from a small gap semiconductor (or semimetal) for lighter rare-earth compounds to metallic (or semimetallic) for heavier rare-earth
compounds.

\begin{figure}%
\centering
\includegraphics[width=0.5\linewidth]{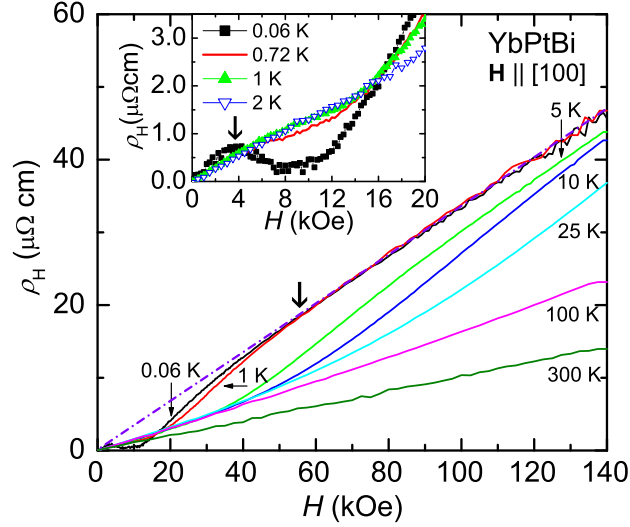}
\caption[Magnetic field dependence of the Hall resistivity for YbPtBi]{Hall resistivity, $\rho_{H}$, of YbPtBi as a function of magnetic field, applied along the [100] direction, at
various temperatures. The arrow pointing to the 0.06\,K curve near 55\,kOe indicates a deviation from linear field dependence of $\rho_{H}$. The dash-dotted line is guide to the eye.
Inset shows the low temperature and low field $\rho_{H}$ at selected temperatures.}
\label{YbPtBiHallH}%
\end{figure}

Figure \ref{YbPtBiHallH} displays the magnetic field-dependent Hall resistivity, $\rho_{H}$, of YbPtBi in magnetic fields up to 140\,kOe at various temperatures. The high temperature
results, obtained in this study (shown as inset to Fig. \ref{YbPtBiHallH1} below), are similar to previous Hall effect measurements above 2\,K \cite{Hundley1997}. Here, the
measurements have been extended to much higher magnetic fields, up to 140\,kOe, and to much lower temperatures, down to 0.06\,K, investigating the phenomena that are related to quantum
criticality. Below 1\,K the $\rho_{H}$ data as function of temperature and magnetic field were taken with the condition that the sample was mounted on a dilution refrigerator cold
stage with very thin layer of GE-varnish. At high temperatures (for $T$\,$\geq$ 0.5\,K), after cleaning the GE-varnish off using ethanol, the sample was mounted on the cold stage of
$^{3}$He option in PPMS with Apiezon N-grease and $\rho_{H}$ was measured. The data, taken from a dilution refrigerator measurements, are in good agreement with the data, taken from
$^{3}$He setup.

The sign of $\rho_{H}$ is positive for all temperatures measured which, as was the case for LuPtBi, is suggestive that hole-type carriers are dominant. Above 100\,K, $\rho_{H}$
follows a linear magnetic field dependence, whereas, for $T\,\leq$ 25\,K, $\rho_{H}$ exhibits a non-linear magnetic field dependence. A clear deviation from the linear magnetic field
dependence of $\rho_{H}$ is shown in Fig. \ref{YbPtBiHallH} and indicated by the heavy arrow on 0.06\,K data. As highlighted in the inset, the overall features of $\rho_{H}$ at
0.06\,K are strongly non-monotonic as a function of magnetic field. The $\rho_{H}$ data manifest distinct features: a local maximum around 4\,kOe and a broad local minimum between 4
and 12\,kOe. These characteristic features broaden for $T$ $>$ 0.4 K.

Figure \ref{YbPtBiHallH1} shows $R_{H}$ of YbPtBi as a function of magnetic field. At high temperatures (inset), $R_{H}$ is almost magnetic field-independent. As temperature is
lowered, a broad local minimum in $R_{H}$ is developed and sharpened. An anomalous low temperature behavior of Hall effect can be clearly seen in $R_{H}$ plot; at base temperature, $T$
= 0.06\,K, the high magnetic field limit of $R_{H}$ ($H$ $\rightarrow$ 140\,kOe) is close to the low magnetic field limit of $R_{H}$ ($H$ $\rightarrow$ 0), but, as magnetic field
increases from $H$ = 0 two features develops a weak slope change near 4\,kOe and a clear minimum around 8\,kOe. Given that similar features are also seen in the MR and
magnetostriction measurements, the anomaly near 4\,kOe can be related to the AFM phase boundary, and the 8\,kOe anomaly tracks the $H^{*}$ line.

\begin{figure*}%
\centering
\includegraphics[width=0.5\linewidth]{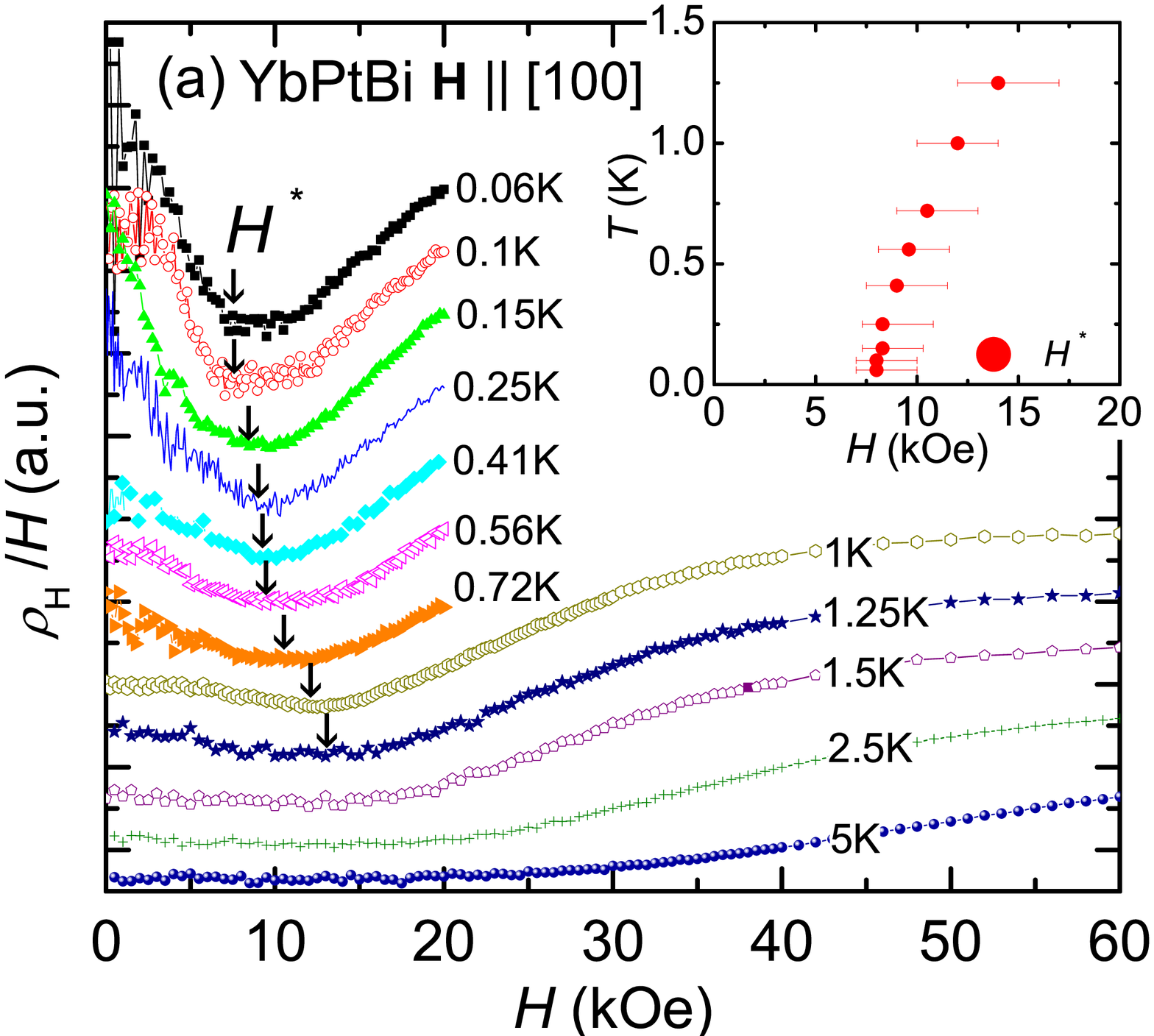}\includegraphics[width=0.5\linewidth]{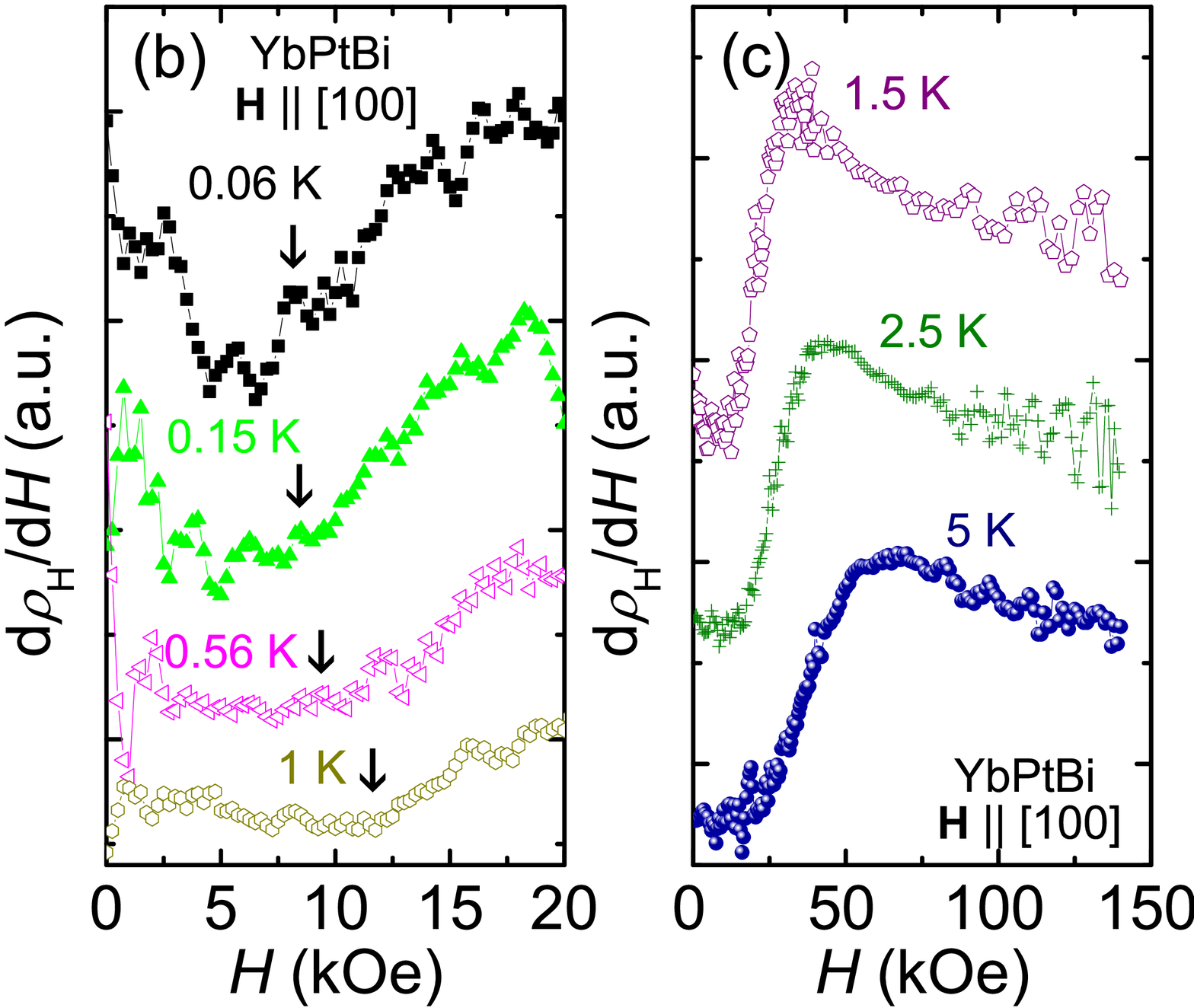}
\caption[Magnetic field dependence of the $\rho_{H}/H$ at selected temperatures]{(a) $\rho_{H}/H$ at selected temperatures. For clarity, the data sets have been shifted by different
amounts vertically. Arrows indicate the position of local minimum. The positions of $H^{*}$ are plotted in the $H-T$ plane in the inset. For comparison, d$\rho_{H}$/d$H$ at selected
temperatures are plotted in (b) and (c). Vertical arrows in (b) indicate the position $H^{*}$ used in (a) as the local minimum.}
\label{YbPtBiHallPhase1}%
\end{figure*}

Figure \ref{YbPtBiHallPhase1} (a) shows $R_{H}$ of YbPtBi at selected low temperatures; the data sets have been shifted by different amounts vertically for clarity. Because of the
poor signal to noise ratio associated with low field measurements, the position of the characteristic feature of the SDW transition can not be determined precisely. The local maximum
in $\rho_{H}$ near 4\,kOe that is clear at 0.06\,K, (inset, Fig. \ref{YbPtBiHallH}) broadens significantly as temperature increases and is no longer visible for $T\,>$ 0.5\,K. The
local minimum, $H^{*}$ $\sim$ 8\,kOe observed at $T$ = 0.06\,K, gradually shifts to higher magnetic fields as temperature increases. For $T$ $>$ 1.25 K, $R_{H}$ does not show the
local minimum. The determined positions of the local minimum are indicated by arrows in Fig. \ref{YbPtBiHallPhase1} (a) and also plotted in the $H-T$ plane in the inset. For
comparison, d$\rho_{H}$/d$H$ curves are plotted in Figs. \ref{YbPtBiHallPhase1} (b) and (c). The local minimum observed in $R_{H}$ is indicated by arrows in (b). In the high field
region, a local maximum in d$\rho_{H}$/d$H$ is observed and shifted to higher fields as temperature increases (Fig. \ref{YbPtBiHallPhase1} (c)). As will be shown below, the positions
of $H^{*}$ agree with the anomalies developed in MR, magnetostriction, specific heat, and TEP measurements.

\begin{figure*}%
\centering
\includegraphics[width=0.5\linewidth]{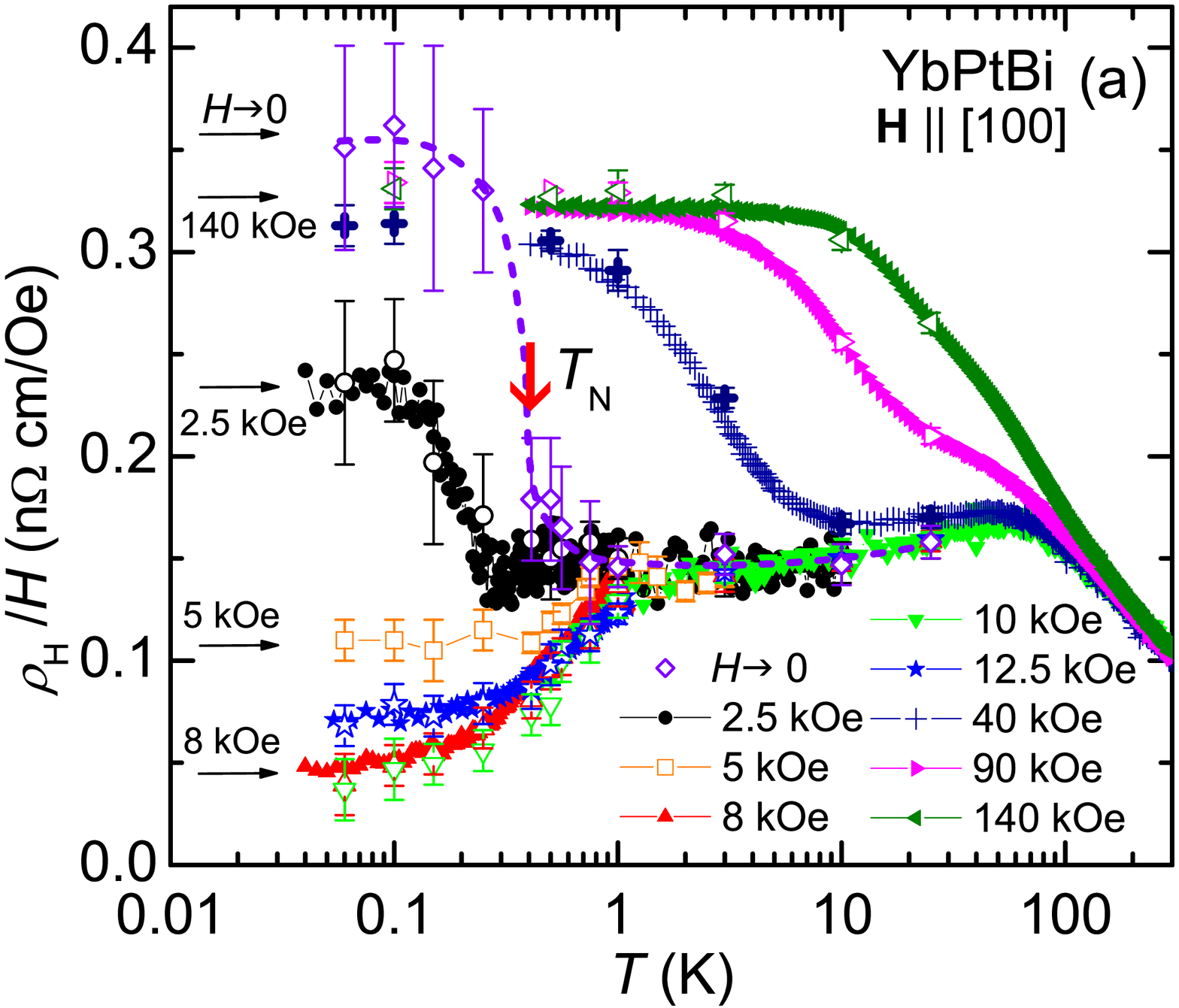}\includegraphics[width=0.5\linewidth]{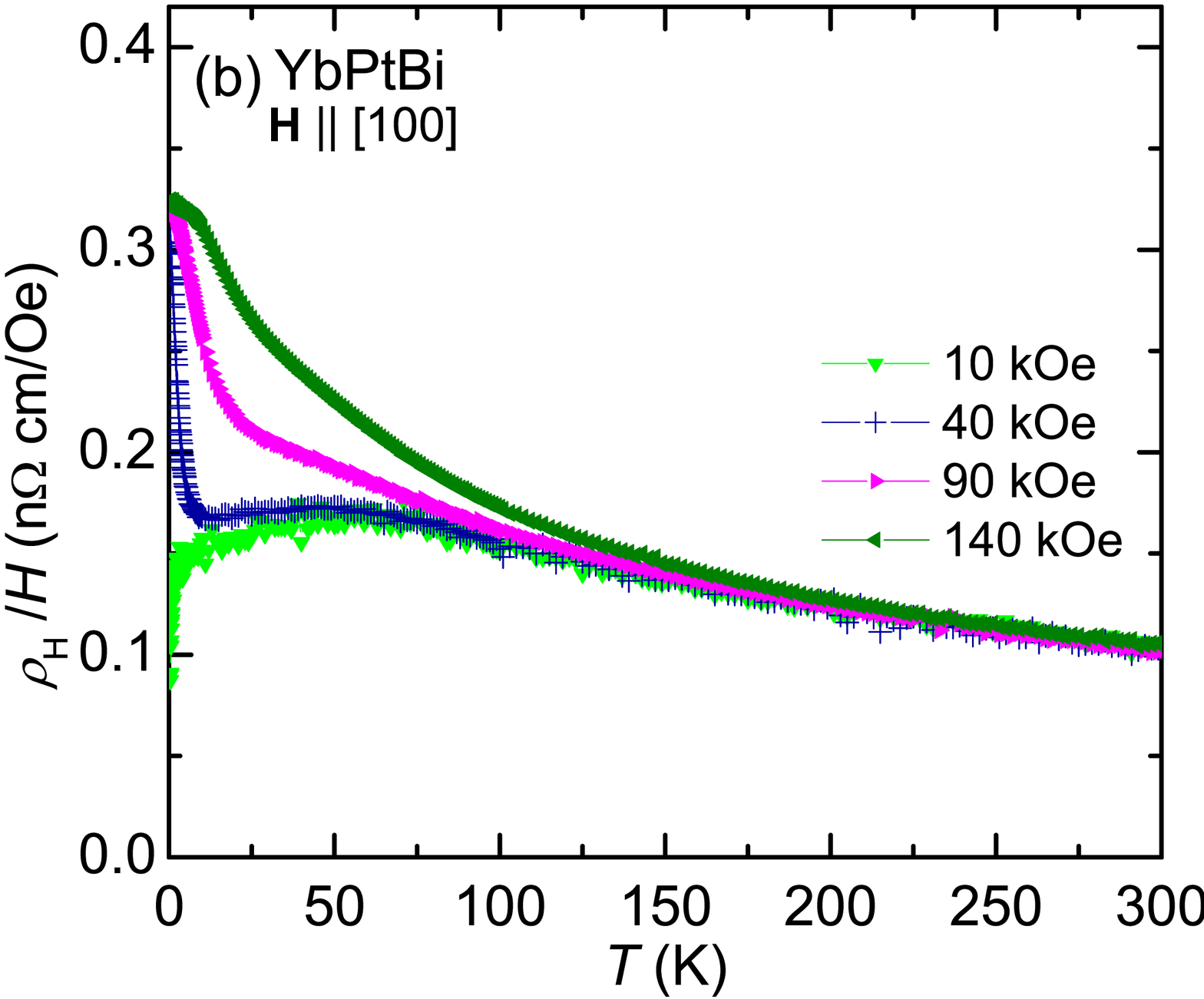}
\caption[Temperature dependence of the Hall coefficient for YbPtBi]{(a) Temperature dependence of the Hall coefficient ($R_{H}$ = $\rho_{H}/H$) of YbPtBi at various magnetic fields,
applied along the [100] direction. Closed- and open-symbols are taken from temperature and field sweeps of $\rho_{H}$, respectively. The open-diamond symbols ($\diamond$) of $R_{H}(T
\rightarrow 0)$ are obtained from the initial slope of $\rho_{H}$ vs. $H$. The dashed-line is guide to the eye. (b) $R_{H}$ for $H$ = 10, 40, 90, and 140 kOe in linear scale of $T$.}
\label{YbPtBiHallT}%
\end{figure*}

In Figs. \ref{YbPtBiHallT} (a) and (b), $R_{H}$ is plotted as a function of temperature at selected magnetic fields, where closed- and open-symbols are taken from temperature and
magnetic field sweeps of $\rho_{H}$, respectively. The $R_{H}$ ($H$ $\rightarrow$ 0) data were obtained by taking the low field limit of d$\rho_{H}$/d$H$; given the weak, low field,
signal of $\rho_{H}$, the error bars for $R_{H}$ ($H$ $\rightarrow$ 0) are large. In the low magnetic field ($H\rightarrow 0$ and 2.5\,kOe) results, $R_{H}$ shows a clear change near
0.4\,K and $\sim$ 70\,K. The steep increase by factor of $\sim$2.3 below 0.4\,K in $R_{H}$ ($H$ $\rightarrow$ 0) agrees with the behavior observed from resistivity measurements and is
consistent with a partial gapping of the Fermi surface (see Fig. \ref{YbPtBiRT2a}). The temperature dependence of $R_{H}$ depends strongly on the applied magnetic field below
$\sim$100\,K, whereas above $\sim$100\,K $R_{H}$ is basically magnetic field-independent for $H$\,$\leq$ 140\,kOe as shown in Fig. \ref{YbPtBiHallT} (b).

As temperature decreases, the zero field limit $R_{H}$ ($H$ $\rightarrow$ 0) data below 10\,K show a very weak temperature dependence and the opening of the SDW gap below $T_{N}$ =
0.4\,K gives rise to an abrupt enhancement of $R_{H}$ ($H$ $\rightarrow$ 0). A steep increase of $R_{H}$ below $T_{N}$ implies a significant carrier density reduction associated with
the Fermi surface gapping. For $H$ = 5\,kOe $R_{H}$ becomes almost temperature-independent below 10\,K. Similar results have been observed in URu$_{2}$Si$_{2}$ compound
\cite{Schoenes1987}. Below $T_{0}$ = 17.5 K, $R_{H}$ of URu$_{2}$Si$_{2}$ increases by factor of 5-20 because of the opening of a gap over the Fermi surface. It should be noted,
though, that since the Hall sample was mounted with GE-varnish, there is a possibility that the steep increase by factor of $\sim$2.3 in $R_{H}$ below 0.4\,K may be altered by strain
(as was the resistivity). Thus, in order to further clarify the actual reduction of carrier density due to the gapping of Fermi surface below $T_{N}$, Hall data and resistivity data
would need to be collected at the same time on a sample secured by contact wires and dental floss (i.e. minimal strain anchoring).

\subsection{Thermoelectric power}%

\begin{figure}%
\centering
\includegraphics[width=0.5\linewidth]{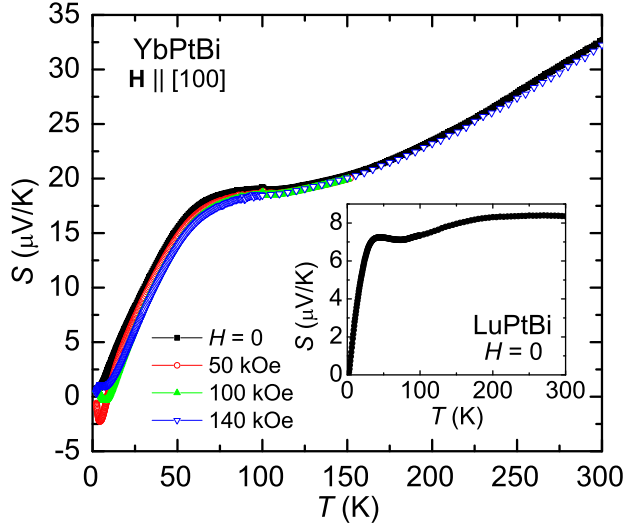}
\caption[High temperature TEP of YbPtBi and LuPtBi]{Temperature-dependent TEP, $S(T)$, of YbPtBi at selected magnetic fields, applied along the [100] direction. Inset shows the zero
field $S(T)$ of LuPtBi.}
\label{YbPtBiST1}%
\end{figure}

The TEP as a function of temperature, $S(T)$, for LuPtBi is plotted in the inset of Fig. \ref{YbPtBiST1}. The positive sign of TEP indicates that holes are dominant carriers which is
consistent with $R_{H}$ results. As temperature increases $S(T)$ increases monotonically, after passing through a broad peak structure around 40\,K probably due to the phonon drag,
and then $S(T)$ gradually increases to 8 $\mu$V/K at 250\,K. Above 250\,K $S(T)$ shows an essentially temperature-independent behavior up to 300\,K. The observed TEP of LuPtBi is not
consistent with the behavior expected from simple metals and the origin of the strong break in slope near 40\,K is unknown at present.

Figure \ref{YbPtBiST1} shows the evolution of $S(T)$ for YbPtBi with magnetic fields applied along the [100] direction. In zero field the observed TEP is positive, indicating that
holes are dominant carriers which is consistent with $R_{H}$ results and with previous TEP results \cite{Hundley1997} above 2\,K. However, the positive sign of TEP for YbPtBi is
opposite to that generally observed in Yb-based HF systems, which is negative due to the location of a narrow Kondo resonance peak slightly below the Fermi energy \cite{Hewson1993}.
The broad shoulder structure, centered around 70\,K, can be associated with excited CEF energy levels of Yb$^{3+}$ ions. This can be also related to the appearance of a high
temperature broad maximum around 70\,K in $\rho_{H}/H$ and an inflection point near 85\,K in $\rho(T)$. In these cases the temperature of the CEF related features corresponds to a
fraction of the CEF splitting (0.4-0.6$\Delta_{CEF}$) as evidenced in many other Ce- and Yb-based compounds and alloys \cite{Maekawa1986, Bickers1987, Ocko2004, Kohler2008}.

\begin{figure*}%
\centering
\includegraphics[width=0.5\linewidth]{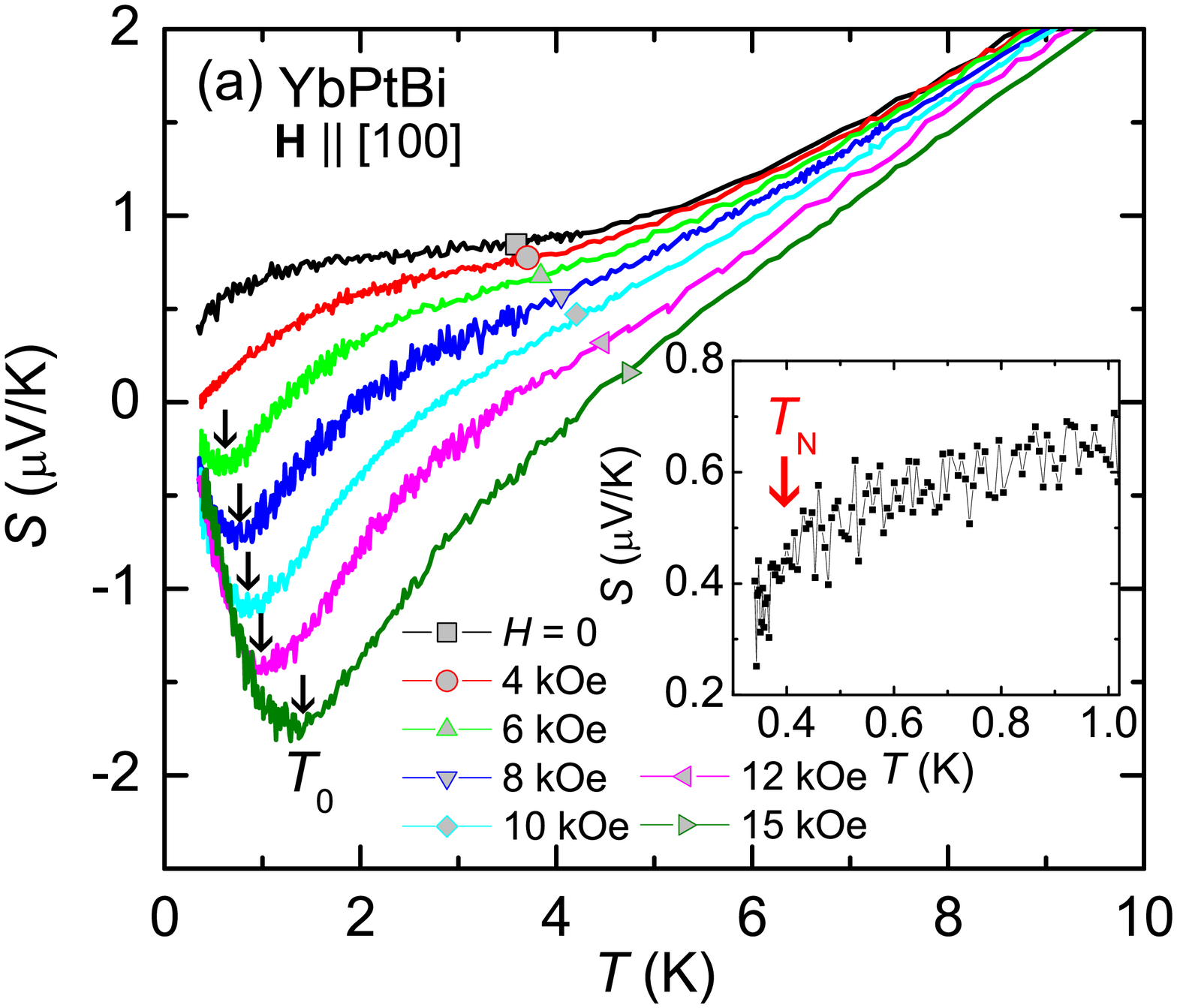}\includegraphics[width=0.5\linewidth]{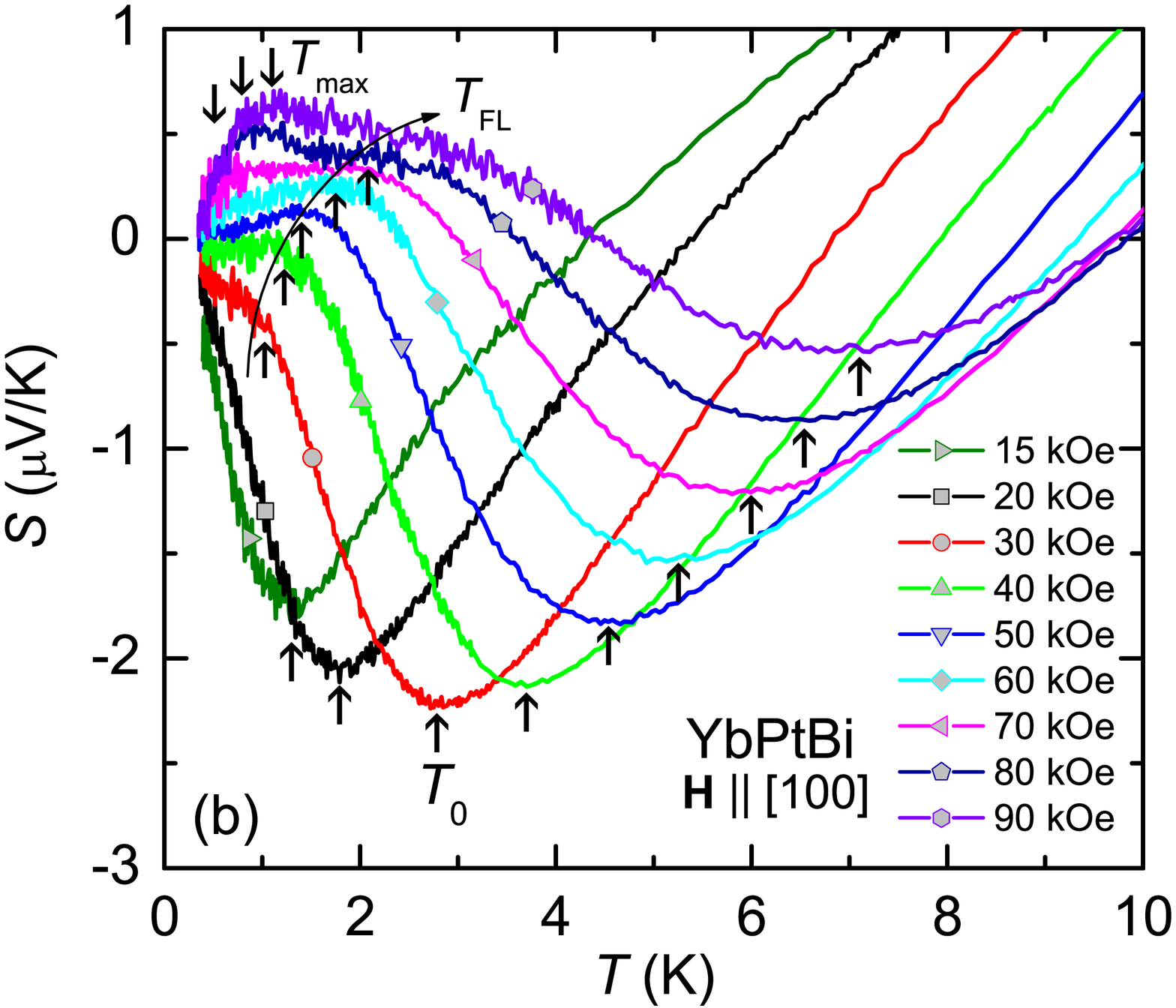}
\caption[Low temperature TEP of YbPtBi]{(a) Low temperature $S(T)$ of YbPtBi at selected magnetic fields for $H$\,$\leq$ 15\,kOe. Vertical arrows indicate a local minimum $T_{0}$.
Inset shows the zero field $S(T)$ below 1\,K. Vertical arrow represents the AFM ordering temperature at which the slope, d$S(T)$/d$T$, is changed. (b) Low-temperature $S(T)$ for 15
$\leq$\,$H$\,$\leq$\, 90\,kOe. Vertical arrows indicate the characteristic features corresponding to a local minimum temperature $T_{0}$, a slope change at $T_{FL}$, and a local
maximum $T_{max}$ for $H$ $\geq$ 70\,kOe.}
\label{YbPtBiST2}%
\end{figure*}

$S(T)$ changes very little with applied magnetic field for $T\,\gtrsim$ 20\,K. For $T\,\lesssim$ 20\,K $S(T)$ shows a rather complex behavior, with the emergence of new broad peak
structures as magnetic field increases. In Figs. \ref{YbPtBiST2} (a) and (b), the low temperature TEP data for YbPtBi are plotted as $S(T)$ vs. $T$ for selected magnetic fields. In
contrast to the high temperature behavior, $S(T)$ data reveal complex and strong magnetic field dependences. In zero field, the sign of TEP is positive down to 0.35 K (the base
temperature of the $^3$He system used) and $S(T)$ exhibits a broad feature around 2\,K. No clear signature of the AFM phase transition near 0.4\,K is observed. As presented in the
inset there is a weak change in slope near $T_{N}$. Generally, for a SDW antiferromagnet such as Cr \cite{Fawcett1988}, the TEP measurements revealed a sudden enhancement due to the
opening a gap below SDW state, similar to what was seen in the zero field limit Hall data ($R_{H}$($H\,\rightarrow$\,0)) in Fig. \ref{YbPtBiHallT}. Unfortunately, at 0.35\,K $S(T)$ is
just starting to change; lower temperature measurements (e.g. in a dilution refrigerator) will be needed to fully define the zero field $S(T)$ feature. When a magnetic field is
applied along the [100] direction, $S(T)$ curves shift toward a negative direction and a local minimum, $T_{0}$, develops for $H\,>$ 5\,kOe. The position of $T_{0}$ continuously
shifts to higher temperature as magnetic field increases up to 90 kOe, indicated by arrows in Figs. \ref{YbPtBiST2} (a) and (b). For 30 $< H\,<$ 70\,kOe, the low temperature behavior
changes significantly; the TEP shows the development of a new, broad feature, $T_{FL}$, below which $S(T)$ $\propto$ $T$ is indicated by arrows in Fig. \ref{YbPtBiST2} (b). For $H\,>$
70\,kOe, an additional local maximum, $T_{max}$, develops with $T_{max}$ $<$ $T_{0}$. The positions of both $T_{FL}$ and $T_{max}$ shift to higher temperature with increasing magnetic
field.

\begin{figure}%
\centering
\includegraphics[width=0.5\linewidth]{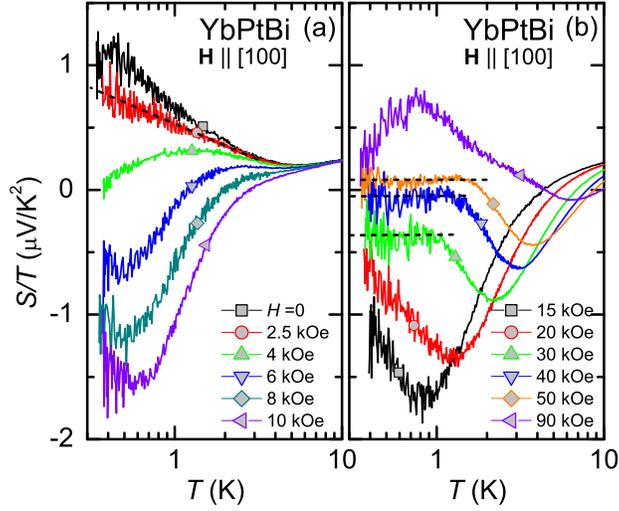}
\caption[$S(T)/T$ of YbPtBi at selected magnetic fields]{Temperature-dependent TEP divided by temperature, $S(T)/T$, in a logarithmic scale; $S(T)/T$ vs. $\log$($T$). (a) The
dashed-line on the curve for $H$ = 2.5\,kOe is guide to eye, representing a logarithmic increase of $S(T)/T$ below 4\,K. (b) Dashed-lines on the curves for $H$ = 30, 40, and 50\,kOe
indicate a saturation of $S(T)/T$ which corresponds to the linear temperature dependence of $S(T)$ below $T_{FL}$, shown in Fig. \ref{YbPtBiST2} (b).}
\label{YbPtBiST3}%
\end{figure}

In order to investigate the low temperature behavior, a plot of $S(T)/T$ is presented in Figs. \ref{YbPtBiST3} (a) and (b) as a function of $\log$($T$) for selected magnetic fields.
In zero field, $S(T)/T$ exhibits a logarithmic temperature dependence between $T_{N}$ and $\sim$ 3\,K. For $H$ = 2.5\,kOe the $\log$($T$) dependence of $S(T)/T$ holds below 4\,K. This
$\log$($T$)-dependence of $S(T)/T$ has been observed for YbRh$_{2}$Si$_{2}$ \cite{Hartmann2010} and YbAgGe \cite{Mun2010b} in the vicinity of the QCP, as a signature of nFL-like
behavior. As magnetic field increases $S(T)/T$ moves toward negative direction for $H\,>$ 4 kOe, and the low temperature behavior changes dramatically. At higher fields, for $H$ = 30,
40, and 50 kOe, and for $T\,<\,T_{FL}$ (Fig. \ref{YbPtBiST2}) $S(T)/T$ = $\alpha$, indicating the onset of FL behavior. For $H$ = 90\,kOe $S(T)/T$ deviates from a constant, indicating
a deviation from FL behavior, due to the development of the local maximum, $T_{max}$, (see Fig. \ref{YbPtBiST2}).

\begin{figure}%
\centering
\includegraphics[width=0.5\linewidth]{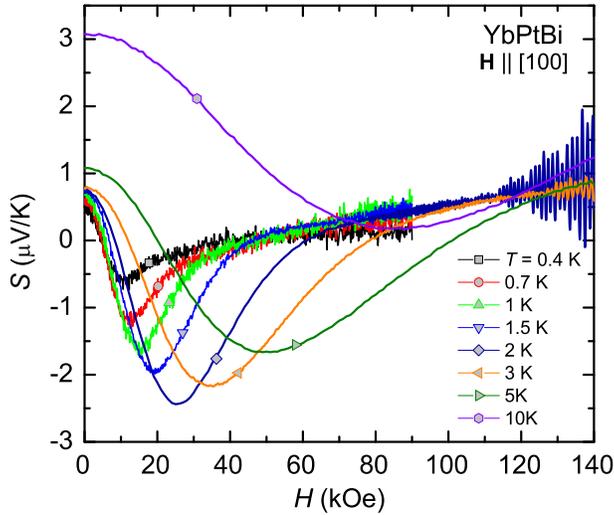}
\caption[$S(H)$ of YbPtBi at selected temperatures]{Magnetic field dependence of TEP, $S(H)$, of YbPtBi at selected temperatures.}
\label{YbPtBiSH1}%
\end{figure}

Figure \ref{YbPtBiSH1} shows the magnetic field dependence of TEP, $S(H)$, for YbPtBi. As magnetic field increase $S(H)$ curves initially decrease steeply and then increase after
passing through a minimum, $H^{*}$. For $H\,>$ 110\,kOe at $T$ = 2\,K, the oscillatory behavior corresponds to quantum oscillations, which is consistent with Shubnikov de Haas (SdH)
results. As temperature increases from 0.4\,K, $H^{*}$ shifts to higher magnetic fields and the absolute TEP value at $H^{*}$ increases up to 2\,K and then decreases. The sign of TEP
changes from positive to negative around $H_{SR}$ = 4.2\,kOe at 0.4\,K and recovers a positive sign near 43\,kOe; both $H_{SR}$ values move to higher magnetic fields with increasing
temperature. For $H\,>$ 100\,kOe and $T\,>$ 10\,K a sign reversal on TEP is no longer visible.

\begin{figure}%
\centering
\includegraphics[width=0.5\linewidth]{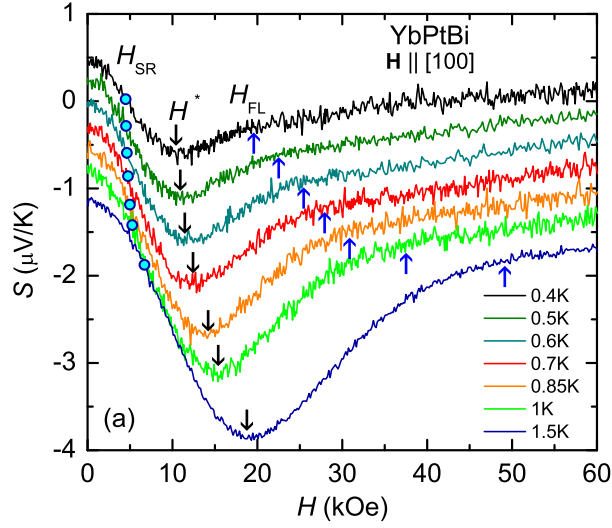}
\caption[$S(H)$ of YbPtBi below 1.5\,K]{$S(H)$ of YbPtBi below 1.5\,K. For clarity, the data sets have been shifted by every -3$\mu$V/K vertically. Solid circles $H_{SR}$ indicate a
sign change of TEP from positive to negative. Down-arrows $H^{*}$ represent the determined position of the local minimum. Up-arrows indicate a slope change, d$S(H)$/d$H$, above which
$S(H)$ follows a linear field dependence.}
\label{YbPtBiSH2}%
\end{figure}

Figure \ref{YbPtBiSH2} shows the $T$ $\leq$ 1.5 K $S(H)$ data, where each $S(H)$ curve is shifted by -0.3 $\mu$V/K, for clarity. In addition to the lower $H_{SR}$ and $H^{*}$, there is
a slope change, $H_{FL}$, near 20\,kOe above which $S(H)$ is linear in magnetic field. The lower sign reversal ($H_{SR}$), the local minimum ($H^{*}$), and the slope change ($H_{FL}$)
on $S(H)$ move to higher magnetic fields with increasing temperatures, indicated by solid circles, down arrows, and up arrows, respectively, in Fig. \ref{YbPtBiSH2}.

\begin{figure}%
\centering
\includegraphics[width=0.5\linewidth]{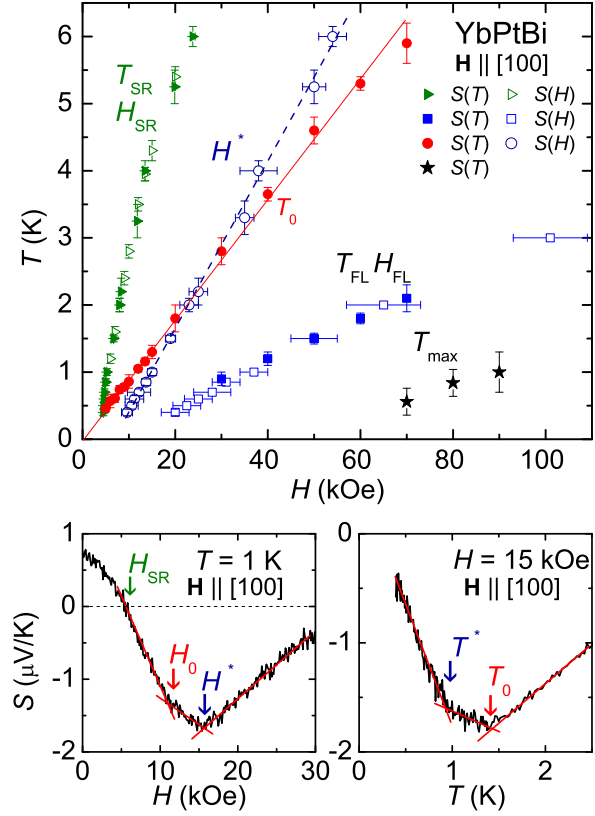}
\caption[$H-T$ phase diagram of YbPtBi constructed from $S(T, H)$]{Features from $S(T, H)$ measurements plotted in $H-T$ diagram: $T_{SR}$ and $H_{SR}$ represents the sign reversal
extracted from the position of $S(T,H)$ = 0; $H^{*}$ marks the position of the local minimum in $S(H)$; $T_{0}$ indicates the position of the local minimum in $S(T)$; $T_{max}$
represents the position of the local maximum developed at low temperatures for $H\,\geq$ 70\,kOe; and $T_{FL}$ and $H_{FL}$ represent the slope change in $S(T)$ (Fig. \ref{YbPtBiST2})
and $S(H)$ (Fig. \ref{YbPtBiSH2}), respectively. Bottom panels show the horizontal and vertical cut through the $H-T$ plane. Left panel: below 30\,kOe $S(H)$ at $T$ = 1\,K hits all
three characteristic lines of $H_{SR}$, $H_{0}$, and $H^{*}$. Right panel: below 2.5\,K $S(T)$ at $H$ = 15\,kOe indicates both $T^{*}$ and $T_{0}$ line. See details in the text.}
\label{YbPtBiTEPPhase}%
\end{figure}

The features, collected from the $S(T)$ and $S(H)$ measurements, are plotted in the $H-T$ plane in Fig. \ref{YbPtBiTEPPhase}. In zero field a weak signal as a small drop near 0.4\,K
is consistent with the $T_{N}$ determined from resistivity (not shown in figure). The sign reversal temperatures determined from $S(T)$ are well matched with the sign reversal fields
determined from $S(H)$, where the higher field sign reversal is not plotted. The line of sign reversal terminates near 4\,kOe by simple linear extrapolation of the data below 1\,K.
The $H^{*}$ line determined from the local minimum in $S(H)$ is not matched with the $T_{0}$ line obtained from the local minimum in $S(T)$. Two lines linearly rise with increasing of
magnetic field.

By carefully examining $S(T)$ and $S(H)$ data, as shown in bottom panels in Fig. \ref{YbPtBiTEPPhase}, there are signatures corresponding to $H^{*}$ and $T_{0}$ in both figures even
though one of features is very weak. Below 30\,kOe $S(H)$ for $T$ = 1\,K (a horizontal cut through the $H-T$ plane) shows a sign change at $H_{SR}$ = 5.6\,kOe, a slope change near
$H_{0}$ = 11\,kOe, and a local minimum around $H^{*}$ = 15\,kOe, where the signature of $H_{0}$ is very weak. Below 2.5\,K $S(T)$ for $H$ = 15\,kOe (a vertical cut through the $H-T$
plane) indicates a slope change around $T^{*}$ = 1\,K and a local minimum near $T_{0}$ = 1.3\,K, where the signature of $T^{*}$ is very weak. Thus, $H^{*}$ line is sensitive to the
magnetic field sweeps and $T_{0}$ is sensitive to the temperature sweeps. Because of the weak signal, $T_{0}$ and $H^{*}$ were taken only from temperature sweeps and magnetic field
sweeps, respectively, and these are plotted in Fig. \ref{YbPtBiTEPPhase}. $T_{0}$ seems to extrapolate to the origin ($T$ = 0 and $H$ = 0) of the $H-T$ plane and $H^{*}$ tends toward
$H$ = 8 kOe at $T$ = 0. The crossover $T_{FL}$ (Fig. \ref{YbPtBiST2} and Fig. \ref{YbPtBiSH2}) is well overlapped with $H_{FL}$ and is almost linear in magnetic fields above 0.4\,K.
As mentioned above, for $H$ = 30, 40, and 50 kOe, the TEP shows a linear temperature dependence, $S(T)$ = $\alpha T$, which is a indication of FL behavior. Between 20 $\sim$ 30\,kOe
the boundary of $T_{FL}$ is overlapped with the boundary of the FL region determined from $T^{2}$-dependence of $\rho(T)$. Therefore, TEP below 0.4\,K is expected to follow $S(T)$ =
$\alpha T$ for $H\,<$ 30\,kOe. The local maximum developed in $S(T)$ for $H\,>$ 70 kOe is plotted in Fig. \ref{YbPtBiTEPPhase} as stars. Because of the very weak TEP signal in this
regime the signature is not discernible in $S(H)$ data. Since the TEP is known to be particularly sensitive to Kondo and CEF effects, the development of $T_{max}$ can be related to
the effect of further CEF splitting via Zeeman effect. In such a high magnetic field the Kondo effect with $T_{K}$ $\sim$ 1\,K for YbPtBi is expected to be suppressed.

\section{Discussion}

\subsection{Antiferromagnetic order}

In zero field the observed $\rho(T)$ below $T_{N}$ depends on the measurements conditions, but the $T_{N}$ remains approximately the same for all cases. Similar behavior has been
reported in Ref. \cite{Movshovich1994}, where $\rho(T)$ data for several rod-shaped samples show either an increase or a decrease below $T_{N}$. The different relative height of
$\rho(T)$ below $T_{N}$ was explained due to the partial gapping of the Fermi surface. In addition, the results of $\rho(T)$, measured by Montgomery arrangement \cite{Montgomery1971},
reveal anisotropy for current directions between along the high temperature [100] and [010] directions, which indicated a broken cubic symmetry below $T_{N}$ \cite{Movshovich1994}. In
this study, for testing the anisotropy with respect to the different current directions, several resistivity samples were cut from a plate-shaped sample with a wire-saw both parallel
to the [100] and [010] crystallographically equivalent direction. The results indicate that the anisotropy of $\rho(T)$ below $T_{N}$ does not depend on the different current
directions but highly depend on the sample mounting conditions. In the earlier studies it has been speculated that the anisotropy was caused either by the highly oriented domains or
by internal stress developed during material growth \cite{Movshovich1994}. In this study, however, the anisotropy is caused by the external parameters and expected to be due to the
external stress (anisotropic pressure), which is consistent with earlier specific heat results \cite{Lacerda1993}. Similar results have been found in cubic chromium (Cr)
\cite{Fawcett1988, Bastow1966}, which is the canonical example of SDW material with $T_{N}$ = 311\,K, that magnetic field cooling and compressive stress cooling profoundly change the
magnetic structure \cite{Bastow1966}. The application of a uniaxial stress ($\sim$ 0.07\,kbar) to a single crystal of Cr, while cooling through $T_{N}$, prohibits the development of
domains with a SDW vector ($\overrightarrow{q}$) parallel to the direction of stress, where the shifts of $T_{N}$ and magnitude of the $\overrightarrow{q}$ vector were detected
\cite{Bastow1966}. In YbPtBi, for stress cooling through $T_{N}$, it is suspected that the anisotropic distortion of the Fermi surface under external strain can cause the radical
variation of the resistivity below $T_{N}$.

One of the interesting aspects of antiferromagnetism in YbPtBi is the rapid suppression of $T_{N}$ by the application of hydrostatic pressure \cite{Movshovich1994}, where a pressure
as low as 1\,kbar suppresses the signature of the phase transition in resistivity measurements. On the other hand, the specific heat measurements has been shown \cite{Lacerda1993}
that the phase transition feature, shown in $C(T)/T$ for the single crystal samples, is completely smeared out for the pressed pellet samples, prepared from ground single crystals,
which were mixed with GE-7301 varnish. In addition to the resistivity results in this study, the drastic difference of the specific heat results between single crystals and pressed
pellet samples suggests that the results of the pressure dependence of resistivity are caused mainly by the external stress applied and also possibly non-hydrostatic components in
pressure experiments.

The temperature dependence of the electrical resistivity shows a sharp rise below $T_{N}$ which is reminiscent of a SDW antiferromagnet Cr \cite{Fawcett1988} and URu$_{2}$Si$_{2}$
\cite{Schoenes1987}. From a simple point of view, we expect that parts of the high temperature Fermi surface disappears when the gap is opened. As shown in Fig. \ref{YbPtBiHallT}, the
opening of the SDW gap below $T_{N}$ gives rise to an abrupt enhancement of $R_{H}$ ($H$ $\rightarrow$ 0), enhanced by roughly a factor of two compared to the value above $T_{N}$. From
the earlier study of the electrical resistivity and specific heat \cite{Movshovich1994}, it has been shown by the analysis of these data, based on BCS theory, that the Fermi surface is
removed roughly 16 \% by the formation of the SDW state. Thus, the steep increase of $R_{H}$ below $T_{N}$ implies a carrier density reduction with Fermi surface nesting of highly
renormalized bands. Although previous neutron scattering experiments have not confirmed AFM order \cite{Robinson1994}, the $\mu$-SR experiments suggested tiny ordered moment
\cite{Amato1992}. Therefore, a SDW ground state is supported by compelling evidence from $\rho(T)$, $C_{p}(T)$, and $R_{H}(T)$ as well as the $\mu$-SR measurements. Note that very
similar results have been observed in URu$_{2}$Si$_{2}$ \cite{Schoenes1987}. The carrier concentration of URu$_{2}$Si$_{2}$ estimated from $R_{H}$ is 0.05 holes per formula unit which
is close to the value of YbPtBi, and about 40 \% of the Fermi surface, calculated from specific heat, is removed by the formation of the hidden ordered state at $T_{0}$ = 17.5\,K
\cite{Maple1986}. Below $T_{0}$, $R_{H}$ of URu$_{2}$Si$_{2}$ increases by factor of 5-20 because of the opening of a gap over the Fermi surface. Recently $\rho_{H}$ measurements in
pulsed magnetic field show that the steep enhancement of $R_{H}$ below $T_{0}$ is completely suppressed across the QCP by order of 40 Tesla \cite{Oh2007}. Similarly the sharp rise of
$R_{H}$ for YbPtBi is completely suppressed near $H_{c}$ (Fig. \ref{YbPtBiHallT}).

\subsection{Quantum criticality}

\begin{figure}%
\centering
\includegraphics[width=0.5\linewidth]{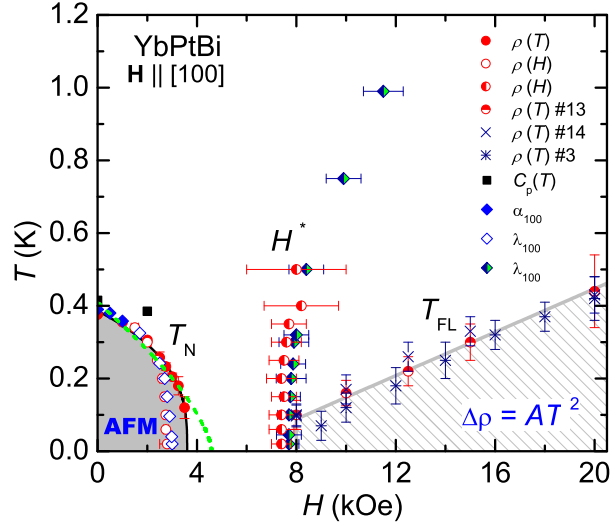}
\caption[$H-T$ phase diagram for YbPtBi along \textbf{H}\,$\parallel$\,100]{$H-T$ phase diagram of YbPtBi in configuration \textbf{H}\,$\parallel$\,[100]. The $T_{N}$ was derived from
d$\rho(T)$/d$T$, d$\rho(H)$/d$H$, $\alpha_{100}$, and $\lambda_{100}$. For the phase boundary the $\rho(T, H)$ results for sample \#13 are only included. The solid line on the AFM
phase boundary represents the fit of equation $T_{N} = [(H-H_{c})/H_{c}]^{0.33}$ to the data. The dashed line represents the fit of equation $T_{N} = [(H-H_{c})/H_{c}]^{2/3}$ to the
data. The $T_{FL}$ represents the upper limit of the $T^{2}$-dependence of $\rho(T)$, where the results of sample \#13, \#14, and \#3 are plotted. The solid line is guide to the eye.
The local maximum of d$\rho(H)$/d$H$ and the local minimum of $\lambda_{100}$ are assigned to $H^{*}(T)$. It should be noted that the width of each of these $H^*$ features decreases
with decreasing temperature as shown by horizontal bars. For each data set the width monotonically decreases with decreasing temperature. }
\label{YbPtBiPhase1}%
\end{figure}

\begin{figure}%
\centering
\includegraphics[width=0.5\linewidth]{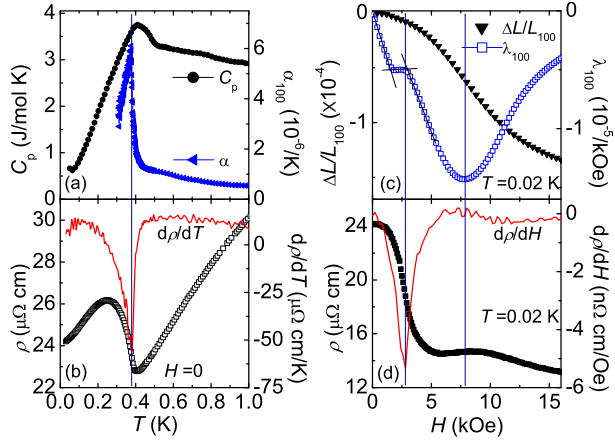}
\caption[Criteria for constructing $H-T$ phase diagram]{Criteria for determining $T_{N}$. (a) Zero field specific heat $C_{p}$ and the coefficient of linear thermal expansion
$\alpha_{100}$. (b) Zero field electrical resistivity $\rho(T)$ and the derivative d$\rho(T)$/d$T$. (c) Linear magnetostriction $\Delta L/L$ and the coefficient $\lambda_{100}$ =
d($\Delta L/L$)/d$H$ at $T$ = 0.02\,K. (d) Magnetoresistivity $\rho(H)$ and the derivative d$\rho(H)$/d$H$ at $T$ = 0.02\,K. Solid lines are guides to the eye.}
\label{YbPtBiPhase2}%
\end{figure}

The results of the low temperature thermodynamic and transport experiments are summarized in the $H-T$ phase diagram shown in Fig. \ref{YbPtBiPhase1}. (For clarity only the
resistivity data from sample \#13 are used to plot AFM phase boundary.) The magnetic field dependence of the AFM phase boundary, $T_{N}$, was mainly determined from the sharp peak
position in d$\rho(T)$/d$T$ and d$\rho(H)$/d$H$ (Fig. \ref{YbPtBiRTPhasediagram}), the sharp peak position in $\alpha_{100}$, and the slope change in $\lambda_{100}$ (Fig.
\ref{YbPtBiMSPhase}). For comparison, the temperatures of the maximum in $C_{p}$ (and the minimum in $\rho(T)$) are higher than those of $\alpha_{100}$ and d$\rho(T)$/d$T$, (Fig.
\ref{YbPtBiPhase2}) but as discussed above, the position of the higher field slope change in $\lambda_{100}$ is well matched with the sharp peak position in d$\rho(H)$/d$H$.

Figure \ref{YbPtBiPhase1} clearly shows that the AFM order can be suppressed to $T$ = 0 by an applied magnetic field of less than 4 kOe. This being said, it is worth discussing that
there is not perfect agreement between the temperature and magnetic field sweep data below 0.2\,K and there is an approximately 0.8\,kOe difference between them at 0.02\,K. No
noticeable hysteresis was observed between the up- and down-sweeps of magnetic field. However, the field dependence of $\rho_{H}$ at 0.06\,K shows clear feature at $H$ = 3.9\,kOe
(inset of Fig. \ref{YbPtBiHallH}), which is close to the AFM boundary determined from the temperature sweeps. It has been shown in earlier studies that the AFM order can be suppressed
by external magnetic field of 3.1\,kOe \cite{Movshovich1994}, which is mainly based on magnetic field sweeps. It is not clear at this point that whether this discrepancy is merely
based on the criteria for determining the $T_{N}$ or whether the AFM order splits into two different phases below 0.2\,K and for $H\,>$ 2\,kOe.

Based on the scaling properties near QCP, the phase transition temperature is expected to follow a characteristic power law dependence; $T_{N}$ $\propto$ (-$r$)$^{\psi}$, where $r$ is
the distance to the QCP and $\psi$ is the exponent \cite{Lohneysen2007}. In Fig. \ref{YbPtBiPhase1} the solid line on the AFM phase boundary represents the best fit of equation $T_{N}$
$\propto$ [$(H - H_{c})/H_{c}$]$^{\psi}$ to the data with $T_{N}(0)$ = 0.38 $\pm$ 0.02\,K, $H_{c}$ = 3.6 $\pm$ 0.2\,kOe, and $\psi$ = 0.33 ($\simeq$ 1/3) $\pm$ 0.03, where the error
bar depends on the fitting range. For (SDW) antiferromagnets with three dimensional critical fluctuations ($d$ = 3) the boundary of the ordered phase varies as $T_{N}$ $\propto$
(-$r$)$^{2/3}$ \cite{Hertz1976, Millis1993}. When the exponent is fixed to $\psi$ = 2/3, the fit curve is represented by a dashed line (Fig. \ref{YbPtBiPhase1}) on the phase boundary
with $T_{N}$ = 0.4\,K and $H_{c}$ = 4.6\,kOe. Apparently, for YbPtBi the AFM phase boundary can be better described with $\psi$ $\simeq$ 1/3, which deviates from the theoretical
prediction for a three dimensional AFM QCP of SDW scenario.

\begin{figure}%
\centering
\includegraphics[width=0.5\linewidth]{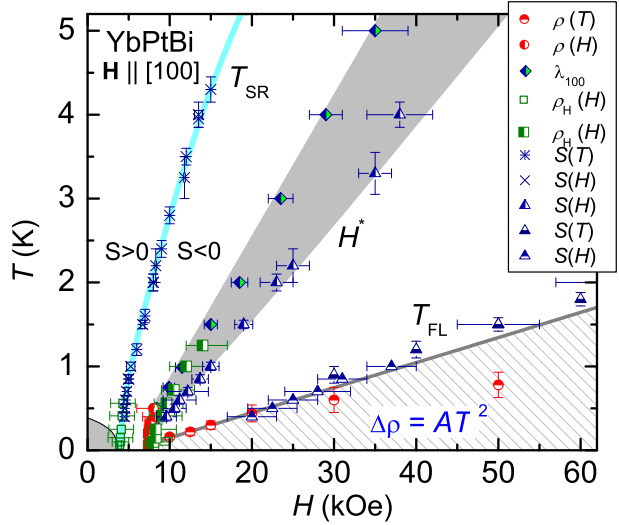}
\caption[High temperature $H-T$ phase diagram for YbPtBi]{High temperature $H-T$ phase diagram for YbPtBi. The $S(T, H)$ = 0 and the slope change from $\rho_{H}/H$ are assigned to
$T_{SR}(H)$. The local maximum in d$\rho(H)$/d$H$, the local minimum of $\lambda_{100}$, the local minimum of $\rho_{H}/H$, and the local minimum in $S(H)$ are assigned to $H^{*}(T)$.
The $T_{FL}$ was derived from the upper limit of the $T^{2}$-dependence of $\rho(T)$ and the upper limit of the $T$-dependence of $S(T)$. The slope change in $S(H)$ is also assigned
to $T_{FL}$. All lines and shaded area are guides to the eye.}
\label{YbPtBiPhase3}%
\end{figure}

In addition to $T_{N}$, measurements indicate a crossover region of $H^{*}(T)$. The features in d$\rho(H)$/d$H$ (Fig. \ref{YbPtBiRTPhasediagram}), $\lambda_{100}$ (Fig.
\ref{YbPtBiMSPhase}), $\rho_{H}/H$ (Fig. \ref{YbPtBiHallPhase1}), and $S(H)$ (Fig. \ref{YbPtBiTEPPhase}), associated with $H^{*}$, are assigned to $H^{*}(T)$ and are plotted in the
$H-T$ plane as shown in Fig. \ref{YbPtBiPhase1} for lower $T$ and $H$ and in Fig. \ref{YbPtBiPhase3} over a wider range. The error bars are rough estimates of the crossover widths,
based on the widths of those features. The width of the $H^{*}$ crossover region is wider as temperature is increased. However, in the zero temperature limit each $H^{*}$ sharpens and
tends to converge near $H^{*}$\,$\sim$\,7.8\,kOe. For other field-induced QCP systems, Ge-doped \cite{Custers2010} and parent YbRh$_{2}$Si$_{2}$ \cite{Paschen2004} and YbAgGe
\cite{Budko2005a}, a similar crossover field has been observed from various thermodynamic and transport measurements. The FL region is consistently inferred from $S(T)$ and $\rho(T)$
data below 30\,kOe; for $H\,>$ 30\,kOe, the FL region determined from $S(T)$ and $S(H)$ is not consistent with the one inferred from $\rho(T)$. Given that $T_{FL}$ represents a cross
over, differences in its value, inferred from different data sets, are not unexpected.

Even though the physical meaning behind the experimental signature is not clear and the primary experimental signature comes from TEP data, there is an another crossover scale of
$T_{SR}$ (Fig. \ref{YbPtBiPhase3}). The lower magnetic field signature in $\rho_{H}/H$, which corresponds to the slope change in $\rho_{H}/H$ emerging from $H_{c}$, overlaps the sign
reversal in $S(T, H)$. Thus, in the $T\,\rightarrow$ 0 limit, $T_{SR}$ is expected to converge to $H_{c}$ by tracking the $\rho_{H}/H$ feature. For YbAgGe this $T_{SR}$ crossover line
has also been observed with similar behavior \cite{Mun2010b}.

One of the interesting issues is the magnetic field modification of the power law dependence of the resistivity (Fig. \ref{YbPtBiRTPhase1}, Fig. \ref{YbPtBiRTPhase2}), $\rho(T)$ =
$\rho_{0}$ + $AT^{n}$, which describes the low temperature quasi-particle behavior. In Fig. \ref{YbPtBiPhase1}, for $H\,>$ 8 kOe, the characteristic scale of $T_{FL}$ marks the upper
limit of the observed $T^{2}$-dependence of the resistivity below which the FL state is stabilized. In Fig. \ref{YbPtBiPhase1} the results for sample \#13, \#14, and \#3 are plotted
and the solid line is guide to eye. The $T_{FL}$ region shrinks quasi-linearly with decreasing magnetic field from the paramagentic state. By using simple linear extrapolation, the
$T_{FL}$ line terminates at $H\,\sim$ 5.2 $\pm$ 0.5\,kOe, based on the results of three samples, which is close to but distinct from $H_{c}$. Below $H$ $\sim$ 8\,kOe, the $\rho(T)$
curve is better fitted to the $T^{1.5}$- than $T^{2}$-dependence, indicating nFL-like behavior (4 kOe $<\,H\,<$ 8\,kOe). A detailed analysis of $\rho(T)$ (Fig. \ref{YbPtBiRTPhase2})
reveals that as magnetic field decreases a nFL-like behavior ($\Delta\rho(T)\,\propto\,T^{1.5}$) of resistivity above the $T^{2}$-region is also observed, which shrinks progressively
towards $H\,\sim\,H_{c}$. Although the question of whether $\rho(T)\,\propto\,T^{2}$ exists at very low temperature down to $H_{c}$\,$\sim$ 4\,kOe is still open (although not strongly
supported by the data), a clear nFL region between 4 and 8\,kOe is strongly indicated.

The observation of these two distinct, low temperature regimes, FL and nFL, in YbPtBi raises the question of whether the FL state survives in the magnetic field range between $H_{c}$
and $H^{*}$ at $T$ = 0 and what is the physical origin of the crossover scale $H^{*}$. The $H^{*}$ line seems to block the extension of FL state below 8\,kOe, but for unambiguous
conclusions it will be necessary to perform high resolution measurements of the resistivity to temperatures even lower than 0.02 K. In any case, it is natural to interpret the
constructed $H-T$ phase diagram as showing that $T_{N}$ is suppressed to $T$ = 0 for $H_{c}\,\leq$ 4\,kOe and the FL state is stabilized for $H$ $\geq$ 8\,kOe. The $T_{SR}$ and $T_{N}$
line vanish at $H_{c}$ and the $H^{*}$ vanishes near the magnetic field of 7.8\,kOe at $T$ $\rightarrow$ 0 which is not directly connected to $T_{N}$. It currently seems likely that
$T_{FL}$ terminates at $H^{*}$ in the zero temperature limit.

\begin{figure*}%
\centering
\includegraphics[width=0.5\linewidth]{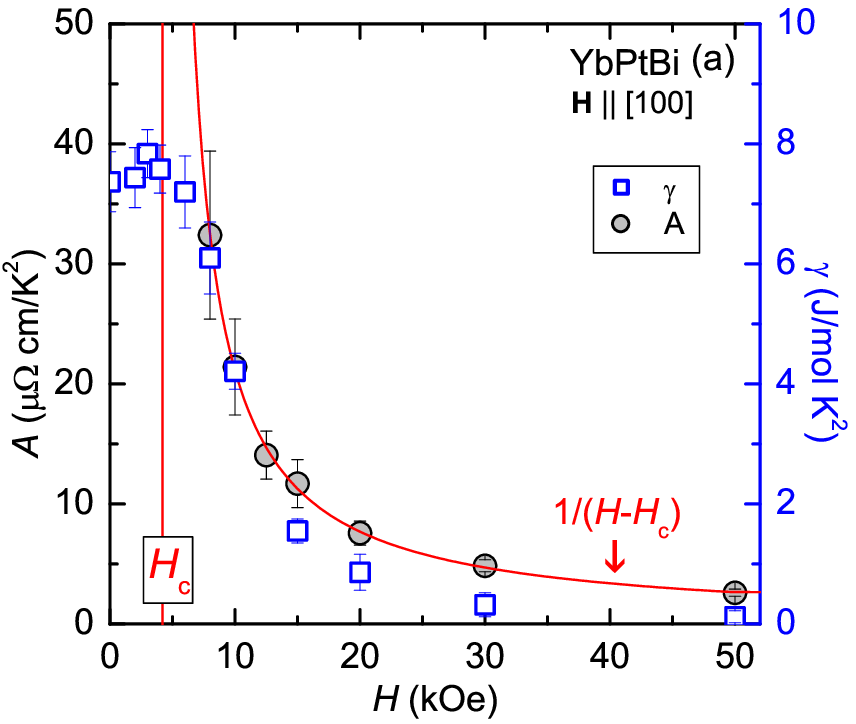}\includegraphics[width=0.5\linewidth]{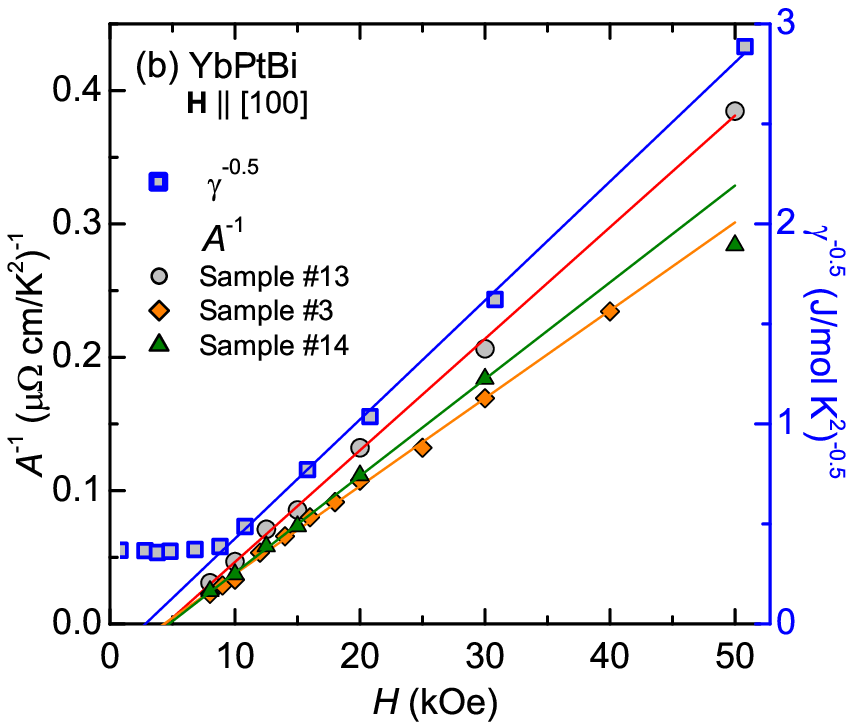}
\caption[Power law analysis of $A$ and $\gamma$ for YbPtBi]{(a) Fermi liquid coefficient $A$ = $\Delta\rho(T)/T^{2}$ and $\gamma$ = $C(T)/T|_{T\rightarrow 0}$. Solid line through the
higher field $A$ values represents a fit of equation, $A$ - $A_{0}$ $\propto$ 1/($H$ - $H_{c}$), performed up to 50\,kOe with the constant offset $A_{0}$ $\simeq$ 0.03
$\mu\Omega$cm/K$^{2}$ and $H_{c}$ = 4.2\,kOe. Vertical line represents the critical field ($H_{c}$). (b) $A^{-1}$ (left axis) vs. $H$ for three different samples (samples \#3, \#13,
and \#14 in Fig. \ref{YbPtBiRTPhase2}) and $\gamma^{-0.5}$ (right axis) vs. $H$. Solid lines represent the linear fit to the data. See text for details.}
\label{YbPtBiKW}%
\end{figure*}

Since $T_{FL}$ seems to be detached from the $T_{N}$, it would be interesting to assess whether the quasi-particle effective mass diverges at the critical field of $H_{c}$ via a
strong magnetic field dependence of the FL coefficients $A$ and $\gamma$. The coefficient $A$ rapidly increases with decreasing magnetic field from the paramagnetic state (Fig.
\ref{YbPtBiKW} (a)). Indeed, the steep variation of the $A$ value can be well described by a scaling analysis with a form of $A(H)$\,-\,$A_{0}$ $\propto$ $(H - H_{c})^{-\beta}$, where
$A_{0}$ is the adjustable parameter, $H_{c}$ is the critical field, and $\beta$ is the exponent. In Fig. \ref{YbPtBiKW} (a) the solid line on $A$ values for sample \#13 represents a
fit of the scaling form, where the fit was performed between 8 and 50\,kOe yielding a critical field $H_{c}$ = 4.2 $\pm$ 0.5\,kOe, an exponent $\beta$ = 1 $\pm$ 0.05, and $A_{0}$
$\simeq$ 0.03 $\mu\Omega$cm/K$^{2}$. The power law dependence of $A$ can be clearly seen, when it is plotted as $A^{-1}$ vs. $H$, as shown in Fig. \ref{YbPtBiKW} (b). From a linear
fit to the data the critical field is obtained to be $H_{c}$\,$\sim$ 4.4\,kOe, which is close to the critical field of power law fit. Similar critical fields for samples \#3 ($H_{c}$
$\simeq$ 4.3\,kOe) and \#14 ($H_{c}$ $\simeq$ 4.2\,kOe) with $\beta$ $\simeq$ 1 can be obtained with the adjustable parameter $A_{0}$. Note that without $A_{0}$ the critical field and
the exponent, obtained from the fit to three different sets of $A$ value, vary between 3.5\,kOe $\leq$\,$H_{c}$\,$\leq$ 4.7\,kOe and 0.92 $\leq$$\,\beta$\,$\leq$ 1.12, respectively,
thus the adjustable parameter $A_{0}$ is necessary to allow the three data sets to converge to the same $H_{c}$ and $\beta$ values in the same magnetic field range, but even with
$A_{0}$ = 0, the value of $H_{c}$ is much closer to $H_{c}$ $\sim$ 4.5\,kOe than to $H^{*}$ $\sim$ 8\,kOe and $\beta$ is closer to 1.0 than to 0.5 or 1.5. Since the $A$ value diverges
at near $\sim$ 4\,kOe, the scattering cross-section between quasi-particles becomes singular at $H_{c}$. The observed divergence of $A$ assigned $H_{c}$ as the QCP and $\beta$ = 1 as
the exponent characterizing quantum criticality. A power law divergence of the $A$ value near QCP has been observed from other field-induced QCP systems such as YbRh$_{2}$Si$_{2}$
\cite{Gegenwart2002}, CeCoIn$_{5}$ \cite{Paglione2003}, and CeAuSb$_{2}$ \cite{Balicas2005} with exponent $\beta$ = 1 or close to 1.

A FL state can be characterized by the Kadowaki-Woods (K-W) ratio \cite{Kadowaki1986}, $A \propto \gamma^{2}$, where $\gamma$ is a direct measure of the effective mass, $m^{*}$, of
quasi-particles. Thus, the dramatic variation of $\gamma$ was also analyzed with a relation of $\gamma(H)$ - $\gamma_{0}$ $\propto$ $(H - H_{c})^{-\beta}$, which is the same form as
$A$, where $\gamma_{0}$ is the adjustable parameter. The power law fit to the $\gamma(H)$, performed between 8 and 50\,kOe, yields a critical field $H_{c}$ = 4.6 $\pm$ 0.4\,kOe, an
exponent $\beta$ = 1 $\pm$ 0.2, and $\gamma_{0}$ = 0.55 J/mol$\cdot$K$^{2}$. Although this analysis gives a consistent critical field with that obtained from the fit of $A$, the
required value of $\gamma_{0}$ = 0.55 J/mol$\cdot$K$^{2}$ is very high. Without $\gamma_{0}$ the fit yields a critical field of 1.5 $\pm$ 0.5 \,kOe and an exponent $\beta$ = 2 $\pm$
0.4. This result can be clearly seen in the $\gamma^{-0.5}$ vs. $H$ plot (Fig. \ref{YbPtBiKW} (b)) which is close to the linear in $H$, and thus $\beta$ $\sim$ 2. In this plot, the
critical field is estimated to be $H_{c}$\,$\sim$ 1.8 $\pm$ 0.5\,kOe from the linear fit to the data. The observed exponents, 1\,$\leq$\,$\beta$\,$\leq$\,2, are striking deviation
from the K-W ratio, where the exponent $\beta$ = 0.5 is expected in FL regime. It is worth noting, though, that $\gamma(H)$ diverges near or below 4.5\,kOe in all cases. Note that
such a deviation from the K-W ratio across the field tuned QCP has also been observed in Ge-doped YbRh$_{2}$Si$_{2}$ \cite{Custers2003}.

\begin{figure*}%
\centering
\includegraphics[width=0.5\linewidth]{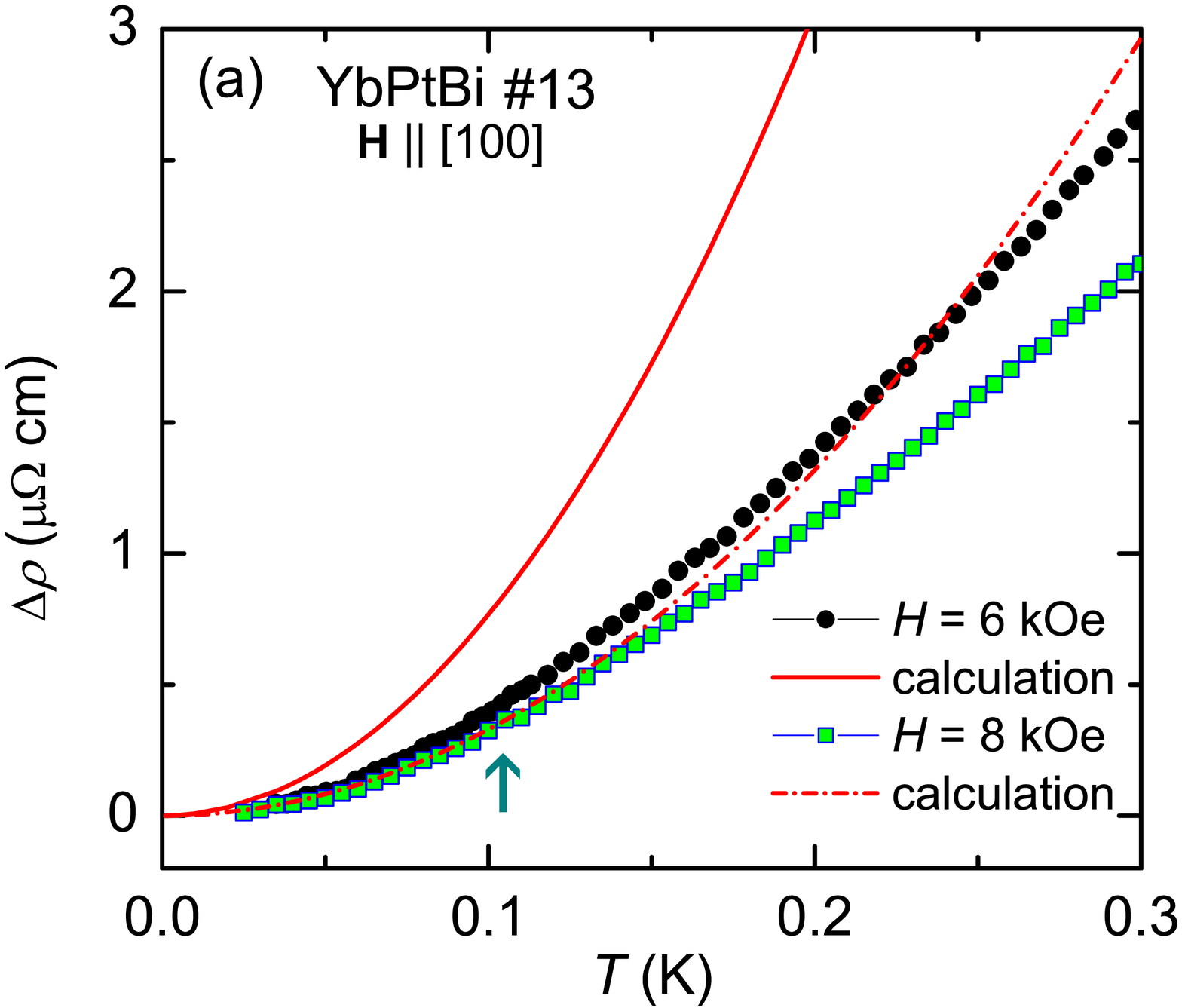}\includegraphics[width=0.5\linewidth]{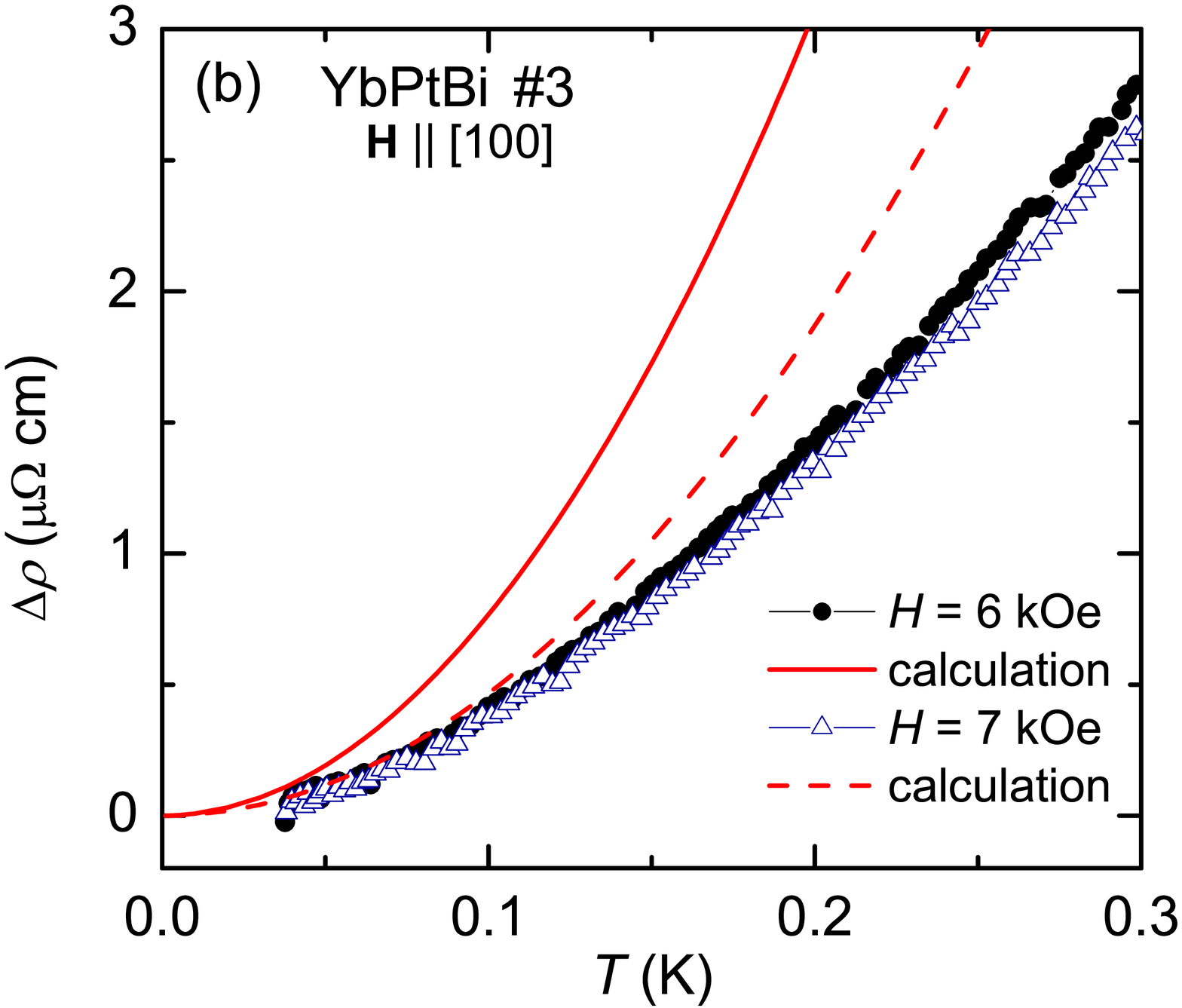}
\caption[Power law dependence of resistivity for YbPtBi]{The temperature dependence of the resistivity, $\Delta\,\rho(T)$ = $\rho(T)$\,-\,$\rho_{0}$, of YbPtBi for (a) sample\,\#13
and (b) sample \#3. (a) The solid and dash-dotted line represent the calculated $\Delta\,\rho(T)$ with $A$ $\simeq$ 76.7 $\mu\Omega$cm/K$^{2}$ for $H$ = 6\,kOe and with $A$ $\simeq$
32.4 $\mu\Omega$cm/K$^{2}$ for $H$ = 8\,kOe, respectively. (b) The solid line and dashed line represent the calculated $\Delta\,\rho(T)$ with $A$ $\simeq$ 76.7\,$\mu\Omega$cm/K$^{2}$
for $H$ = 6\,kOe and with $A$ $\simeq$ 46.7 $\mu\Omega$cm/K$^{2}$ for $H$ = 7\,kOe, respectively. The $A$ values used to generate $\Delta\,\rho(T)$ were obtained from the power law
fit ($A$\,$\propto$\,1/($H-H_{c}$)) to the $A$ values shown in Fig. \ref{YbPtBiKW}. See text for details.}
\label{YbPtBiRsimulation}%
\end{figure*}

To clarify the observed, anomalous power law dependence of resistivity below 8\,kOe, the measured resistivity was compared to the predicted $T^{2}$-dependence of resistivity based on
the power law analysis of $A$ values. In Fig. \ref{YbPtBiRsimulation} the measured resistivity for samples \#13 and \#3, together with the calculated resistivity curves, are plotted
after subtracting $\rho_{0}$ value ($\Delta\,\rho(T)$). For $H$ = 6, 7, and 8\,kOe, predicted $A$ values, obtained from the power law fit ($A$\,$\propto$\,1/($H-H_{c}$), Fig.
\ref{YbPtBiKW}) to the experimental $A$ values, are used to generate $\Delta\,\rho(T)$ curves. For sample \#13 as shown in Fig. \ref{YbPtBiRsimulation} (a), the measured
$\Delta\,\rho(T)$ for $H$ = 8\,kOe is in good agreement with the calculated $\Delta\,\rho(T)$ below $\sim$ 0.11\,K (indicated by arrow), whereas the observed $\Delta\,\rho(T)$ for $H$
= 6\,kOe can not be reproduced by the predicted $\Delta\,\rho(T)$ fundamentally due to the large, predicted $A$ value used. For sample \#3 (Fig. \ref{YbPtBiRsimulation} (b)), the
calculated curves for both $H$ = 6 and 7\,kOe shows no agreement with the measured $\Delta\,\rho(T)$. Therefore, as shown in Fig. \ref{YbPtBiKW}, there seems to be a disruption of
high field FL behavior near $H^{*}$ ($\sim$ 8\,kOe) rather than going down to $H_{c}$ ($\sim$ 4\,kOe). This analysis suggests that the $T^{1.5}$ dependence of resistivity below 8 kOe
can be from the electronic system entering a new kind of state such as nFL state that has less scattering. This result is consistent with the behavior of $\gamma(H)$ which clearly
shows a deviation from the power law dependence below 8\,kOe (Fig. \ref{YbPtBiKW}). It is worth noting that there is the possibility of having $T^{2}$ dependence of the resistivity
between 4 and 8 kOe at very low temperatures, i.e. below 0.08 K, as discussed earlier. In this field region the obtained $A$ value for $T^{2}$ fit is far smaller than the predicted
$A$ value. Thus, although the FL state might be stabilized below 8 kOe, the electronic state would be distinct from that above 8 kOe.

\begin{figure*}%
\centering
\includegraphics[width=0.5\linewidth]{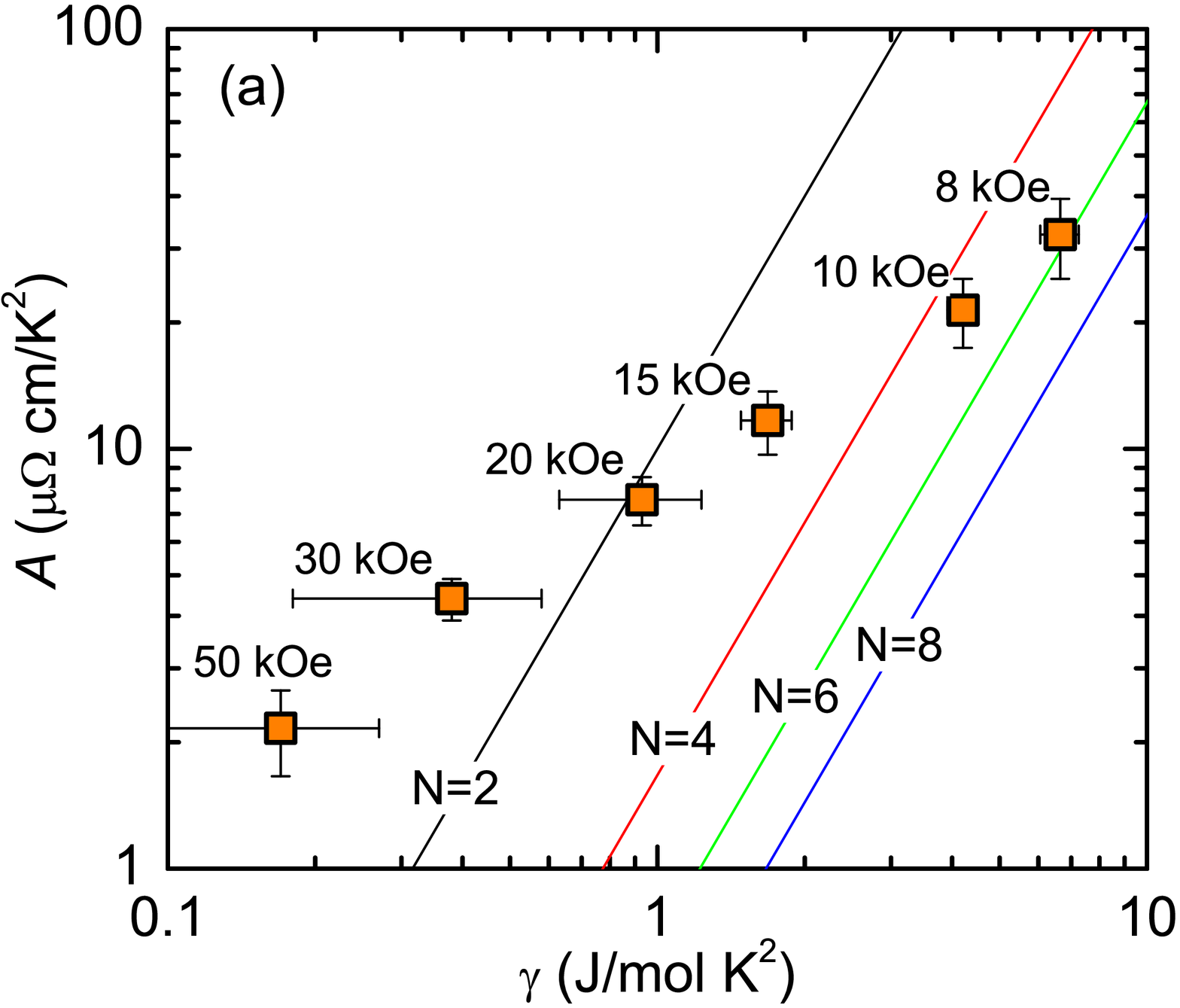}\includegraphics[width=0.5\linewidth]{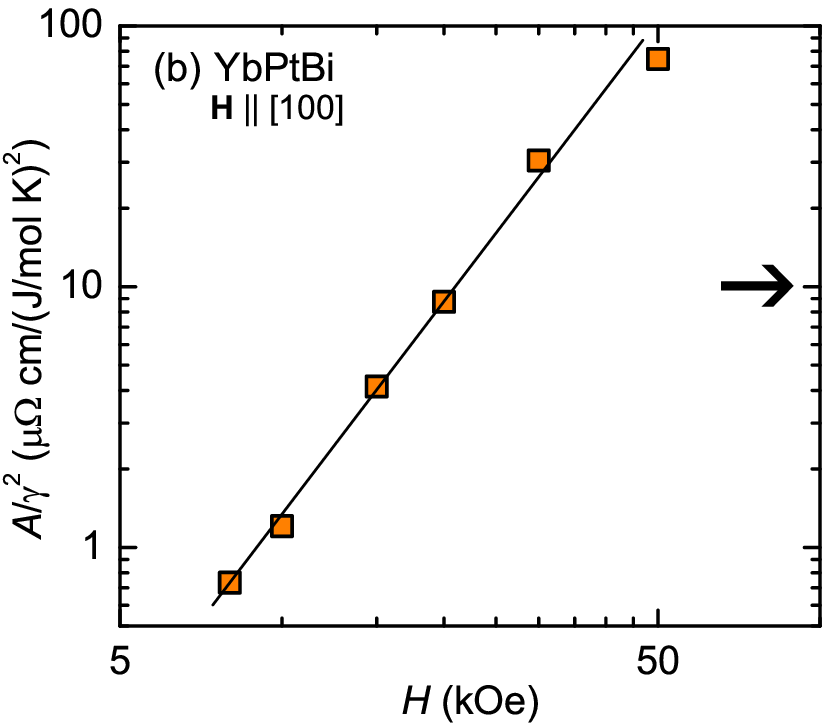}
\caption[Kadowaki-Woods ratio for YbPtBi]{(a) $\log-\log$ plot of $A$ vs. $\gamma$. Solid lines represent the Kadowaki-Woods (K-W) ratio for different ground state degeneracy
\cite{Tsujii2005} for $N$ = 2 - 8. (b) $A/\gamma^{2}$ vs. $H$, where the horizontal arrow indicates the K-W ratio for $N$ = 2. Solid line is guide to the eye.}
\label{YbPtBiKW1}%
\end{figure*}

It has been shown that the $A/\gamma^{2}$ ratio depends on the ground state degeneracy \cite{Tsujii2005}. A clear dependence of the $A/\gamma^{2}$ ratio on the degeneracy, $N$, is
shown in Fig. \ref{YbPtBiKW1} (a). The experimental $A/\gamma^{2}$ ratio continuously shifts from high degeneracy (near $N$ = 6 at 8\,kOe) toward low degeneracy ($N$ = 2 at 20\,kOe).
A clear variation of K-W ratio in the presence of magnetic field is better seen when $A/\gamma^{2}$ is directly plotted as a function of magnetic field (Fig. \ref{YbPtBiKW1} (b)); the
ratio, $A/\gamma^{2}$, continuously increases as the magnetic field increases. In zero field and at ambient pressure, it has been shown \cite{Movshovich1994a} that the K-W ratio for
YbPtBi is located close to the $N$ = 8 curve (not plotted in Fig. \ref{YbPtBiKW1} (a)). Because of the AFM order, the $A$ value at ambient pressure was estimated by linearly
extrapolating pressure dependence of $A$ values between 4 and 19\,kbar \cite{Movshovich1994a}. In this pressure range the resistivity data followed $\Delta\rho(T)$ = $AT^{2}$ below
0.3\,K. The observed behavior of K-W ratio suggests that the variation of $A/\gamma^{2}$ values is due to magnetic field induced changes in $N$, a supposition that seems plausible
because the ground state CEF degeneracy in zero field can be lifted by applied magnetic field.

However there are several points about K-W scaling and YbPtBi that need to be considered. First, in zero field the ground state degeneracy of YbPtBi should be $N$ = 2 (doublet) or $N$
= 4 (quartet) in cubic CEF \cite{Lea1962}. Based on this we would expect $N$ = 4 at most, not 6 or 8. Second, the K-W ratio not only depends on the degeneracy but also on the carrier
concentration, $n$, as $n^{-4/3}$ \cite{Tsujii2005, Jacko2009}, which is an important correction in low carrier density systems. Thus, it is necessary to consider the carrier density
for lower carrier systems. Although a single band model will ultimately be inadequate for YbPtBi, it does provide a useful starting point; when the carrier density, 0.04 hole per
formula unit (in a single band model) for YbPtBi at 300\,K, is considered, the $N$ = 2, 4, 6, and 8 manifold shown in Fig. \ref{YbPtBiKW1} (a) shifts downward with the $N$ = 2 line
falling well below the data. Thus the carrier concentration within a single band model can not explain the observed behavior of K-W ratio. For YbPtBi the K-W ratio may depend on CEF
splitting, low carrier density, and details of the multiple Fermi surfaces.

The multiband nature of YbPtBi is clearly evidenced from quantum oscillations \cite{Mun2010a} (the analysis of the quantum oscillation is beyond the scope of this paper) and can be
supported from the TEP results. Many metals, including HF compounds, have shown correlations between $C(T)/T$ and $S(T)/T$ in the zero temperature limit, linking these two quantities
via the dimensionless ratio, $q = \frac{SN_{A}e}{\gamma T}\sim\pm$1, where $N_{A}$ is the Avogadro number and the constant $N_{A}e$ is called the Faraday number \cite{Behnia2004}. At
finite temperature, near 0.4\,K, this relation seems not to be relevant for YbPtBi. Taking the values of $S(T)/T$ = 1.2 $\mu$V/K at the onset of $T_{N}$ and $\gamma$ = 7.4
J/mol$\cdot$K$^{2}$ yields $q$ = 0.015. Since the dimensionless ratio holds for a single carrier per formula unit, generally a larger $q$ value is expected when the carrier density is
as low as this is; the carrier density of 0.04 hole per formula unit implies $q$ = -25. Therefore $S(T)/T$ $\sim$ -20\,$\mu$V/K$^{2}$ is expected for $\gamma$ = 7.4
J/mol$\cdot$K$^{2}$. As seen in Fig. \ref{YbPtBiST3} the absolute value of $S(T)/T$ up to 8\,kOe is considerably lower than this value, where $\gamma$ remains the same order of
magnitude. Therefore, the low carrier density of YbPtBi can not, by itself, provide a natural explanation for this small magnitude of $q$. This again points toward the multiband nature
of this material as a likely explanation. In order to clearly address this issue, further experimental investigations are required below 0.35\,K. In multiband metals, the TEP for each
band can be positive or negative, therefore, in principle, the absolute value of the weighted sum of the overall TEP could be considerably reduced, compared to the single band
picture. When the same amount of entropy is carried by each type of carrier a reduction of $S(T)/T$ is expected. Therefore, in addition to the ground state degeneracy and carrier
concentration, the multiband (multi-Fermi surface) effect and/or the strong anisotropy of the Fermi surfaces should be considered in the K-W ratio as well as the $q$ value. It is
worth noting that a deviation from K-W relation and $q$-ratio has been observed in semi-metallic HF system CeNiSn and such a deviation has been qualitatively explained by considering
carrier density \cite{Behnia2004, Terashima2002}. However, the low carrier density of YbPtBi, on its own, can not explain the observed behaviors.


Based on the scaling analysis of $A$ for magnetic field higher than $H^{*}$, the quasi-particle mass shows a power law divergence near $H_{c}$. However, the experimentally observed
$\gamma$ is essentially constant for $H\,<$ 8\,kOe (close to $H^{*}$) (Fig. \ref{YbPtBiCp2} (a)). An intriguing question to raise is if the QCP is at $H_{c}$, what is the physical
origin of the crossover line $H^{*}(T)$, which seems to cut off the divergence of quasi-particle mass enhancement; and why do specific heat measurements indicate no pronounced nFL
behavior, -$\log(T)$ or $\sqrt{T}$, for $H\,\geq\,H_{c}$ down to lowest temperature measured? The resistivity results reveal a nFL state with $\Delta\rho(T)$ = $T^{1.5}$ and the TEP
measurements indicate a logarithmic temperature dependence, $S(T)$\,$\propto$ -$\log$($T$), for $H\,<\,H_{c}$ and $T\,>\,T_{N}$. Based on these transport results one should ask
whether an extended regime of nFL state is caused by purely quantum fluctuations or whether other effects, such as magnetic field induced metamagnetic-like state or the modification
of the CEF ground state with a characteristic field of $H^*$, need to be considered.

The $H-T$ phase diagram, constructed from several experimental results, for YbPtBi will now be compared to the other Yb-based, field-induced QCP systems; YbRh$_{2}$Si$_{2}$
\cite{Gegenwart2002}, Ge-doped YbRh$_{2}$Si$_{2}$ \cite{Custers2010, Custers2003}, and YbAgGe \cite{Budko2005a}. Each of these systems shows AFM order being suppressed to $T$ = 0 by
an external magnetic field and beyond a given critical field a FL state, exists below a $T_{FL}$ crossover. However, the details of characteristic crossover scales, such as $H^{*}$,
are different. Note that the crossover scale $H^{*}$ used in this paper represents the $T^{*}$ used in the references. For YbRh$_{2}$Si$_{2}$ $H^{*}$ has been interpreted as a
characteristic energy scale below which the quasi-particles are break down, involving a Fermi surface volume change from small to large across the QCP \cite{Paschen2004}. The sign
reversal in TEP, $T_{SR}$, has been observed from both YbRh$_{2}$Si$_{2}$ \cite{Hartmann2010} and YbAgGe \cite{Mun2010b} across the quantum critical region. Whereas the $T_{SR}$ for
YbAgGe emerges at the critical field and persists up to high temperature, the $T_{SR}$ for YbRh$_{2}$Si$_{2}$ exists inside the AFM region and terminates at the critical field as the
system is tuned through the QCP. For YbPtBi, considering these two crossovers, $H^{*}$ and $T_{SR}$, the constructed phase diagram is more similar to YbAgGe.

For both YbRh$_{2}$Si$_{2}$ and YbAgGe the resistivity, specific heat, and thermoelectric power in the vicinity the QCP manifest a clear $\Delta\rho(T)$\,$\propto$\,$T$,
$C(T)/T$\,$\propto$\,-$\log(T)$, and $S(T)/T$\,$\propto$\,-$\log(T)$ behaviors as signatures of strong quantum fluctuations, which can be understood within the conventional SDW
scenario with $z$ = 2 and $d$ = 2 \cite{Hertz1976, Millis1993, Paul2001}, and are also compatible with the unconventional Kondo breakdown scenario \cite{Coleman2001, Senthil2004,
Paul2008, Kim2010}. Note that the dimensionality of these systems needs to be clarified. For YbPtBi no consistent nFL behavior is observed in thermodynamic and transport measurements:
the resistivity measurements show a $T^{1.5}$-dependence between $T_{SR}$ and $H^{*}$ in which the strongest signature (longest temperature range of this power law) is observed near
$H^{*}$, the specific heat shows a -$\log(T)$ dependence over only limited temperature range, and thermoelectric power measurements shows a -$\log(T)$ dependence below the critical
field. In the paramagnetic region, for Ge-doped \cite{Custers2003} and parent YbRh$_{2}$Si$_{2}$ \cite{Gegenwart2002} a divergence of the effective mass at the QCP has been inferred
from the power law analysis of the FL coefficients of $A$. For YbPtBi a power law analysis of the $A$-coefficient shows an indication of divergence at the critical field, however the
specific heat remains finite (and near constant) for $H\,<\,H^{*}$ at which the divergence nature of the effective mass is essentially cut off. For YbAgGe the power law dependence of
these coefficients has not been analyzed.

The biggest difference between YbRh$_{2}$Si$_{2}$ and YbAgGe is that the crossover scales, $H^{*}$ and $T_{FL}$, are detached from the AFM phase boundary ($T_{N}$) for YbAgGe, whereas
$T_{N}$, $H^{*}$, and $T_{FL}$ terminate at the QCP for YbRh$_{2}$Si$_{2}$. Interestingly the $H^{*}$ for Ge-doped YbRh$_{2}$Si$_{2}$ is also detached from $T_{N}$. When the nFL
region is considered, a wide nFL region, determined from $\Delta\rho(T)$\,$\propto$\,$T$, is robust for YbAgGe \cite{Niklowitz2006} and Ge-doped YbRh$_{2}$Si$_{2}$ \cite{Custers2010},
in contrast to the field-induced QCP in YbRh$_{2}$Si$_{2}$ of which the FL behavior is recovered when $T_{N}\,\rightarrow\,0$. From this point of view the constructed $H-T$ phase
diagram of YbPtBi is similar to that of YbAgGe and Ge-doped YbRh$_{2}$Si$_{2}$. For YbAgGe, the two crossover scales, $T_{SR}$ and $H^{*}$, are evidenced from thermodynamic and
transport measurements, where the wide nFL region has been seen between these two crossovers, which is similar to that of YbPtBi. Note the for Ge-doped YbRh$_{2}$Si$_{2}$ a $T_{SR}$
line has not been identified.

\begin{figure*}%
\centering
\includegraphics[width=0.5\linewidth]{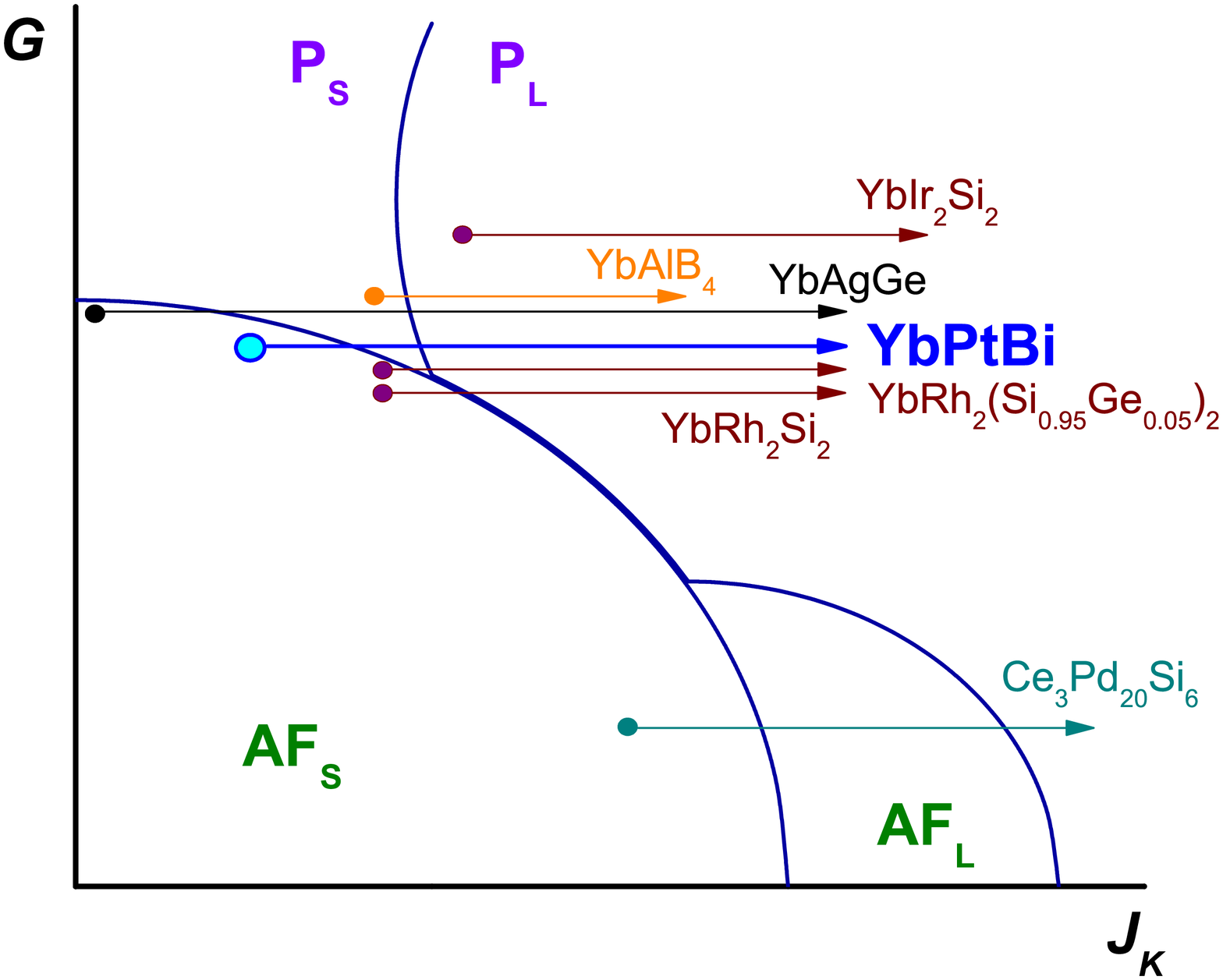}
\caption[Global Phase Diagram]{Global phase diagram for heavy fermion materials, adopted from Refs. \cite{Si2006, Si2010}, displaying the combined effects of Kondo coupling ($J_K$) and
magnetic frustration ($G$). The large circle is the hypothetical location of YbPtBi. In order to destroy the AFM order, a small applied magnetic field ($\sim$4 kOe) is required.
Beyond the AFM QCP, the $T^{1.5}$ dependence of resistivity is observed passing through a finite magnetic field range (4 kOe $<$ $H$ $<$ 8 kOe). For $H \geq$ 8 kOe, passing through
the $f$-electron localized-to-delocalize line, the $T^{2}$ dependence of resistivity is clearly observed.}
\label{YbPtBiGlobal}%
\end{figure*}

However, there are remaining questions when YbAgGe is compared to other systems. In the zero temperature limit, both $H^{*}$ and $T_{FL}$ terminate at the same field for Ge-doped and
pure YbRh$_{2}$Si$_{2}$, whereas $T_{FL}$ for YbAgGe is detached from $H^{*}$. For YbPtBi it is reasonable to assume that $T_{FL}$ terminates at or near $H^{*}$ at $T$ = 0. In a
simple point of view, YbPtBi is very similar to YbAgGe with regards to the crossover scales of $T_{SR}$ and $H^{*}$ and is close to that of Ge-doped YbRh$_{2}$Si$_{2}$
\cite{Custers2010} with regards to the $H^{*}$ and $T_{FL}$. Therefore, YbPtBi can be located between YbAgGe and Ge-doped YbRh$_{2}$Si$_{2}$ (closer to the Ge-doped
YbRh$_{2}$Si$_{2}$) in the global phase diagram \cite{Si2006,Si2010}  as shown in Fig. \ref{YbPtBiGlobal}. The AFM order in YbPtBi can be suppressed to $T$ = 0 by applying magnetic
field of $\sim$4 kOe (AFM state in Fig. \ref{YbPtBiGlobal}). Further increasing magnetic field the electrical resistivity follows $T^{1.5}$ dependence between $\sim$4 kOe and $\sim$ 8
kOe in which the paramagnetic, small Fermi surface, phase (possibly spin liquid phase) can be formed in YbPtBi ($P_S$ region in Fig. \ref{YbPtBiGlobal}) where the frustration effect,
caused by the faced centered cubic structure, may give rise to the spin liquid state. For $H$ $\geq$ 8 kOe, after passing through the $f$-electron localized-to-delocalize line in Fig.
\ref{YbPtBiGlobal}, the $T^{2}$ dependence of resistivity is clearly observed. By following the global phase diagram, the crossover line $H^{*}$ in YbPtBi corresponds to the
$f$-electron localized-to-delocalize line and the nFL state can be based on the spin liquid state. It needs to be clarified what are the characteristics of spin liquid phase in a
metallic system. In order to clarify the proposed Doniach-like diagram, further theoretical and experimental work are needed. It has to be noted that our results for YbPtBi appear to
be in discord with the suggested effect of dimensionality alone on the placement of the material in the global phase diagram \cite{Coleman2012, Custers2012}.

\section{Summary and Conclusion}

The $H-T$ phase diagram of YbPtBi has been constructed by low temperature thermodynamic and transport measurements. In zero field the strength of the anomaly developed in $\rho(T)$
below $T_{N}$ is sensitive to the strain, but the relevant physics of the field tuned quantum criticality remains the same for magnetic field applied along
\textbf{H}\,$\parallel$\,[100] up to 140\,kOe. The AFM order can be suppressed to $T$ = 0 by external magnetic field of $H_{c}\,\leq$ 4\,kOe and the temperature dependence of the
resistivity indicates the recovery of the FL state (clearly) for $H\,\geq$ 8\,kOe. The two well separated crossover scales, $T_{SR}$ and $T^{*}$, have been found, where these
crossover lines show a tendency of converging toward to $H_{c}$\,$\sim$\,4\,kOe and $H^{*}$\,$\sim$\,7.8\,kOe in the zero temperature limit. Although no clear nFL behavior is observed
in the specific heat measurements in the vicinity of the critical field, the electrical resistivity shows anomalous temperature dependence, $\rho(T)$ $\propto$ $T^{1.5}$, as a
signature of nFL behavior, between these two crossovers and $S(T)/T$ exhibits a logarithmic temperature dependence for $H\,<\,H_{c}$ above the AFM ordering temperature. The observed
$\gamma$ is finite below $H$\,$\sim$ 8\,kOe and the quasi-particle scattering cross-section, $A$, indicates a power law divergence as $A\,\propto\,1/(H-H_{c})$ upon approaching the
critical field from paramagnetic state. As magnetic field decrease from higher field side the power law dependence of both $A$ and $\gamma$ show a disruption below $H^{*} \sim$
8\,kOe. The constructed $H-T$ phase diagram and the details of the quantum criticality in YbPtBi turn out to be complicated.

\begin{acknowledgments}
SLB, PCC, and EDM thank Qimiao Si for critical reading of the manuscript and useful comments. Work performed at the Ames Laboratory was supported by the U.S. Department of Energy,
Office of Basic Energy Science, Division of Materials Sciences and Engineering. Ames Laboratory is operated for the U.S. Department of Energy by Iowa State University under Contract
No. DE-AC02-07CH11358. Work at the National High Magnetic Field Laboratory is supported by NSF Cooperative Agreement No. DMR-0654118 and by the State of Florida. Work at Occidental
College was supported by the National Science Foundation under DMR-1006118.
\end{acknowledgments}

\clearpage

\clearpage

\begin{figure}%
\centering
\includegraphics[width=0.5\linewidth]{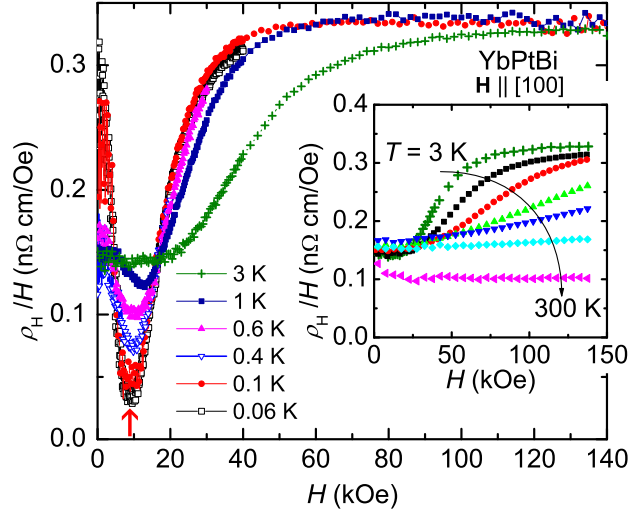}
\caption[Magnetic field dependence of the Hall coefficient for YbPtBi]{Hall coefficient, $R_{H}$ = $\rho_{H}/H$, of YbPtBi as a function of magnetic field, applied along the [100]
direction, at various temperatures. The arrow near 8\,kOe indicates the position of local minimum shown in $R_{H}$ at $T$ = 0.06\,K. Inset shows the high temperature $R_{H}$ for $T$ =
3, 5, 10, 25, 50, 100, and 300\,K (top to bottom).}
\label{YbPtBiHallH1}%
\end{figure}%

\clearpage

\clearpage

\begin{thebibliography}{99}


\bibitem{Canfield1991}
P. C. Canfield, J. D. Thompson, W. P. Beyermann, A. Lacerda, M. F. Hundley, E. Peterson, Z. Fisk, and H. R. Ott, J. Appl. Phys. {\bf 70}, 5800 (1991).


\bibitem{Fisk1991}
Z. Fisk, P. C. Canfield, W. P. Beyermann, J. D. Thompson, M. F. Hundley, H. R. Ott, E. Felder, M. B. Maple, M. A. Lopez de la Torre, P. Visani,
and C. L. Seaman, Phys. Rev. Lett. {\bf 67}, 3310 (1991).


\bibitem{Movshovich1994}
R. Movshovich, A. Lacerda, P. C. Canfield, J. D. Thompson, and Z. Fisk, Phys. Rev. Lett. {\bf 73}, 492 (1994).


\bibitem{Amato1992}
A. Amato, P. C. Canfield, R. Feyerherm, Z. Fisk, F. N. Gygax, R. H. Heffner, D. E. MacLaughlin, H. R. Ott, A. Schenck, J. D. Thompson, Phys. Rev. B {\bf 46}, 3151 (1992).


\bibitem{Robinson1994}
R. A. Robinson, A. Purwanto, M. Kohgi, P. C. Canfield, T. Kamiyama, T. Ishigaki, J. W. Lynn, R. Erwin, E. Peterson, R. Movshovich, Phys. Rev. B
{\bf 50}, 9595 (1994).


\bibitem{Hundley1997}
M. F. Hundley, J. D. Thompson, P. C. Canfield, and Z. Fisk, Phys. Rev. B {\bf 56}, 8098 (1997).


\bibitem{Hertz1976}
J. A. Hertz, Phys. Rev. B {\bf 14}, 1165 (1976).


\bibitem{Millis1993}
A. J. Millis, Phys. Rev.B {\bf 48}, 7183 (1993).


\bibitem{Moriya1995}
T. Moriya and T. Takimoto, J. Phys. Soc. Jpn. {\bf 64}, 960 (1995).


\bibitem{Gegenwart2008}
P. Gegenwart, Q. Si and F. Steglich, Nature Physics {\bf4}, 186 (2008).



\bibitem{Coleman2001}
P. Coleman, C. P\'{e}pin, Q. Si, and R. Ramazashvili, J. Phys. Condens. Matter {\bf 13}, R723 (2001).


\bibitem{Si2001}
Q. Si, S. Rabello, K. Ingersent, and J. L. Smith, Nature (London) {\bf 413}, 804 (2001).


\bibitem{Si2003}
Q. Si, S. Rabello, K. Ingersent, and J. L. Smith, Phys. Rev. B {\bf 68}, 115103 (2003).


\bibitem{Senthil2004}
T. Senthil, M. Vojta, and S. Sachdev, Phys. Rev. B {\bf 69}, 035111 (2004).


\bibitem{Paul2008}
I. Paul, C. P\'{e}pin, and M. R. Norman, Phys. Rev. B {\bf 78}, 035109 (2008).


\bibitem{Gegenwart1998}
P. Gegenwart, C. Langhammer, C. Geibel, R. Helfrich, M. Lang, G. Sparn, F. Steglich, R. Horn, L. Donnevert, A. Link, and W. Assmus, Phys. Rev.
Lett. {\bf 81}, 1501 (1998).


\bibitem{Julian1996}
S.R. Julian, C. Pfleiderer, F. M. Grosche, N. D. Mathur, G. J. McMullan, A. J. Diver, I. R. Walker, and G. G. Lonzarich, J. Phys.: Condens. Matter
{\bf 8}, 9675 (1996).


\bibitem{Schroder1998}
A. Schr\"{o}der, G. Aeppli, E. Bucher, R. Ramazashvili, and P. Coleman, Phys. Rev. Lett. {\bf 80}, 5623 (1998).


\bibitem{Schroder2000}
A. Schr\"{o}der, G. Aeppli, R. Coldea, M. Adams, O. Stockert, H. v. L\"{o}hneysen, E. Bucher, R. Ramazashvili, and P. Coleman, Nature (London)
{\bf 407}, 351 (2000).


\bibitem{Paschen2004}
S. Paschen, T. L\"{u}hmann, S. Wirth, P. Gegenwart, O. Trovarelli, C. Geibel, F. Steglich, P. Coleman, and Q. Si, Nature {\bf 432}, 881 (2004).


\bibitem{Trovarelli2000}
O. Trovarelli, C. Geibel, S. Mederle, C. Langhammer, F. M. Grosche, P. Gegenwart, M. Lang, G. Sparn, and F. Steglich, Phys. Rev. Lett. {\bf85},
626 (2000).


\bibitem{Gegenwart2002}
P. Gegenwart, J. Custers, C. Geibel, K. Neumaier, T. Tayama, K. Tenya, O. Trovarelli, and F. Steglich, Phys. Rev. Lett. {\bf 89}, 056402 (2002).


\bibitem{Gegenwart2007}
P. Gegenwart, T. Westerkamp, C. Krellner, Y. Tokiwa, S. Paschen, C. Geibel, F. Steglich, E. Abrahams, and Q. Si, Science {\bf 315}, 969 (2007).


\bibitem{Friedemann2009}
S. Friedemann, T. Westerkamp, M. Brando, N. Oeschler, S. Wirth, P. Gegenwart, C. Krellner, C. Geibel, and F. Steglich, Nature Phys. {\bf5}, 465
(2009).


\bibitem{Budko2004}
S. L. Bud'ko, E. Morosan, and P. C. Canfield, Phys. Rev. B {\bf69}, 014415 (2004).


\bibitem{Budko2005}
S. L. Bud'ko, E. Morosan, and P. C. Canfield, Phys. Rev. B {\bf71}, 054408 (2005).


\bibitem{Budko2005a}
S. L. Bud'ko, V. Zapf, E. Morosan, and P. C. Canfield, Phys. Rev. B {\bf72}, 172413 (2005).


\bibitem{Niklowitz2006}
P. G. Niklowitz, G. Knebel, J. Flouquet, S. L. Bud'ko, and P. C. Canfield, Phys. Rev. B {\bf73}, 125101 (2006).


\bibitem{Tokiwa2006}
Y. Tokiwa, A. Pikul, P. Gegenwart, F. Steglich, S. L. Bud'ko, and P. C. Canfield, Phys. Rev. B {\bf73}, 094435 (2006).


\bibitem{Schmiedeshoff2011}
G. M. Schmiedeshoff, E. D. Mun, A. W. Lounsbury, S. J. Tracy, E. C. Palm, S. T. Hannahs, J.-H. Park, T. P. Murphy, S. L. Bud'ko, and P. C. Canfield, Phys. Rev. B {\bf 83}, 180408
(2011).


\bibitem{Mun2010b}
Eundeok Mun, Sergey L. Bud'ko, and Paul C. Canfield, Phys. Rev. B {\bf 82}, 174403 (2010).


\bibitem{Custers2010}
J. Custers, P. Gegenwart, C. Geibel, F. Steglich, P. Coleman, and S. Paschen, Phys. Rev. Lett. {\bf 104}, 186402 (2010).


\bibitem{Si2006}
Qimiao Si, Physica B {\bf 378-380}, 23 (2006).


\bibitem{Si2010}
Qimiao Si, Phys. Status Solidi B {\bf 247}, 476 (2010).


\bibitem{Coleman2010}
P. Coleman, Phys. Status Solidi B {\bf 247}, 506 (2010).


\bibitem{Doniach1977}
S. Doniach, Physica B {\bf 91}, 231 (1977).


\bibitem{Canfield1992}
P. C. Canfield and Z. Fisk, Philos. Mag. B {\bf 65}, 1117 (1992).


\bibitem{Canfield2010}
P. C. Canfield, \textit{Solution growth of intermetallic single crystals}: \textit{a beginner's guide} (Book Series on Complex Metallic Alloys 2010, 93-111, World Scientific
Publishing Co. Pte. Ltd.).


\bibitem{Schmiedeshoff2006}
G. M. Schmiedeshoff, A. W. Lounsbury, D. J. Luna, S. J. Tracy, A. J. Schramm, S. W. Tozer, V. F. Correa, S. T. Hannahs, T. P. Murphy, E. C. Palm, A. H. Lacerda, S. L. Bud'ko, P. C.
Canfield, J. L. Smith, J. C. Lashley, and J. C. Cooley, Rev. Sci. Instrum. {\bf 77}, 123907 (2006).


\bibitem{Mun2010}
E. D. Mun, S. L. Bud'ko, M. S. Torikachvili, and P. C. Canfield, Meas. Sci. Technol. {\bf 21}, 055104 (2010).


\bibitem{Lacerda1993}
A. Lacerda, R. Movshovich, M. F. Hundley, P. C. Canfield, D. Arms, G. Sparn, J. D. Thompson, Z. Fisk, R. A. Fisher, N. E. Phillips, and H.-R. Ott, J. Appl. Phys. {\bf 73}, 5415 (1993).


\bibitem{Myers1999}
K. D. Myers, S. L. Bud'ko, I. R. Fisher, Z. Islam, H. Kleinke, A. H. Lacerda, and P. C. Canfield, J. Magn. Magn. Mater. {\bf 205}, 27 (1999).


\bibitem{Fawcett1988}
E. Fawcett, H. L. Alberts, V. Yu. Galkin, D. R. Noakes, and J. V. Yakhmi, Rev. Mod. Phys. {\bf60}, 209 (1988).


\bibitem{Elliott1972}
R. J. Elliott, \textit{Magnetic Properties of Rare Earth Metals} (Plenum press, London and New York, 1972).


\bibitem{McWhan1967}
D. B. McWhan and T. M. Rice, Phys. Rev. Lett. {\bf 19}, 846 (1967).

\bibitem{Flouquet1988}
J. Flouquet, P. Haen, F. Lapierre, C. Fierz, A. Amato, and D. Jaccard, J. Magn. Magn. Mat. {\bf 76-77}, 285 (1988).

\bibitem{Jaccard1998}
D. Jaccard, E. Vargoz, K. Alami-Yadri, and H. Wilhelm, Rev. High Pressure Sci. Technol. {\bf 7} 412 (1998).


\bibitem{Stewart2001}
G. R. Stewart, Rev. Mod. Phys. {\bf 73}, 797 (2001). and Rev. Mod. Phys. {\bf 78}, 743 (2006).


\bibitem{Torikachvili2007}
M. S. Torikachvili, S. Jia, E. D. Mun, S. T. Hannahs, R. C. Black, W. K. Neils, D. Martien, S. L. Bud'ko, and P. C. Canfield, Proc. Natl. Acad. Sci. U.S.A. {\bf 104}, 9960 (2007).


\bibitem{Budko2008}
S. L. Bud'ko, J. C. Frederick, E. D. Mun, P. C. Canfield, and G. M. Schmiedeshoff, J. Phys.: Condens. Matter {\bf 20}, 025220 (2008).


\bibitem{Schmiedeshoff2009}
G. M. Schmiedeshoff, A. W. Lounsbury, D. J. Luna, E. W. Okraku, S. J. Tracy, S. L. Bud'ko, and P. C. Canfield, J. Phys.: Conf. Ser. {\bf 150}, 042177  (2009).


\bibitem{Ziman1960}
J. M. Ziman, \textit{Electrons and Phonons} (Clarendon Press, Oxford, 1960).


\bibitem{Morelli1996}
D. T. Morelli, P. C. Canfield, and P. Drymiotis, Phys. Rev. B {\bf 53}, 12896 (1996).


\bibitem{Jung2001}
M. H. Jung, T. Yoshino, S. Kawasaki, T. Pietrus, Y. Bando, T. Suemitsu, M. Sera, and T. Takabatake, J. Appl. Phys. {\bf 89}, 7631 (2001).


\bibitem{Schoenes1987}
J. Schoenes, C. Sch\"{o}nenberger, J. J. M. Franse, and A. A. Menovsky, Phys. Rev. B {\bf 35}, 5375 (1987).


\bibitem{Hewson1993}
A. C. Hewson, \textit{The Kondo Problem to Heavy Fermions} (Cambridge: Cambridge University Press, 1993)


\bibitem{Maekawa1986}
S. Maekawa, S. Kashiba, M. Tachiki, and S. Takahashi, J. Phys. Soc. Jpn. {\bf55}, 3194 (1986).


\bibitem{Bickers1987}
N. E. Bickers, D. L. Cox, and J. W. Wilkins, Phys. Rev. B {\bf36}, 2036 (1987).


\bibitem{Ocko2004}
M. O\v{c}ko, J.L. Sarrao, and \v{Z}. \v{S}imek, J. Magn. Magne. Mater. {\bf284}, 43 (2004).


\bibitem{Kohler2008}
U. K\"{o}hler, N. Oeschler, F. Steglich, S. Maquilon and Z. Fisk, Phys. Rev. B {\bf 77}, 104412 (2008).


\bibitem{Hartmann2010}
S. Hartmann, N. Oeschler, C. Krellner, C. Geibel, S. Paschen, F. Steglich, Phys. Rev. Lett. {\bf104}, 096401 (2010).


\bibitem{Montgomery1971}
H. C. Montgomery, J. Appl. Phys. {\bf 42}, 2971 (1971).


\bibitem{Bastow1966}
T. J. Bastow and R. Street, Phys. Rev. {\bf141}, 510 (1966).


\bibitem{Maple1986}
M. B. Maple, J. W. Chen, Y. Dalichaouch, T. Kohara, C. Rossel, M. S. Torikachvili, M. W. McElfresh, and J. D. Thompson, Phys. Rev. Lett. {\bf 56}, 185 (1986).


\bibitem{Oh2007}
Y. S. Oh, K. H. Kim, P. A. Sharma, N. Harrison, H. Amitsuka, and J. A. Mydosh, Phys. Rev. Lett. {\bf 98}, 016401 (2007).


\bibitem{Lohneysen2007}
H. v. L\"{o}hneysen, A. Rosch, M. Vojta, and P. W\"{o}lfle, Rev. Mod. Phys. {\bf 79}, 1015 (2007).


\bibitem{Paglione2003}
J. Paglione, M. A. Tanatar, D. G. Hawthorn, E. Boaknin, R. W. Hill, F. Ronning, M. Sutherland, L. Taillefer, C. Petrovic, and P. C. Canfield, Phys. Rev. Lett. {\bf 91}, 246405 (2003).


\bibitem{Balicas2005}
L. Balicas, S. Nakatsuji, H. Lee, P. Schlottmann, T. P. Murphy, and Z. Fisk, Phys. Rev. B {\bf 72}, 064422 (2005).


\bibitem{Kadowaki1986}
K. Kadowaki and S. B. Woods, Solid State Commun. {\bf 58}, 507 (1986).


\bibitem{Custers2003}
J. Custers, P. Gegenwart, H. Wilhelm, K. Neumaier, Y. Tokiwa, O. Trovarelli, C. Geibel, F. Steglich, C. P\'{e}pin, and P. Coleman, Nature (London) {\bf 424}, 524 (2003).


\bibitem{Tsujii2005}
N. Tsujii, H. Kontani, and K. Yoshimura, Phys. Rev. Lett. {\bf 94}, 057201 (2005).


\bibitem{Movshovich1994a}
R. Movshovich, A. Lacerda, P. C. Canfield, J. D. Thompson, and Z. Fisk, J. Appl. Phys. {\bf 76}, 15 (1994).


\bibitem{Lea1962}
K. R. Lea, M. J. M. Leask, and W. P. Wolf. J. Phys. Chem. Solids {\bf 23}, 1381 (1962).


\bibitem{Jacko2009}
A. C. Jacko, J. O. Fj\ae restad, and B. J. Powell, Nature Physics {\b 5}, 422 (2009).


\bibitem{Mun2010a}
Eundeok Mun, Ph. D. thesis, Iowa State University (2010).


\bibitem{Behnia2004}
K. Behnia, D. Jaccard, and J. Flouquet, J. Phys.: Condens. Matter {\bf 16}, 5187 (2004).


\bibitem{Terashima2002}
T. Terashima, C. Terakura, S. Uji, H. Aoki, Y. Echizen and T. Takabatake, Phys. Rev. B {\bf 66}, 075127 (2002).


\bibitem{Paul2001}
I. Paul and G. Kotliar, Phys. Rev. B {\bf 64}, 184414 (2001).


\bibitem{Kim2010}
K.-S. Kim and C. P\'{e}pin, Phys. Rev. B {\bf 81}, 205108 (2010).


\bibitem{Coleman2012}
Piers Coleman, Nature Materials {\bf 11}, 185 (2012).


\bibitem{Custers2012}
J. Custers, K-A. Lorenzer, M. M\"{u}ller, A. Prokofiev, A. Sidorenko, H. Winkler, A. M. Strydom, Y. Shimura, T. Sakakibara, R. Yu, Q. Si, and S. Paschen, Nature Materials {\bf 11},
189 (2012).


\end{thebibliography}
\end{document}